\documentclass{article}     

\usepackage{etoolbox}
\newbool{arxiv}

\booltrue{arxiv}   

\ifbool{arxiv}{
    \usepackage[margin=1.8in]{geometry}
    \usepackage[authoryear]{natbib}
}

\usepackage{appendix}

\usepackage{amsthm}
\usepackage{amsmath}
\usepackage{natbib}
\usepackage[colorlinks,citecolor=blue,urlcolor=blue,filecolor=blue,backref=page]{hyperref}
\usepackage{graphicx}

\usepackage{microtype}
\usepackage{graphicx}
\usepackage{subfigure}
\usepackage{booktabs} 
\usepackage{siunitx}
\usepackage{hyperref}
\usepackage{xargs}[2008/03/08]
\usepackage{xfrac}

\usepackage{blindtext}
\usepackage[inline]{enumitem}

\usepackage{refstyle}
\usepackage{varioref} 

\usepackage{amssymb}
\usepackage{amsfonts}
\usepackage{mathtools}
\usepackage{colonequals}
\usepackage{algpseudocode, algorithm} 
\usepackage{xargs} 

\usepackage{mathrsfs}  

\usepackage{pdfpages} 

\usepackage[]{graphicx}
\usepackage[]{color}
\makeatletter
\def\maxwidth{ %
  \ifdim\Gin@nat@width>\linewidth
    \linewidth
  \else
    \Gin@nat@width
  \fi
}
\makeatother

\definecolor{fgcolor}{rgb}{0.345, 0.345, 0.345}

\usepackage{framed}
\makeatletter
 {\par\unskip\endMakeFramed%
 \at@end@of@kframe}
\makeatother

\definecolor{shadecolor}{rgb}{.97, .97, .97}
\definecolor{messagecolor}{rgb}{0, 0, 0}
\definecolor{warningcolor}{rgb}{1, 0, 1}
\definecolor{errorcolor}{rgb}{1, 0, 0}
\newenvironment{knitrout}{}{} 

\usepackage{alltt}

\notbool{arxiv}{
    \startlocaldefs
}

\def\ind#1{\mathbb{I}\left(#1\right)}
\def\evalat#1#2{\left.#1\right|_{#2}}
\def\fracat#1#2#3{\left.\frac{#1}{#2}\right\vert_{#3}}
\def\iid{\overset{iid}{\sim}}
\def\indep{\overset{indep}{\sim}}
\def\expect#1#2{\underset{#1}{\mathbb{E}}\left[#2\right]}

\def\abs#1{\left|#1\right|}
\def\norm#1{\left\Vert#1\right\Vert}
\def\norminf#1{\left\Vert#1\right\Vert_{\infty}}

\def\sumkm{\sum_{\k=1}^{\kmax - 1}}
\def\linop{\mathcal{L}} 

\DeclareMathOperator*{\argmax}{\mathrm{argmax}}
\DeclareMathOperator*{\argmin}{\mathrm{argmin}}
\DeclareMathOperator*{\esssup}{\mathrm{esssup}}

\def\etaopt{\hat\eta} 
\def\x{x}   
\def\t{t}   
\def\tp{t_{p}}   
\def\z{z}   
\def\g{g}   
\def\k{k}   
\def\n{n}   
\def\b{b}   
\def\l{l}   
\def\i{i}   
\def\d{d}   
\def\s{s}   
\def\nuk{\nu_{\k}}   
\def\const{C}   
\def\lnu{\tilde{\nu}}   
\def\lnuk{\tilde{\nu}_{\k}}   
\def\lnumean{\eta^{\mu}}   
\def\lnusd{\eta^{\sigma}}   
\def\hess#1{H_{#1}}   
\def\crosshessian{\hat J} 
\def\infl{\Psi}   

\def\etabeta{\eta_{\beta}}  
\def\etanu{\eta_{\nu}}  
\def\etanuk{\eta_{\nuk}}  
\def\etaz{\eta_{\z}}  
\def\etaglob{\eta_{\gamma}}  
\def\etalocal{\eta_{\ell}}  
\def\etaoptglob{\etaopt_{\gamma}}  
\def\etaoptlocal{\etaopt_{\ell}}  
\def\etaoptz{\etaopt_{\z}}  
\def\hessopt{\hat{H}}  

\def\p{\mathcal{P}}   
\def\q{\mathcal{Q}}   
\def\qk{\q_k}   
\def\ptil{\tilde{\p}}   
\def\qtil{\tilde{\q}}   
\def\pstick{\p_{\mathrm{stick}}} 
\def\pbetaprior{\p_{\mathrm{base}}}   
\def\lqgrad#1{{\nabla_\eta \log \qtil}\left(#1\right)}   
\def\lqgradbar#1{\mathcal{S}({#1})}
\def\lqgradbark#1{\mathcal{S}_k({#1})}
\def\pbase{\p_{0}}   
\def\palt{\p_{1}}   
\def\pbasetil{\ptil_{0}}   
\def\palttil{\ptil_{1}}   

\def\normdist#1{\mathcal{N}\left(#1\right)}   
\def\gem{\mathrm{GEM}(\alpha)}
\def\KL#1{\mathrm{KL}\left(#1\right)}   
\def\normalwishart#1{\mathcal{N}\mathcal{W}\left(#1\right)}   
\def\betadist#1{\mathrm{Beta}\left(#1\right)}   
\def\KLglobal{\mathrm{KL}_{\mathrm{glob}}}   

\def\etalin{\etaopt^{\mathrm{lin}}}
\def\etalinglobal{\etaopt_\gamma^{\mathrm{lin}}}
\def\glin{\g^{\mathrm{lin}}}

\def\N{N}   
\def\K{K}   
\def\kmax{{\K_{\mathrm{max}}}}   
\def\etadim{{D_{\eta}}}     
\def\betadim{{D_{\beta}}}   
\def\ngh{N_{\mathrm{GH}}}   
\def\thetadim{{D_{\theta}}} 

\def\etadom{\Omega_{\eta}}
\def\thetadom{\Omega_{\theta}}

\def\betadom{\Omega_{\beta}}

\def\linf{{L_{\infty}}}
\def\ball{\mathcal{B}}
\def\ballclosed{\overline{\ball}}

\def\mathtxt#1{\quad\textrm{#1}\quad}%
\def\mathand{\quad\textrm{and}\quad}%
\def\mathwhere{\quad\textrm{where}\quad}%
\def\constdesc#1{\textrm{(}\const\textrm{ does not depend on }#1\textrm{)}}
\def\assuitemref#1#2{\assuref{#1} (\itemref{#2})}%

\newcommand{\gclusters}{\g_{\text{cl},\tau}} 
\newcommand{\gclustersabbr}{\g_{\text{cl}}} 
\newcommand{\gclusterspredabbr}{\g_{\text{pred,cl}}} 
\newcommand{\gcoclustering}{\g_{\text{cc}}} 
\newcommand{\laplacianevsum}{\g_{\text{ev}}} 
\newcommand{\gloci}{\g_{\text{loci}}} 
\newcommand{\gadmix}{\g_{\text{admix}}} 
\newcommand{\nclusters}{G_{\text{cl}}} 

\newcommand{\ngenes}{N} 
\newcommand{\ntimepoints}{M} 
\newcommand{\regmatrix}{A}
\newcommand{\timeindx}{T} 

\newcommand{\nindiv}{N} 
\newcommand{\nloci}{L} 
\newcommand{\latentpop}{\beta} 
\newcommand{\latentadmix}{\pi} 
\def\categoricaldist#1{\mathrm{Categorical}\left(#1\right)}
\newcommand{\phiworstcase}{\phi_{\mathrm{wc}}} 

\theoremstyle{plain}
\newtheorem{lem}{Lemma}
\newtheorem{thm}{Theorem}

\newtheorem{assu}{Assumption}
\newtheorem{cor}{Corollary}

\theoremstyle{definition}
\newtheorem{examplex}{Example}
\newenvironment{ex}
  {\pushQED{\qed}\examplex}
  {\popQED\endexamplex}

\theoremstyle{definition}
\newtheorem{defnx}{Definition}
\newenvironment{defn}
    {\pushQED{\qed}\defnx}
    {\popQED\enddefnx}

\newcommand{\seeproof}[1]{(See \proofref{#1} \proofpageref[vref]{#1}.)}
\newcommand{\proofof}[1]{\noindent{\bf Proof of #1.}}

\newref{event}{
    name=Event~, %
    names=Events~, %
    Name=Event~,
    Names=Events~
    }

\newref{item}{
    name=Item~, %
    names=Items~, %
    Name=Item~,
    Names=Items~
    }

\newref{fig}{
    name=Figure~, %
    Name=Figure~
    }

\newref{tab}{
    name=Table~, %
    Name=Table~
    }

\newref{sec}{
    name=Section~, %
    Name=Section~,
    names=Sections~,
    Names=Sections~,
    }

\ifbool{arxiv}{
    \newref{app}{
        name=Appendix~, %
        Name=Appendix~,
        names=Appendices~,
        Names=Appendices~,
        }
} {
    \newref{app}{
        name=Supplement~, %
        Name=Supplement~,
        names=Supplements~,
        Names=Supplements~,
        }
}

\newref{eq}{
    name=Eq.~, %
    Name=Eq.~,
    names=Eqs.~, %
    Names=Eqs.~
    }

\newref{fig}{
    name=Figure~, %
    Name=Figure~,
    names=Figures~, %
    Names=Figures~,
    }

\newref{def}{
    name=Definition~, %
    Name=Definition~,
    names=Definitions~, %
    Names=Definitions~
    }

\newref{assu}{
    name=Assumption~, %
    Name=Assumption~,
    names=Assumptions~, %
    Names=Assumptions~,
    }

\newref{cond}{
    name=Condition~, %
    Name=Condition~,
    names=Conditions~, %
    Names=Conditions~
    }

\newref{prop}{
    name=Proposition~, %
    Name=Proposition~,
    names=Propositions~, %
    Names=Propositions~
    }

\newref{lem}{
    name=Lemma~, %
    Name=Lemma~,
    names=Lemmas~, %
    Names=Lemmas~
    }

\newref{ex}{
    name=Example~, %
    Name=Example~,
    names=Examples~,
    Names=Examples
    }

\newref{cory}{
    name=Corollary~, %
    Name=Corollary~,
    names=Corollaries~, %
    Names=Corollaries~
    }

\newref{thm}{
    name=Theorem~, %
    Name=Theorem~,
    names=Theorems~, %
    Names=Theorems~
    }

\newref{proof}{
    name=Proof~, %
    Name=Proof~
    }

\newref{conj}{
    name=Conjecture~, %
    Name=Conjecture~
    }

\newref{algr}{
    name=Algorithm~, %
    Name=Algorithm~,
    names=Algorithms~, %
    Names=Algorithms~,
    }

\notbool{arxiv}{
    \endlocaldefs
}

\notbool{arxiv}{
    \pubyear{2022}
    \volume{TBA}
    \issue{TBA}
    \firstpage{1}
    \lastpage{34}
}

\begin{document}


\newcommand{\SimPathologicalRTwoFig}{

\begin{knitrout}
\definecolor{shadecolor}{rgb}{0.969, 0.969, 0.969}\color{fgcolor}\begin{figure}[!h]

{\centering \includegraphics[width=0.980\linewidth,height=0.784\linewidth]{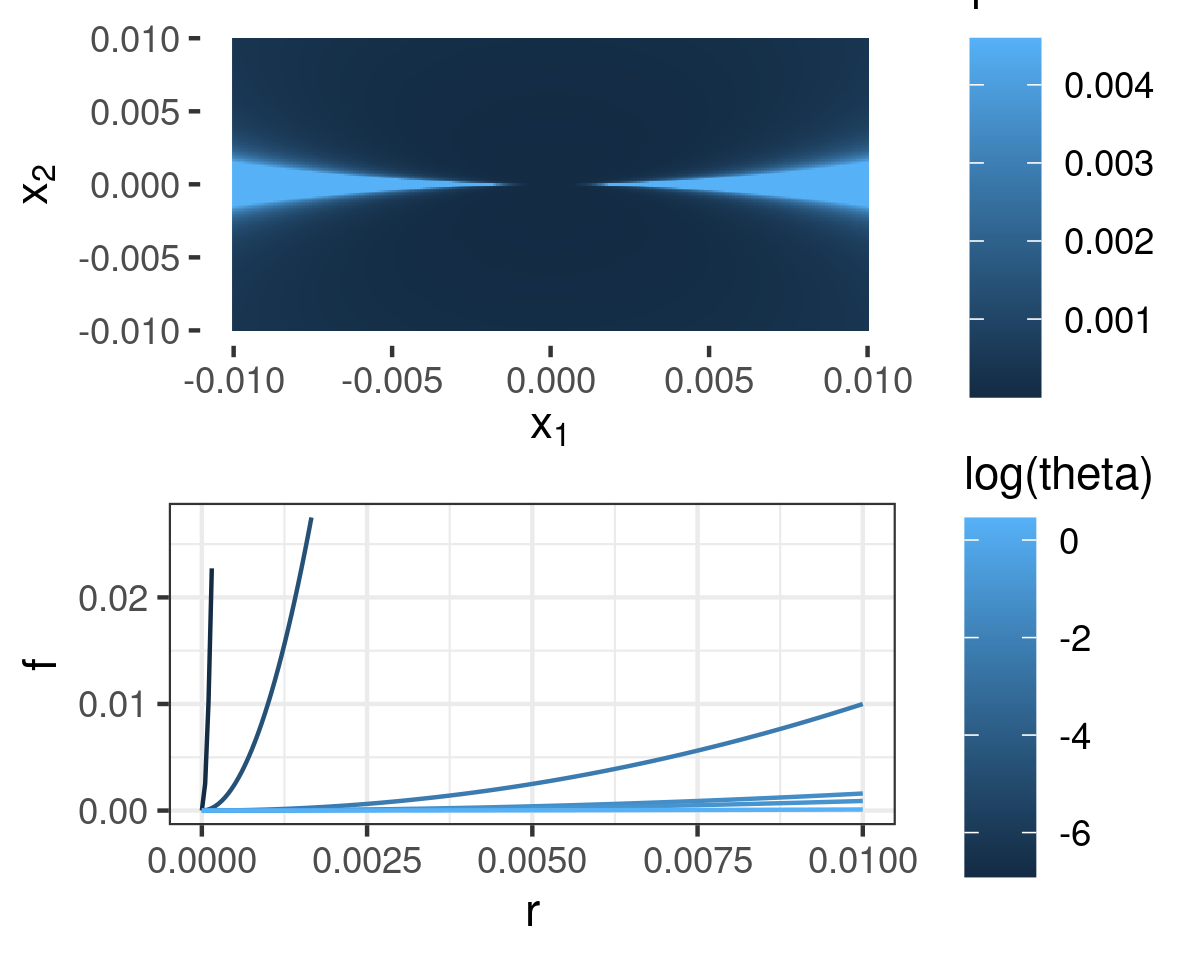} 

}

\caption[A plot of $f(x_1, x_2)$ from \exref{r2_pathological}]{A plot of $f(x_1, x_2)$ from \exref{r2_pathological}.}\label{fig:r2_pathological}
\end{figure}

\end{knitrout}
}

\newcommand{\SimPositivePertFig}{

\begin{knitrout}
\definecolor{shadecolor}{rgb}{0.969, 0.969, 0.969}\color{fgcolor}\begin{figure}[!h]

{\centering \includegraphics[width=0.980\linewidth,height=0.784\linewidth]{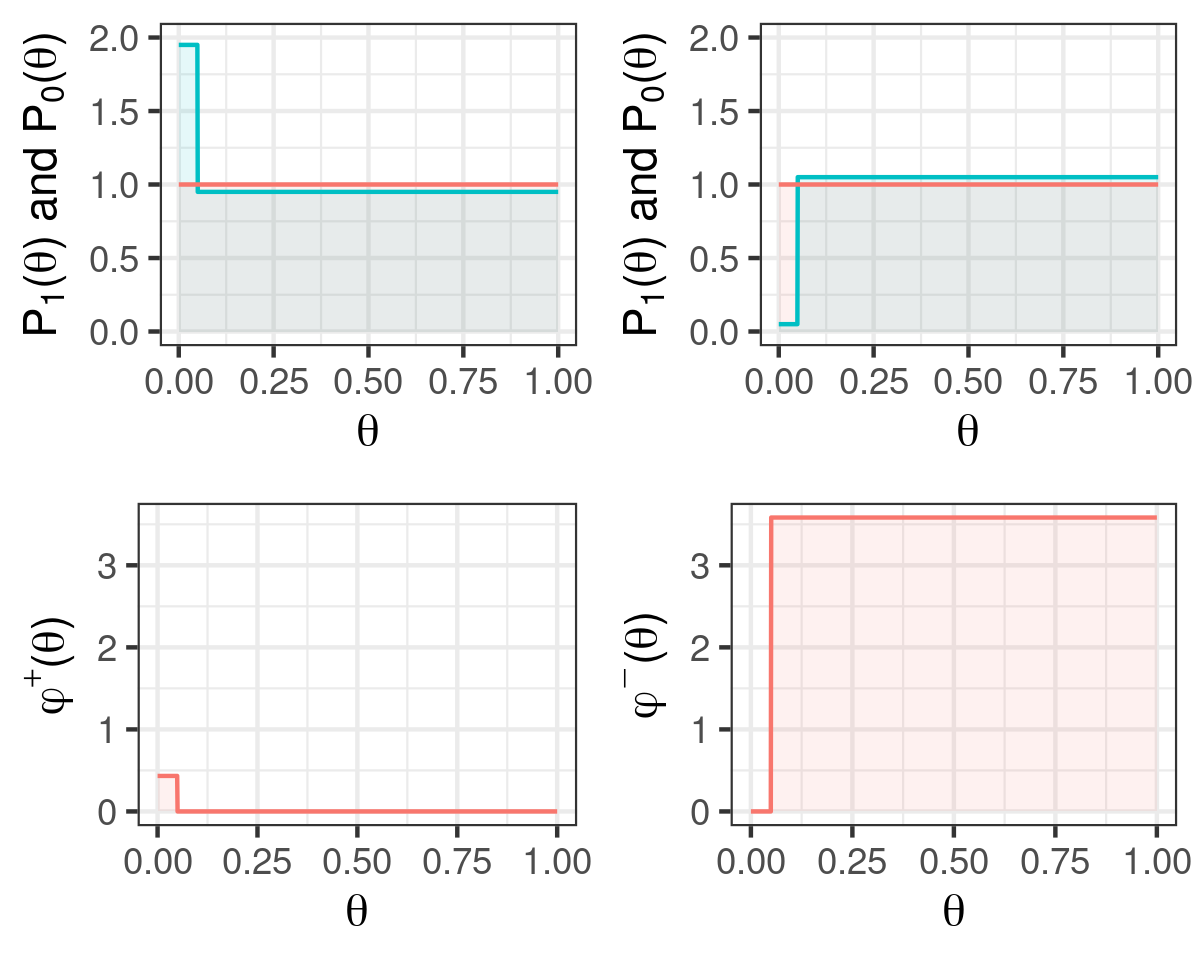} 

}

\caption{A plot of the perturbations from \exref{positive_pert_large} with $p=2$ and $\epsilon=0.05$.  Positive $\phi$ can only add mass, so to remove a small amount of mass requires adding mass everywhere else and re-normalizing, resulting in a large perturbation according to $\norm{\cdot}_p$.}\label{fig:positive_pert}
\end{figure}

\end{knitrout}
}

\newcommand{\FunctionPathsFig}{

\begin{knitrout}
\definecolor{shadecolor}{rgb}{0.969, 0.969, 0.969}\color{fgcolor}\begin{figure}[!h]

{\centering \includegraphics[width=0.980\linewidth,height=0.470\linewidth]{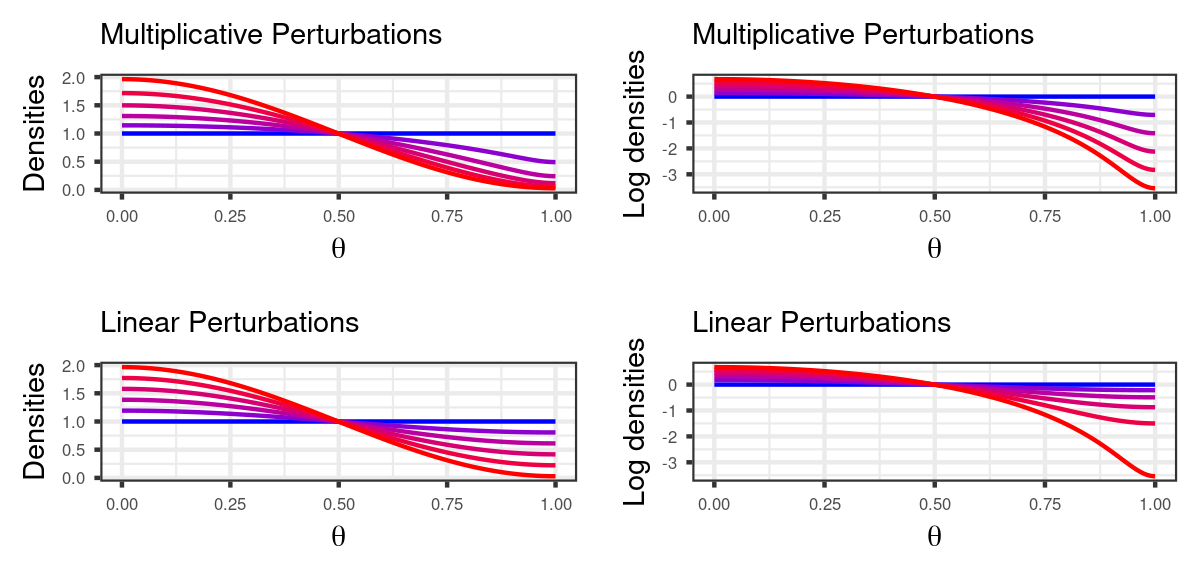} 

}

\caption[Multiplicative and linear mixture paths between two densities]{Multiplicative and linear mixture paths between two densities.}\label{fig:path}
\end{figure}

\end{knitrout}
}

\newcommand{\FunctionPathsMultFig}{

\begin{knitrout}
\definecolor{shadecolor}{rgb}{0.969, 0.969, 0.969}\color{fgcolor}\begin{figure}[!h]

{\centering \includegraphics[width=0.980\linewidth,height=0.274\linewidth]{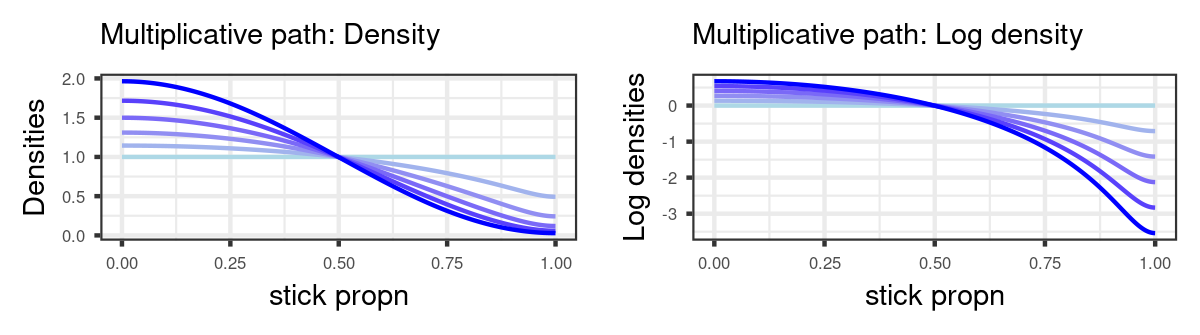} 

}

\caption[Multiplicative mixture paths between two densities]{Multiplicative mixture paths between two densities.}\label{fig:mult_path}
\end{figure}

\end{knitrout}
}

\newcommand{\FunctionPathsLinFig}{

\begin{knitrout}
\definecolor{shadecolor}{rgb}{0.969, 0.969, 0.969}\color{fgcolor}\begin{figure}[!h]

{\centering \includegraphics[width=0.980\linewidth,height=0.274\linewidth]{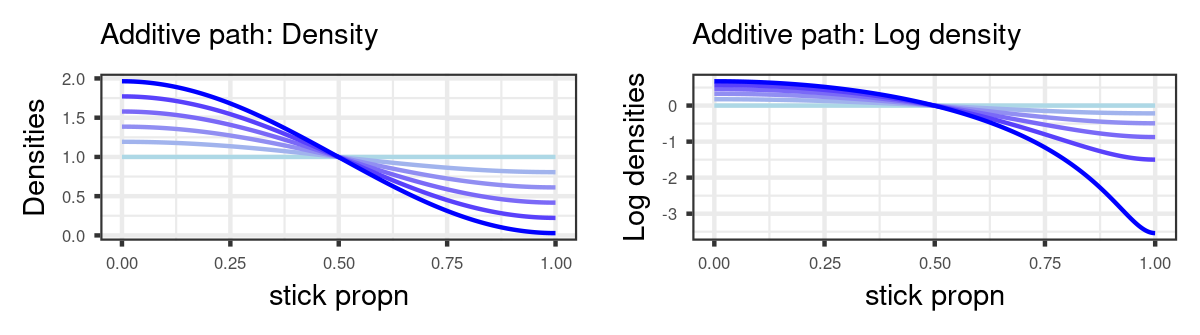} 

}

\caption[Linear mixture paths between two densities]{Linear mixture paths between two densities.}\label{fig:lin_path}
\end{figure}

\end{knitrout}
}

\newcommand{\FunctionBallFig}{

\begin{knitrout}
\definecolor{shadecolor}{rgb}{0.969, 0.969, 0.969}\color{fgcolor}\begin{figure}[!h]

{\centering \includegraphics[width=0.980\linewidth,height=0.274\linewidth]{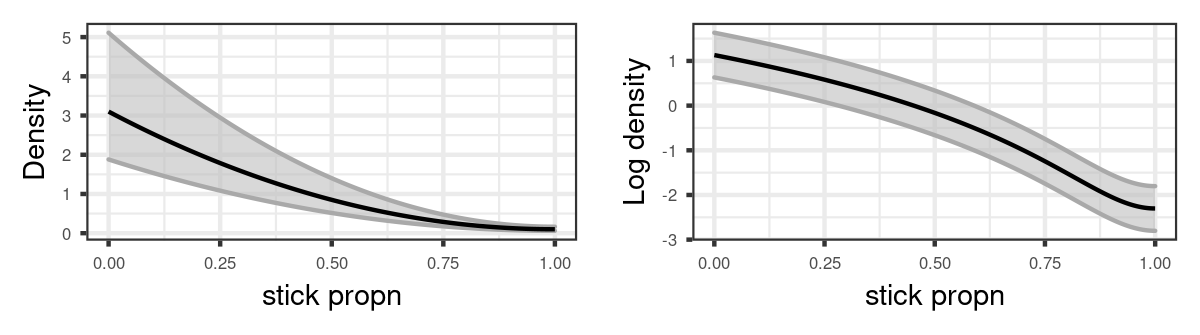} 

}

\caption[An multiplicative ball $\ball_\phi(\delta)$]{An multiplicative ball $\ball_\phi(\delta)$.}\label{fig:func_ball}
\end{figure}

\end{knitrout}
}

\newcommand{\FunctionDistFig}{

\begin{knitrout}
\definecolor{shadecolor}{rgb}{0.969, 0.969, 0.969}\color{fgcolor}\begin{figure}[!h]

{\centering \includegraphics[width=0.980\linewidth,height=0.784\linewidth]{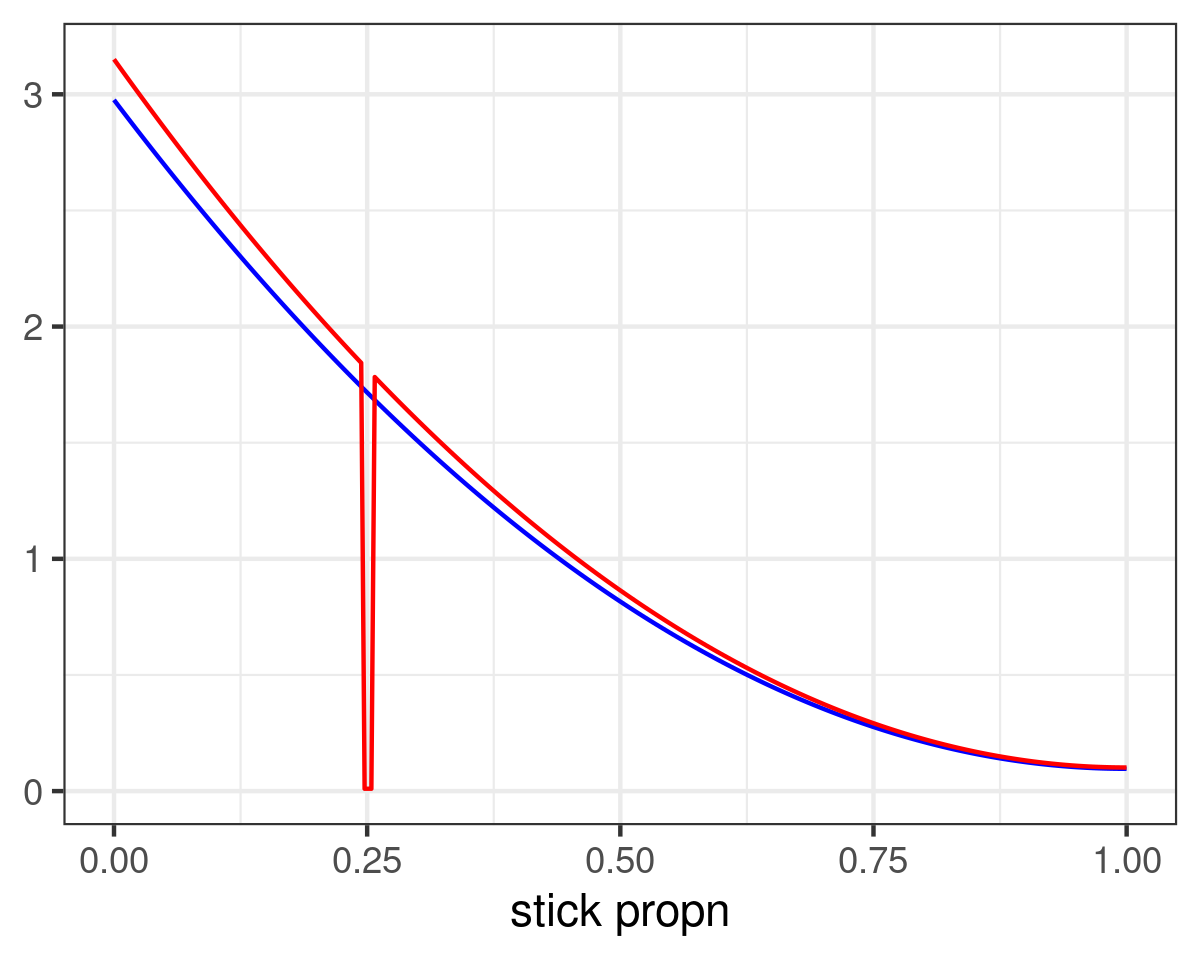} 

}

\caption[Two densities that are distant according to reverse KL divergence and $\norminf{\cdot}$ but close according to $\norm{\cdot}_p$ for $p \in [1, \infty)$]{Two densities that are distant according to reverse KL divergence and $\norminf{\cdot}$ but close according to $\norm{\cdot}_p$ for $p \in [1, \infty)$.}\label{fig:func_dist}
\end{figure}

\end{knitrout}
}

\newcommand{\LinfExamplesFig}{

\begin{knitrout}
\definecolor{shadecolor}{rgb}{0.969, 0.969, 0.969}\color{fgcolor}\begin{figure}[!h]

{\centering \includegraphics[width=0.980\linewidth,height=0.314\linewidth]{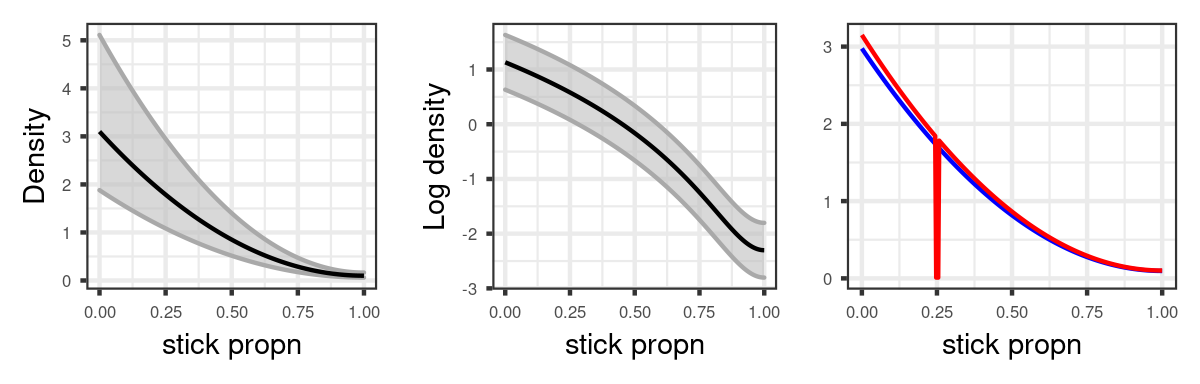} 

}

\caption[Left two]{Left two: A multiplicative ball $\ball_\phi(\delta)$.  Right: Two densities that are distant according to reverse KL divergence and $\norminf{\cdot}$ but close according to $\norm{\cdot}_p$ for $p \in [1, \infty)$.}\label{fig:linf_examples}
\end{figure}

\end{knitrout}
}

\title{Evaluating Sensitivity to the Stick-Breaking Prior in Bayesian Nonparametrics}

\ifbool{arxiv}{
    \author{
      \hspace{4em}
      Ryan Giordano \thanks{Equal contribution author}
      \thanks{Department of EECS, MIT}

      \and
      Runjing Liu \footnotemark[1]
      \thanks{Department of Statistics, UC Berkeley}

      \hspace{4em}
      \and
      Michael I.~Jordan \footnotemark[3]
      \and
      Tamara Broderick \footnotemark[2]
    }

    \maketitle

Bayesian models based on the Dirichlet process and other stick-breaking
priors have been proposed as core ingredients for clustering, topic
modeling, and other unsupervised learning tasks.
However, due to the flexibility of these models, the consequences of
prior choices can be opaque. And so prior specification can be
relatively difficult. At the same time, prior choice can have a
substantial effect on posterior inferences.
Thus, considerations of robustness need to go
hand in hand with nonparametric modeling.  In the current paper, we tackle
this challenge by exploiting the fact that variational Bayesian methods,
in addition to having computational advantages in fitting complex
nonparametric models, also yield sensitivities with respect to
parametric and nonparametric aspects of Bayesian models.  In particular,
we demonstrate how to assess the sensitivity of conclusions to the choice
of concentration parameter and stick-breaking distribution for inferences
under Dirichlet process mixtures and related mixture models.  We provide
both theoretical and empirical support for our variational approach to
Bayesian sensitivity analysis.

} {
    \runtitle{Evaluating Sensitivity to the Stick-Breaking Prior in BNP}
    \runauthor{Giordano, Liu, Jordan, and Broderick}

    \newcommand\authors{
        \begin{aug}
            \author{\fnms{Ryan} \snm{Giordano}\thanksref{addr1,t1}},
            \author{\fnms{Runjing} \snm{Liu}\thanksref{addr2,t1}},
            \author{\fnms{Michael I.} \snm{Jordan}\thanksref{addr2}},
            \and
            \author{\fnms{Tamara} \snm{Broderick}\thanksref{addr1}}

            \address[addr1]{Department of EECS, MIT,
            77 Massachusetts Ave., 38-401,
            Cambridge, MA 02139}

            \address[addr2]{Department of Statistics,
            367 Evans Hall, UC Berkeley,
            Berkeley, CA 94720}

            \thankstext{t1}{Equal contribution. }
        \end{aug}
    }

    \begin{frontmatter}
    \authors{}

    \begin{abstract}
    
Bayesian models based on the Dirichlet process and other stick-breaking
priors have been proposed as core ingredients for clustering, topic
modeling, and other unsupervised learning tasks.
However, due to the flexibility of these models, the consequences of
prior choices can be opaque. And so prior specification can be
relatively difficult. At the same time, prior choice can have a
substantial effect on posterior inferences.
Thus, considerations of robustness need to go
hand in hand with nonparametric modeling.  In the current paper, we tackle
this challenge by exploiting the fact that variational Bayesian methods,
in addition to having computational advantages in fitting complex
nonparametric models, also yield sensitivities with respect to
parametric and nonparametric aspects of Bayesian models.  In particular,
we demonstrate how to assess the sensitivity of conclusions to the choice
of concentration parameter and stick-breaking distribution for inferences
under Dirichlet process mixtures and related mixture models.  We provide
both theoretical and empirical support for our variational approach to
Bayesian sensitivity analysis.

    \end{abstract}


    \end{frontmatter}
}

\section{Introduction}\seclabel{introduction}
Scientists and engineers working in a wide range of fields are often interested
in inferring the number of clusters in a given data set, as well as inferring
which data points belong together. Such inferential questions can be posed
naturally within a Bayesian nonparametric (BNP) framework, building on tools
such as the Dirichlet process~\citep{ferguson:1973:bayesian,
sethuraman:1994:constructivedp}. The Dirichlet process has two useful attributes
that have made it be suggested as a natural model of clustering phenomena.
First, it is a combinatorial stochastic process, exhibiting discrete structure
that allows multiple data points to be associated with the same underlying value
of a parameter.  Second, its nonparametric nature means that the number of
unique parameter values generally grows with the size of the data set,
accommodating growth in the number of inferred clusters as data accrue.  Such
growth is appropriate in many real-world settings; for example, we might expect
to keep discovering new species as we examine more individual organisms, and we
might expect to discover more topics as we read more articles in a scientific
literature. Finally, the overall Bayesian framework in which the Dirichlet
process is embedded allows clustering to be treated as one aspect of a larger
inferential problem.  In particular, the Dirichlet process can be flexibly
incorporated into more complex models that exhibit other forms of structure,
including hierarchical, spatio-temporal, and topological structure.

Although the BNP framework offers flexibility, it is important to recognize that
it is not a black-box method. As with any Bayesian methodology, the deployment
of a BNP model involves choices of hyperparameters.  Often, these choices are
made for reasons of mathematical or computational convenience.  Indeed, the
nonparametric nature of BNP models can  make it particularly difficult to
express prior belief subjectively.  For example, the latent frequencies of
clusters provided by the Dirichlet process are obtained by recursively removing
beta-distributed fractions of probability mass from the unit interval. The use
of the beta distribution is motivated by its mathematical tractability under
recursion and by the fact that it yields a form of conditional conjugacy that
can be exploited by Gibbs sampling. These are appealing properties, but it is
difficult to imagine justifying this specific choice subjectively, particularly
given that observable consequences of the choice are indirect.  Even having
accepted the beta distribution as a choice of convenience, there remains the
problem of choosing the parameter $\alpha$ associated with this distribution.
The implications of this choice are again difficult to assess subjectively.  In
practice the choice is often made based on previous applications or by simply
employing a heuristic~\citep[Chapter 23]{teh:2006:hdp, gelman:2013:bda}.

In summary, it is important to recognize that there will exist many possible
values of $\alpha$, and many possible forms of stick-breaking prior, that might
correspond to one's prior beliefs, but which the Dirichlet process framework and
other complex BNP models bundle in a way that makes it difficult to understand
and to specify {\em a priori}. Choices of convenience are therefore made, and,
unfortunately, these choices can change the results of a data analysis. For
instance, $\alpha$ has a direct, proportional relationship to the number of
clusters obtained asymptotically in draws from the Dirichlet process. Thus the
number of clusters inferred at any particular data size may depend strongly on
$\alpha$.  If our scientific conclusions varied substantially because of such
dependence, we might worry that these conclusions were driven not by the data
and meaningful prior beliefs but instead by our arbitrary or default choices. It
behooves us, then, to check how sensitive our conclusions are to these choices.

The outputs of Bayesian inference arise not just from a model and collection of
data but also via the use of some posterior approximation. Accordingly, when we
assess sensitivity, we should assess the sensitivity of this full procedure to
our model choices. In the current paper we focus on Dirichlet process mixture
(DPM) models and Variational Bayesian (VB) posterior approximations based on
reverse Kullback-Leibler (KL) divergence. VB methods have several favorable
properties that motivate their use in the DPM setting. First, they exhibit fast
computational scaling due to their use of gradient-based optimization. Second,
they avoid the label-switching problem exhibited by MCMC in the mixture-model
setting~\citep{jasra:2005:mcmclabelswitch}. Third, their implementation has
become increasingly straightforward due to automatic differentiation
tools~\citep{ranganath:2013:black, kucukelbir:2016:advi}. Finally, and of
particular interest in the current paper, the variational formulation makes it
possible to compute closed-form derivative-based expansions of posterior
distributions as a function of model
hyperparameters~\citep{giordano:2018:covariances}. Thus VB provides a natural
pathway to quantifying the robustness of Bayesian inference.

Concretely, with a fully specified model and inference procedure in hand in the
setting of DPM models, we can ask how sensitive some quantity of interest is to
the choices of $\alpha$ and the stick-breaking distributions. One option is to
propose a number of potential $\alpha$ values, compute the variational
approximation at each $\alpha$ value, and report our quantity of interest for
each $\alpha$ value. We might similarly assess sensitivity to the stick-breaking
distribution over a range of distributional choices. There are at least two
major issues with this proposal: (1) while VB is a relatively fast form of
approximate Bayesian inference in general, it may still be prohibitively
expensive to have to re-run it many times, and (2) it is unclear how best to
choose a collection of $\alpha$ and (especially) the stick-breaking distribution
values---and how many to choose.

In this work, we address these challenges by making full use of the variational
nature of VB methodology.  We show how to approximate the nonlinear dependence
of the VB optimum on prior choices using a first-order Taylor series expansion.
We build on the local robustness tools developed by
\cite{giordano:2018:covariances} for VB and \cite{gustafson:1996:local} for the
exact posterior and MCMC approximations. To enable their application to DPM
models, we solve a number of open problems: (1) we establish that the optimal VB
parameters are a continuously differentiable function of $\alpha$ and a
particular parameterization of the stick-breaking form; (2) we show that the
sensitivity of the VB approximation to functional prior perturbations takes the
form of an integral against a computationally tractable \textit{influence
function}---and illustrate how the influence function can provide an
interpretable summary of the effect of arbitrary changes to the prior density;
(3) to justify using linear approximations over a ball describing different
stick-breaking densities, we show that our method is a \textit{uniformly} good
approximation by establishing Fr{\'e}chet differentiability; (4) we show how to
compute our approximation efficiently in high-dimensional problems; and (5) we
establish the accuracy, practicality, and computational efficiency of our
approximation for a variety of models that use stick-breaking, and for various
quantities of interest in both clustering and topic modeling.

Our ambition is not to draw conclusions concerning the robustness of BNP
procedures in general, nor even of the DPM model in particular.
Rather, we offer an easy-to-use computational tool to quickly and automatically
assess the sensitivity to prior specification of VB approximations in a
particular problem at hand. Though we focus on demonstrating the effectiveness
of our methods on the canonical DPM model, our methods apply immediately to any
discrete BNP model that admits a truncated stick-breaking approximation. Our
ambition, then, is to encourage and empower researchers to explore the
robustness of a wide array of datasets and models, including the DPM, but also
other stick-breaking variants.

Even further, despite the present paper's focus on BNP, we develop theory that
applies directly to all VB approximations based on reverse KL divergence.
Indeed, the formation and analysis of our approximation depends only on the
implicit function theorem, and so could be readily extended to VB approximations
based on other divergence measures. Thus, though our discussion and experiments
will focus on BNP applications, we hope that the present work can serve as a
template for the development and analysis of similar local robustness tools in
other popular applications of VB.

The remainder of the paper is organized as follows. We briefly review related
work in \secref{related_work}.  In \secref{model}, we review the stick-breaking
construction of the Dirichlet process and our chosen variational approximation.
In \secref{local_sensitivity}, we derive the form of local prior robustness
measures for VB approximations.  We consider functional perturbations to the
stick-breaking density in \secref{influence_function}, and define the influence
function from which we can construct influential and worst-case perturbations.
In \secref{computing_sensitivity}, we address scalability and other
computational considerations for computing local sensitivity on real
applications. In \secref{results}, we apply our tools to assess the sensitivity
of BNP models in several data analysis problems.

\section{Related Work}
\seclabel{related_work}
Evaluating sensitivity to prior choices is typically a desirable step of applied
Bayesian data analysis \citep[Chapter 6]{gelman:2013:bda}, and a central aim of
Bayesian robustness is to provide methods and metrics to measure sensitivity of
posterior quantities to variations in the model \citep{insua:2000:robust}. Our
approach to robustness quantification falls in the category of ``local
robustness'' techniques, which are based on differential approximations to model
sensitivity \citep{gustafson:2000:localrobustness}.  The contrasting set of
``global robustness'' techniques avoid differential approximation, but are
computationally expensive or infeasible in all but special cases
\citep{sivaganesan:2000:globallocal}.

In the present work, we study the robustness of a user's problem-specific
posterior quantities of interest, such as the expected number of distinct
clusters, or the membership of a particular cluster (as in, e.g.,
\citet{gustafson:1996:local}).  In contrast, other work attempts to measure the
sensitivity of the entire posterior using, for example, the Wasserstein
distance or the largest change within an expressive class of posterior
expectations (e.g., \citet{roos:2015:sensitivity, ghaderinezhad:2019:stein}). In
other words, we study the robustness of particular posterior conclusions rather
than attempting to measure the robustness, in some sense, of the entire
posterior.

%

Our focus on VB contrasts with much of the previous Bayesian local robustness
literature.  For posteriors that are approximated via MCMC, the derivatives of
local robustness must be approximated with potentially noisy sample covariances
(e.g., \citet{gustafson:1996:marginal}). In contrast, the VB optima that we
study admit closed-form derivatives via the implicit function theorem. As an
optimization procedure, the evaluation of the sensitivity of VB estimates
inherits a long tradition of robustness methods in frequentist statistics (e.g.
\citep{jaeckel:1972:infinitesimal, cook:1986:assessment, hampel:2011:robust}), a
connection which is explored in \citet{giordano:2018:covariances}. Our work
extends \citet{giordano:2018:covariances} by providing more easily verifiable
sufficient conditions for Theorem 2 of \citet{giordano:2018:covariances} and
proving results for nonparametric perturbations to the functional form of the
prior, including continuous Fr{\'e}chet differentiability (and
non-differentiability).  Our theoretical improvements on
\citet{giordano:2018:covariances}  apply to any VB approximation based on
reverse KL divergence, not only BNP models.  Ultimately, our theoretical work
amounts to an application of the implicit function theorem
\citep{krantz:2012:implicit}, and a similar approach to ours could yield
comparable results for VB approximations based on other divergences
(e.g, \citet{li:2016:reyni, liu:2016:stein, ambrogioni:2018:wasserstein}).

Many authors have considered the potential sensitivity of discrete BNP posterior
quantities to prior specification. Typically, such work relies either on the
existence of closed-form solutions or on running multiple MCMC chains with
different prior choices (e.g., \citet{barajas:2009:densitysens,
saha:2019:geometricsens}).  This work has shown that alternatives to the DPM
may exhibit improved robustness properties \citep{barrios:2013:bnp,
lijoi:2007:reinforcement, canale:2017:pitmanyor}.  In the present work, we take
the DPM as our starting point only in order to demonstrate our robustness
methodology on a well-known and canonical choice of BNP prior. We hope that our
methods could act as a supplement to the computationally or analytically
intensive techniques employed by the aforementioned papers to quantify
robustness.  Indeed, our techniques should apply directly to VB approximations
of any discrete BNP prior that admits a truncated approximation
\citep{doshi:2009:ibpvariational, roychowdhury:2015:gammastick,
campbell:2019:truncated}.

A final distinction between our work and much of the prior Bayesian local
robustness literature is underscored by comparison with
\citet{Basu:2000:robustnessBNP}, a work that also employs local robustness
(applied to MCMC) to measure sensitivity to the concentration parameter of a DPM
prior specification.  Unlike \citet{Basu:2000:robustnessBNP}, who considers the
norm of the derivative to be a measure of robustness {\em per se} (following,
for example, \citet{basu:1996:local, gustafson:1996:local}), we focus on the
ability of our linear approximation to {\em extrapolate} to alternative priors.
In this spirit, we hope that our work provides tools for quickly and
interactively exploring the space of subjectively reasonable prior alternatives,
without committing researchers to a single robustness measure chosen more for
mathematical convenience than intuitive validity.

\section{The Model and Variational Approximation}
\seclabel{model}

    \subsection{A stick-breaking model for clustering}
    \seclabel{model_bnp}
    Consider a standard Bayesian nonparametric generative model for clustering, with
observed data $\x = (\x_n)_{n=1}^{N}$. We assume a countable infinity of latent
components, with frequencies $\pi = (\pi_1, \pi_2, \ldots)$, such that $\pi_\k
\in [0,1]$ for all $k \in \{1, 2, \ldots\}$, and $\sum_{\k} \pi_\k = 1$. For the
$n$th data point, the vector $\z_\n = (\z_{\n1}, \z_{\n2}, \ldots)$ is an
indicator vector; $\z_{\n\k} = 1$ represents the assignment of the
$\n$th data point to the $\k$th component, with all other vector elements set
equal to zero. We generate $\z_{\n\k} = 1$ with probability $\pi_\k$, i.i.d.\
across $\n$. To generate the $\x_n$, we assume the $\k$th component is
characterized by a component-specific parameter, $\beta_\k \in \betadom
\subseteq \mathbb{R}^{\betadim}$, and that a data point arising from component
$k$ is generated as $\p(\x_n \vert \beta_\k)$. Then
$
\p(\x_n \vert \z_\n, \beta) =
    \prod_{\k=1}^\infty \p(\x_n \vert \beta_\k)^{\z_{\n\k}}.
$ The $\beta_\k$ in turn are generated i.i.d.\ from a prior
$\pbetaprior(\beta_\k)$. For instance, in a Gaussian mixture model, $\beta_\k$
could be a vector representing the mean and covariance of a Gaussian
distribution.

It remains to place a prior on the component frequencies $\pi$. We will focus on
stick-breaking priors for $\pi$, so we first replace $\pi$ with a stick-breaking
representation. Let $\nu = (\nu_1, \nu_2, \ldots)$ represent proportions: $\nuk
\in [0, 1]$. Take \begin{align}\eqlabel{stick_breaking}
\pi_\k := \nuk \prod_{\k' < \k} (1 - \nu_{\k'}).
\end{align}
We then define a stick-breaking prior by placing a prior on the $\nuk$. Fix a
density, $\pstick(\cdot)$, with respect to the Lebesgue measure on $[0,1]$ and
let $\nuk\iid\pstick(\nuk)$ for $\k \in \{1, 2, \ldots\}$. A common choice of
$\pstick$ is $\mathrm{Beta}(1, \alpha)$, with \emph{concentration parameter}
$\alpha > 0$. With this choice, the $\pi$ are distributed according to the
size-biased weights associated with the atoms of a draw from a Dirichlet
process. This particular beta stick-breaking prior is often favored due to its
convenient mathematical properties and ease of use in inference.

\noindent \textbf{Posterior quantities of interest.}
In theory, with our generative model and observed data in hand, we can find the
Bayesian posterior $\p(\beta, \z, \nu | \x)$ and report any posterior summaries
of interest. For instance, the posterior $\p(\beta, \z, \nu | \x)$ induces a
posterior distribution on the number of clusters $\nclusters(\z)$, where
\emph{clusters} are components to which at least one data point has been
assigned:
\begin{align*}
  \nclusters(\z) := \sum_{k=1}^{\infty} \ind{ \left(\sum_{n=1}^{N}
  \z_{\n\k}\right) > 0},
\end{align*}
where $\ind{\cdot}$ is the indicator function taking value $1$ when the argument
is true and $0$ otherwise.

In practice, though, neither the posterior nor the posterior summary is readily
accessed. An approximation must be used instead.

    \subsection{Variational approximation}
    \seclabel{model_vb}
    To assess the sensitivity of a procedure in practice, we need to consider the
approximate Bayesian inference algorithm used as well. Here we focus on a
variational Bayes approximation due to \citet{blei:2006:vi_for_dp}.

Variational Bayes (VB) posits a class of tractable distributions over the model
parameters and chooses the  element of this class that minimizes the reverse
Kullback-Leibler (KL) divergence to the exact posterior.  One approach to apply
VB to Dirichlet process stick-breaking models assumes $\nu_\kmax = 1$ for all
distributions in the variational class and some truncation level $\kmax$. Let
$\zeta$ collect the first $\kmax - 1$ elements of $\nu$, the first $\kmax$
elements of $\beta$, and the first $\kmax$ elements of $\z_\n$ across $n$. In
what follows, then, we effectively consider the reverse KL divergence to the
posterior marginal $\p(\zeta \vert \x)$. By setting $\kmax$ sufficiently large,
one can make this truncation as accurate as desired.

Mean-field VB is a particularly popular VB variant where the tractable
approximating distributions $\q$ factorize over the parameters. In our case,
then, we consider approximations of the form
\begin{align}\eqlabel{vb_mf}
\q(\zeta \vert \eta) =
    \left( \prod_{\k=1}^{\kmax - 1} \q(\nuk \vert \eta) \right)
    \left( \prod_{\k=1}^{\kmax} \q(\beta_\k \vert \eta) \right)
    \left( \prod_{\n=1}^{\N} \q(\z_{\n} \vert \eta) \right),
\end{align}
where $\eta \in \etadom \subseteq \mathbb{R}^{\etadim}$ represents
\emph{variational parameters} that determine the factors of the $\q$
distribution. When the observation likelihood $\p(\x_n \vert \beta_\k)$ is
conditionally conjugate with the component-parameter prior
$\pbetaprior(\beta_\k)$, no further assumptions are needed on the form of
$\q(\beta_\k \vert \eta)$; one can show that it will take the form of the
conjugate exponential family after the KL optimization
\citep{blei:2017:vi_review}. Similarly, when $\pstick$ is a beta distribution,
no further assumptions are needed on $\q(\nuk \vert \eta)$; it will take a beta
form. However, since we will consider non-beta forms of $\pstick$, we must
specify a more generic approximation---one that will work even when conditional
conjugacy does not hold. To that end, we first transform the $\nuk$
to a value that is unbounded and then use a Gaussian approximation.
Define the logit-transformed stick-breaking proportions $\lnuk$:
\begin{align*}
  \lnu_\k := \log(\nu_\k) - \log(1 - \nu_\k)
  \quad \Leftrightarrow \quad
  \nuk = \frac{\exp(\lnu_\k)}{1 + \exp(\lnu_\k)}.
\end{align*}
We take $\q(\lnuk \vert \eta)$ to be a normal distribution, which induces a
logit-normal distribution on $\nuk$. We approximate all resulting integrals over
$\q(\lnuk \vert \eta)$, as in the KL objective for VB or in our later
sensitivity calculations, with Gauss-Hermite (GH) quadrature; see
\appref{gh_quadrature}.

GH quadrature yields an approximation, which we call $\KL{\eta}$, to the
full KL, $\KL{\q(\zeta \vert \eta) || \p(\zeta \vert \x)}$. We minimize that
approximation to perform approximate posterior inference:
\begin{align}
\eqlabel{kl_def}
\KL{\q(\zeta \vert \eta) || \p(\zeta \vert \x)}
={}    \expect{\q(\zeta \vert \eta)}{
        \log \q(\zeta \vert \eta) - \log\p(\x, \zeta)} + \log\p(\x) \\
\eqlabel{vb_optimization}
\etaopt :={} \argmin_{\eta \in \etadom} \KL{\eta} \mathwhere
\KL{\eta} \approx{} \KL{\q(\zeta \vert \eta) || \p(\zeta \vert \x)}.
\end{align}
Our final approximation to the marginal posterior $\p(\zeta \vert \x)$ is
$\q(\zeta \vert \etaopt)$.

\noindent \textbf{Posterior quantities of interest.} To approximate any
functional of the exact posterior, we apply the equivalent functional to
$\q(\zeta \vert \etaopt)$. For instance, the approximation to the posterior
expected number of clusters among the $N$ observed data points is
\begin{align} \eqlabel{num_clust_vb}
\expect{\q(\zeta \vert\etaopt)}{\nclusters(\z)} =
\expect{\q(\z\vert\etaopt)}{\nclusters(\z)} =
\sumkm \left(1 -  \prod_{\n=1}^\N
    (1 - \expect{\q(\z_\n \vert \etaoptz)}{\z_{\n\k}})\right).
\end{align}

We will see examples in \secref{results} where our quantity of interest is (a)
the expected posterior number of clusters in the observed data, (b) the expected
posterior number of clusters in a new set of (as yet unobserved) data, (c) some
aspect of a co-clustering matrix, or (d) the topic assignments of certain data
points. In all of these cases, as in \eqref{num_clust_vb}, we are able to
express our (approximate) posterior quantity of interest as a smooth function
$g$ of the optimized variational parameters $\etaopt$: $g(\etaopt)$. Indeed, as
we will discuss in \secref{local_sensitivity}, our methods and results  apply to
any quantity of interest that can be written as a smooth function of $\etaopt$.

Once we have an (approximate) posterior quantity of interest, we can ask how
this quantity would change---and whether our substantive scientific conclusions
would change---if we had made reasonably different prior choices.

\section{A Local Approximation for Sensitivity}
\seclabel{local_sensitivity}
We would like to understand how our quantity of interest $g(\etaopt)$ changes
when the concentration parameter or, more generally, the stick-breaking density
$\pstick$ changes. To efficiently compute these changes, we use a
first-order Taylor series approximation in the optimal VB parameters. In this
section, we first present the Taylor series and then show how to compute its
terms.

\noindent \textbf{Sensitivity to the concentration parameter.} First, we show
how to approximate the sensitivity of $g(\etaopt)$ to the choice of
concentration parameter $\alpha$. Let $\etaopt(\alpha)$ represent the value of
$\etaopt$ for a particular choice of $\alpha$. For our approximation, we choose
some initial value $\alpha_0$ of the concentration parameter and solve the
optimization problem to compute $\etaopt(\alpha_0)$. We then approximate
$\etaopt(\alpha)$ with the linear approximation $\etalin(\alpha)$, and in turn
approximate $g(\etaopt(\alpha))$ with $g(\etalin(\alpha))$:
\begin{align}\eqlabel{alpha_perturbation}
\etalin(\alpha) :=
    \etaopt(\alpha_0) +
    \fracat{d\etaopt(\alpha)}{d\alpha}{\alpha_0} (\alpha - \alpha_0)
\mathand
g(\etaopt(\alpha)) \approx g(\etalin(\alpha)).
\end{align}
If $\alpha \mapsto \etaopt(\alpha)$ is continuously differentiable, and $g$ is
sufficiently smooth, then we expect $g(\etaopt(\alpha)) \approx
g(\etalin(\alpha))$ when $\abs{\alpha - \alpha_0}$ is small. We will show in
\thmref{bnp_deriv} below that the map $\alpha \mapsto \etaopt(\alpha)$ is
continuously differentiable for our chosen VB approximation.

\noindent \textbf{Sensitivity to the stick-breaking density.} Next, we show how
to approximate the sensitivity of $g(\etaopt)$ to the choice of concentration
stick distribution $\pstick$. Technically, perturbations of $\alpha$ are
perturbations of $\pstick$. But here we consider more general perturbations of
the form of $\pstick$, potentially outside the beta class. To define our
perturbations, let $\ptil$ represent a potentially unnormalized (but
normalizable) density with respect to Lebesgue measure; the same notation
without the tilde will give the normalized density. Now start from an initial
setting of $\pstick$ at $\pbase$; we will typically start from Dirichlet-process
stick-breaking; i.e., $\pbase = \betadist{1,\alpha_0}$ for some $\alpha_0$. Then
take any Lebesgue-measurable function $\phi(\cdot)$ on $[0,1]$. We consider a
range of alternative (potentially unnormalized) stick-breaking forms
$\ptil(\cdot \vert \t)$ defined on $[0,1]$ by
\begin{equation} \eqlabel{mult_perturbation}
	\log \ptil(\cdot \vert \t) = \log \pbase(\cdot) + \t \phi(\cdot).
\end{equation}
Note that the perturbation applies equally to every stick break $\nuk$. This
style of multiplicative functional perturbation was proposed by
\citet{gustafson:1996:local}; we deviate from \citet{gustafson:1996:local} by
considering VB (rather than MCMC) approximations and by allowing $\phi$ to take
on negative values.

If we now let $\etaopt(t)$ represent the value of $\etaopt$ for a particular
choice of $\ptil(\cdot \vert \t)$, we can form an approximation
analogous to \eqref{alpha_perturbation}:
\begin{align} \eqlabel{taylor_series_t}
%
\etalin(\t) :=
    \etaopt(0) +
    \fracat{d\etaopt(t)}{dt}{\t=0} (\t - 0)
\mathand
g(\etaopt(\t)) \approx g(\etalin(\t)).
\end{align}
As in the case of expansions with respect to $\alpha$, \eqref{taylor_series_t}
is useful only if the map $t \mapsto \etaopt(t)$ is continuously differentiable
for the chosen $\phi$.  As we will show in \thmref{bnp_deriv} below, a
sufficient condition for differentiability is given in terms of the following
norm on the perturbation $\phi$.
\begin{equation} \eqlabel{infty_norm}
    \textrm{Define }
	\norminf{\phi} := \esssup_{\nu_0 \sim \pbase} \abs{\phi(\nu_0)}
    \textrm{ and }
	\quad \ball_\phi(\delta) := \left\{ \phi: \norminf{\phi} <
\delta \right\}.
\end{equation}
The set of priors that arise by considering functional perturbations $\phi \in
\ball_\phi(\delta)$ live in a multiplicative band around the original prior,
$\pbase$, as shown in \figref{linf_examples}.  \Thmref{bnp_deriv} below states
that $\t \mapsto \etaopt(\t)$ is continuously differentiable whenever
$\norminf{\phi} < \infty$.  So, for sufficiently smooth $\g$, we expect the
approximation \eqref{taylor_series_t} to be good for small $\t$, given a
particular choice of $\phi$ with $\norminf{\phi} < \infty$.

The functional perturbation given in \eqref{mult_perturbation} is useful
because, if we consider any other distribution $\palt$ for $\pstick$, we can
continuously warp $\pbase$ to $\palt$ by setting $\phi(\cdot) = \log \left(
\palt(\cdot) / \pbase(\cdot) \right)$ so long as $\palt \ll \pbase$; i.e.,
$\palt$ is absolutely continuous with respect to $\pbase$.
We will see in \secref{influence_function} that we can compute an
\emph{influence function} to provide an interpretable summary of the effect of
arbitrary changes $\phi$.  Using the influence function and the
$\norminf{\cdot}$ norm, we are able to find a worst-case choice of $\phi$ in
$\ball_\phi(\delta)$.

However, we note that restricting to $\norminf{\phi} < \infty$ limits the kinds
of alternative priors $\palt$ that can be formed using
\eqref{mult_perturbation}. Although we show in \lemref{pert_invariance} of
\appref{diffable_nonparametric} that functional perturbations with
$\norminf{\phi} < \infty$ yield valid priors, the converse is not true: there
exist valid priors $\palt$ such that the corresponding $\norminf{\phi} =
\infty$. For instance, perturbing the beta stick-breaking form by changing
$\alpha$ provides a counterexample since the log of the beta density is
unbounded below; see \exref{beta_inf_norm} of \appref{diffable_nonparametric}
for more details.
The limited expressiveness of $\ball_\phi(\delta)$ may
at first seem like a shortcoming of the perturbation given by
\eqref{mult_perturbation}. However, we show in \secref{influence_function} that,
among a class of potential functional perturbations such as those
proposed by \citet{gustafson:1996:local},
only the one we defined in \eqref{mult_perturbation} is {\em Fr{\'e}chet
differentiable}---and thus can be used to safely reason about worst-case
$\phi$.

\noindent \textbf{Computing the terms in the Taylor series.}  It remains to show
that $\alpha \mapsto \etaopt(\alpha)$ and $\t \mapsto \etaopt(\t)$ are
continuously differentiable, and to provide a computable formula for the
derivative.
Differentiability naturally requires some regularity conditions on the VB
parameterization and on the optimum.  We state sufficient conditions in the
following \assuref{kl_opt_ok}, which is satisfied for any local optimum of a
smooth, unconstrained parameterization of the variational approximation.

\begin{assu}\assulabel{kl_opt_ok}
Assume that:
%
\begin{enumerate*}[label=(\arabic*)]
    \item \itemlabel{kl_diffable} the map $\eta \mapsto \KL{\eta}$ is twice
    continuously differentiable at $\etaopt$;

    \item\itemlabel{kl_hess} the Hessian matrix $\fracat{\partial^2 \KL{\eta}}
    {\partial \eta \partial \eta^T} {\etaopt}$ is non-singular; and

    \item \itemlabel{kl_opt_interior} there exists an open ball $\ball_\eta
    \subseteq \mathbb{R}^\etadim$ such that $\etaopt \in \ball_\eta \subseteq
    \etadom$.
\end{enumerate*}
\end{assu}
%
Our next result establishes the differentiability of $\etaopt$ and provides
a computable formula for the derivative.
\begin{thm}\thmlabel{bnp_deriv}
Let \assuref{kl_opt_ok} hold for the VB approximation given in
\secref{model_vb}.  Either take $\varepsilon = \t$ under the perturbation given
by $\log \ptil(\nuk \vert \t) = \log \pbase(\nuk) + \t \phi(\nuk)$ with
$\norminf{\phi} < \infty$, or take $\varepsilon = \alpha - \alpha_0$ in a
perturbation to the concentration parameter $\alpha$ of the unnormalized beta
distribution $\log \ptil(\nuk \vert \alpha) = \alpha \log(1 - \nuk)$. Then the
map $\varepsilon \mapsto \etaopt(\varepsilon)$ is continuously differentiable at
$\varepsilon = 0$ with derivative
\begin{align}
\eqlabel{bnp_vb_eta_sens}
\fracat{d \etaopt(\varepsilon)}{d \varepsilon}{\varepsilon=0} ={}&
    - \hessopt^{-1} \crosshessian, \mathwhere
    \rho_\k(\nuk) := \fracat{\partial \log \ptil(\nu_k \vert \varepsilon)}
            {\partial \varepsilon}{\varepsilon=0}
            ,
\\
\hessopt :={}& \fracat{\partial^2 \KL{\eta}}
                      {\partial \eta \partial \eta^T}
                      {\eta = \etaopt},
\quad \lqgradbar{\zeta \vert \eta} :={}
\fracat{\partial \log \q(\zeta \vert \eta)}{\partial \eta}{\eta}, \textrm{ and}\\
%
\eqlabel{bnp_vb_crosshessian}
\crosshessian :={}&
    \frac{\partial}{\partial\eta}
    \expect{\q(\zeta \vert \eta)}{
        \sum_{k=1}^{\kmax-1}
        \rho_\k(\nuk)
    }
    \Bigg|_{\eta = \etaopt}
={}
    \expect{\q(\zeta \vert \etaopt)}{
          \lqgradbar{\zeta \vert \etaopt}
          \sum_{k=1}^{\kmax-1}
          \rho_\k(\nuk)}.
\end{align}
\end{thm}
\begin{proof}
The result follows from \thmref{etat_deriv} of \appref{diffable_parametric},
which states general conditions for the differentiability of VB optima.  We show
in \appref{diffable_concentration, diffable_nonparametric} that the conditions
of \thmref{etat_deriv} are satisfied in the case of our present BNP problem. The
equivalence of the expressions for $\crosshessian$ follows by differentiating
through the expectation; see \lemref{logq_continuous} of \appref{proofs} for
more details.
\end{proof}

\Eqref{bnp_vb_eta_sens} requires computation of two terms: $\hessopt^{-1}$ and
$\crosshessian$.  Typically, $\crosshessian$, which is a derivative of a
variational expectation, is straightforward to evaluate: the requisite
expectation is evaluated either in closed form or approximated numerically;
then, in either case, an application of automatic differentiation provides the
gradient \citep{baydin:2018:automatic}. Forming and inverting or factorizing
$\hessopt$ can present a challenge due to its high dimensionality---it has
dimensions $\etadim \times \etadim$, where $\etadim$ is the dimension of $\eta$.
However, in many cases---including the BNP problem that is our focus---we can
take advantage of model sparsity to efficiently compute \eqref{bnp_vb_eta_sens}
(see \secref{computing_sensitivity}), and our experiments confirm that we can
compute $\fracat{d \etaopt(\varepsilon)}{d \varepsilon}{\varepsilon=0}$ much
more efficiently than re-optimizing the VB objective directly
(\secref{compute_time}). Moreover, the savings increase dramatically when we
are interested in a range of $\varepsilon$ values because $\fracat{d
\etaopt(\varepsilon)}{d \varepsilon}{\varepsilon=0}$ can be re-used to for any
chosen value of $\varepsilon$.

%

\section{The Influence Function and Worst-Case Functional Perturbations}
\seclabel{influence_function}
We next show how to find influential and worst-case functional perturbations to
the stick-breaking density. We start by showing how to compute an influence
function to summarize the effect of different choices of $\phi$. Using the
influence function, we are able to design stick-breaking densities that produce
a large change in a quantity of interest, including computing the worst-case
perturbation in $\ball_\phi(\delta)$. To justify such uses of the influence
function, we prove that, for multiplicative perturbations and the $\infty$-norm,
the VB objective is Fr{\'e}chet differentiable---i.e., that it admits a
uniformly good linear approximation in a neighborhood of the null perturbation.
Finally, we show that our Fr{\'e}chet differentiability result is unique among a
broad class of alternative choices of functional perturbation.

\noindent \textbf{The influence function and worst-case perturbations.}
We begin by defining the influence function $\infl$ and discussing its
usefulness for understanding the effect of functional perturbations $\phi$.
Suppose we have a one-dimensional, differentiable quantity of interest,
$\g(\cdot): \etadom \mapsto \mathbb{R}$, and are considering various alternative
priors as given by $\phi$ in \eqref{mult_perturbation}.  Under the approximation
in \eqref{taylor_series_t}, the dependence of $\g(\etalin(\t))$ on $\phi$ is
not simple if $\g(\cdot)$ is non-linear.  However, for a particular choice of
$\phi$, by applying the chain rule with \thmref{bnp_deriv}, we can derive a
fully linear approximation $\g(\etaopt(\t)) \approx \g(\etaopt) + \fracat{d
\g(\etaopt(\t))}{d\t}{\t=0}(\t - 0)$.  The advantage of linearizing $\g$ in this
way is that the map $\phi \mapsto \fracat{d \g(\etaopt(\t))}{d\t}{\t=0}$ has a
particularly simple form, as given by the following result.
\begin{cor}\corylabel{etafun_deriv_form_stick}
Under the conditions of \thmref{bnp_deriv}, using \eqref{mult_perturbation} with
$\norminf{\phi} < \infty$ and $\varepsilon = \t$, let $\g(\cdot): \etadom \mapsto
\mathbb{R}$ denote a continuously differentiable, real-valued function of
interest.  Define the \emph{influence function} $\infl: [0,1] \mapsto
\mathbb{R}$:
\begin{align} \eqlabel{infl_defn_bnp}
\infl(\cdot) :={}&
    - \sum_{k=1}^{\kmax-1} \fracat{d g(\eta)}{ d \eta^T}{\etaopt} \hessopt^{-1}
        \lqgradbark{\cdot \vert \etaopt}
        \qk(\cdot \vert \etaopt),
\end{align} where $\lqgradbark{\cdot \vert \etaopt}$ and $\qk(\cdot \vert
\etaopt)$ replace $\q(\zeta \vert \eta)$ with just the factor of $\q$ for
$\nu_k$.
Then the derivative in \eqref{bnp_vb_eta_sens} can be written as
\begin{align} \eqlabel{vb_eta_infl_sens_bnp}
\fracat{d \g(\etaopt(\t))}{d \t}{0} ={}&
    \int_0^1 \infl(\nu_0) \phi(\nu_0) d\nu_0.
\end{align}
\end{cor}
\begin{proof}
The form of the influence function is given by the chain rule, gathering
terms in \eqref{bnp_vb_eta_sens}, and re-writing the variational expectation as
an integral over $[0,1]$. We establish an analogous general result for
general VB approximations in \coryref{etafun_deriv_form} of
\appref{diffable_nonparametric}, specializing to the BNP case in
\exref{infl_univariate} of \appref{diffable_nonparametric}.
\end{proof}
By choosing perturbations $\phi$ that align with the influence function, we can
form priors that we expect to be influential for the function of interest,
$\g(\cdot)$.  For example, in our experiments of \secref{results}, we show that
by choosing $\phi$ to be a Gaussian bump aligned with particularly
high-magnitude positive or negative values of the influence function, one can
ensure a large positive or negative gradient, and hence a large predicted
change.

Further, with \coryref{etafun_deriv_form_stick} in hand, we can find a
closed-form expression for the worst-case choice of $\phi \in
\ball_\phi(\delta)$, which is essentially a VB analogue to \citet[Result
11]{gustafson:1996:local}.
\begin{cor}\corylabel{etafun_worst_case_stick}
Under the conditions of \coryref{etafun_deriv_form_stick},
\begin{align*}
\sup_{\phi \in \ball_\phi(\delta)}
    \fracat{d g(\etaopt(\t))}{d \t}{0} =
        \delta \int \abs{\infl(\nu_0)} \mu(d\nu_0),
\end{align*}
and the supremum is achieved at the perturbation
$\phi^*(\cdot) = \delta \, \mathrm{sign}\left(\infl(\cdot)\right)$.
\end{cor}
\begin{proof}
The result follows immediately from applying H{\"o}lder's inequality to
\eqref{vb_eta_infl_sens_bnp}. We establish a similar but much more general
result for VB approximations with general choices of model and
parameters in \coryref{etafun_worst_case} of \appref{diffable_worst_case}. The
present result is a special case using \exref{infl_univariate} of
\appref{diffable_worst_case}.
\end{proof}

In our experiments of \secref{results}, we use
\coryref{etafun_deriv_form_stick, etafun_worst_case_stick} to choose influential
perturbations, and then use the partially linearized \eqref{taylor_series_t} to
make predictions about the effect of the perturbations.

\LinfExamplesFig{}

\noindent \textbf{Multiplicative perturbations are continuously Fr{\'e}chet
differentiable.}
The influence function provides a succinct summary of the effect of all
perturbations $\phi \in \ball_\phi(\delta)$, which we might hope to be accurate
for sufficiently small $\delta$. However, the accuracy of our approximation
within $\ball_\phi(\delta)$ is not guaranteed by \thmref{bnp_deriv} alone.
Specifically, \thmref{bnp_deriv} states only that, for a {\em particular}
direction $\phi$, $\t \mapsto \etaopt(\t)$ is continuously differentiable---i.e.
that, for a fixed $\phi$, one can make $\t$ sufficiently small so that the error
$\abs{\etaopt(\t) - \etalin(\t)}$ goes to zero faster than $\t$. But, if we
write $\etaopt(\t\phi)$ and $\etalin(\t \phi)$ to make the dependence on $\phi$
explicit, then \thmref{bnp_deriv} does not imply that for a fixed $\delta$ (no
matter how small), the worst-case error $\sup_{\phi \in \ball_\phi(\delta)}
\abs{\etaopt(\phi) - \etalin(\phi)}$ is bounded, much less that it goes to zero
faster than $\delta$.

Thus, to be assured that the influence function is a meaningful summary of the
effect of all  $\phi \in \ball_\phi(\delta)$, we wish to establish that the
linear approximation given by \eqref{taylor_series_t} is uniformly accurate over
all $\phi$ of interest within a sufficiently small neighborhood of the zero
function.  Specifically, observing that $\phi$ is a point in the Banach space
$L_\infty$ \citep[Theorem 5.2.1]{dudley:2018:real}, we wish to establish that
the map $\phi \mapsto \etaopt(\phi)$ from $L_\infty$ to $\mathbb{R}^\etadim$ is
\textit{Fr{\'e}chet differentiable}, as we now formally define.%
\footnote{
Fr{\'e}chet differentiability is sometimes referred to as ``bounded''
differentiability. In addition to Fr{\'e}chet, two other notions of
differentiability are common in statistics: Hadamard (i.e., compact)
differentiability, and Gateaux (i.e., weak, or directional) differentiability.
In each case, the derivative is given by the same linear operator, but comes
with different accuracy guarantees, of which Fr{\'e}chet is the strongest.
Consequently, our \thmref{eta_phi_deriv_stick} below implies both Hadamard and
Gateaux differentiability as a consequence of Fr{\'e}chet differentiability. See
\citet{averbukh:1967:theory} for a general mathematical treatment of
differentiability in Banach spaces, or \citet{reeds:1976:thesis} for a treatment
intended for statisticians.}

\begin{defn}\deflabel{diffable_classes}
    (Fr{\'e}chet differentiability,
    \citep[Definition 4.5]{zeidler:2013:functional})
Let $B_1$ and $B_2$ denote Banach spaces, and let $\ball_1 \subseteq B_1$ define
an open neighborhood of $\phi_0 \in B_1$.
A function $f: \ball_1 \mapsto B_2$ is {\em Fr{\'echet} differentiable} at
$\phi_0$ if there exists a  bounded linear operator, $f^{\mathrm{lin}}: B_1
\mapsto B_2$, such that, for $\phi \in B_1$,
\begin{align*}
f(\phi) - f(\phi_0) - f_{\phi_0}^{\mathrm{lin}}(\phi - \phi_0) =
    o(\norm{\phi - \phi_0})
    \quad \textrm{as }\norm{\phi - \phi_0} \rightarrow 0.
\end{align*}
The function $f$ is {\em continuously Fr{\'echet} differentiable} if the map
$\phi_0 \mapsto f_{\phi_0}^{\mathrm{lin}}(\cdot)$ is continuous as a map from
from $\ball_1$ to the space of all continuous linear operators from $B_1$ to
$B_2$ equipped with the operator norm.
\end{defn}
By \citet[Proposition 4.8]{zeidler:2013:functional}, if a function is
Fr{\'e}chet differentiable, then the linear operator $f^{\mathrm{lin}}$ is given
precisely by the directional derivative $d f(\phi_0 + t (\phi - \phi_0)) / d t$.
Thus, if $\phi \mapsto \etaopt(\phi)$ is Fr{\'e}chet differentiable, its
derivative is given by \coryref{etafun_deriv_form_stick}.  Fr{\'e}chet
differentiability guarantees that, for sufficiently small $\delta$, the error of
the linear approximation given by \coryref{etafun_deriv_form_stick} does not
blow up in the ball $\ball_\phi(\delta)$.

We emphasize that Fr{\'e}chet differentiability is neither sufficient nor
necessary for a derivative to be useful.  For example, it is possible in
principle for a function to be Fr{\'e}chet differentiable but still have a very
large finite second derivative, and so fail to extrapolate meaningfully to any
alternatives one cares about.  Conversely, if a function fails to be Fr{\'e}chet
differentiable, the derivative may still perform well in particular directions,
including that chosen by \coryref{etafun_worst_case_stick}.  Nevertheless,
Fr{\'e}chet differentiability is a strong local result, and provides some
assurance that one can use results such as \coryref{etafun_worst_case_stick}
without uncovering pathological behavior.

Finally, then, we prove that our perturbation is continuously Fr{\'e}chet
differentiable.

\begin{thm}\thmlabel{eta_phi_deriv_stick}
Under the conditions of \thmref{bnp_deriv}, the map $\phi \mapsto \etaopt(\phi)$
is well-defined and continuously Fr{\'e}chet differentiable in a neighborhood of
the zero function as a map from $\linf$ to $\mathbb{R}^\etadim$, with the
derivative given in \coryref{etafun_deriv_form_stick}. \end{thm}
\begin{proof}
Our result here is a special case of our general result for VB approximations
given in \thmref{eta_phi_deriv} of \appref{diffable_worst_case}.
\end{proof}

\noindent \textbf{Many other functional perturbations and norms are not Fr{\'e}chet differentiable.}
So far we have focused on the multiplicative functional perturbations in
\eqref{mult_perturbation} combined with the infinity norm in \eqref{infty_norm}.
We now ask whether we could perform a similar analysis for other functional
perturbations. We show that, of the perturbations proposed by
\citet{gustafson:1996:local}, only multiplicative perturbations yield
Fr{\'e}chet differentiable VB optima.

Specifically, \citet{gustafson:1996:local} examines general perturbations, from
initial prior $\pbase$ to alternative $\palt$, that take the following
form---with $\theta$ a parameter $\theta \in \thetadom \subseteq
\mathbb{R}^{\thetadim}$ and $p \in [1, \infty)$:
\begin{align}\eqlabel{p_pert_simple_bnp}
\ptil(\theta \vert \tp) :=
    \left((1 - \tp)\pbase(\theta)^{1/p} +
    \tp \frac{1}{p}\palt(\theta)^{1/p} \right)^{p}.
\end{align}
Again, let $\phi$ represent the perturbation, now with:
\begin{align}\eqlabel{phi_lp_norm_bnp}
\phi(\theta \vert \palt, p) :={}
    \palt(\theta)^{1/p} - \pbase(\theta)^{1/p} \mathand
\norm{\phi}_p :={} \left(\int_0^1 \abs{\phi(\theta)}^p d \theta \right)^{1/p}.
\end{align}
The limit $p \rightarrow \infty$ recovers our multiplicative perturbation in
\eqref{mult_perturbation} with infinity norm in \eqref{infty_norm}. The choice
$p=1$ recovers a purely additive perturbation. \citet[Result
2]{gustafson:1996:local} states that $\norm{\phi}_p < \infty$ ensures that the
corresponding $\ptil(\theta \vert \tp)$ can be normalized, strongly motivating
using the $\norm{\cdot}_p$ norm with the perturbation given by
\eqref{p_pert_simple_bnp}.

Our next theorem shows that the reverse KL divergence is discontinuous in
$\norm{\cdot}_p$ for $p < \infty$. Since Fr{\'e}chet differentiability implies
continuity \citep[Proposition 4.8 (d)]{zeidler:2013:functional},
\thmref{kl_discontinuous_main} implies that it is impossible to derive an
analogue of \thmref{eta_phi_deriv_stick} for perturbations of the form in
\eqref{p_pert_simple_bnp} with the norms in
\eqref{phi_lp_norm_bnp}.~\footnote{Hadamard differentiability also implies
continuity, so our \thmref{kl_discontinuous_main} also implies Hadamard
non-differentiability \citep[Section 3]{averbukh:1967:theory}. In general, a
functional may be Gateaux differentiable but discontinuous, though in such cases
there are necessarily directions in which the derivative provides an arbitrarily
poor approximation to the behavior of the functional (see \citet[Example
1.19]{averbukh:1967:theory}). In the present case, as we discuss in
\appref{diffable_lp}, there exist pointwise negative priors in every
$\norm{\cdot}_p$ neighborhood of $\pbase$ for $p < \infty$, so even establishing
Gateaux differentiability (i.e., the mere existence of a directional derivative
in every direction) requires a somewhat artificial extension of the KL divergence
to accommodate pointwise negative prior densities. }
\begin{thm}\thmlabel{kl_discontinuous_main}
Let $\mu$ denote a measure on the Borel sets of some domain $\thetadom$, with
$\mu$ absolutely continuous with respect to the Lebesgue measure, and let
$\q(\theta)$ and $\pbase(\theta)$ denote densities with respect to $\mu$.
Without loss of generality, assume that $\q(\theta) > 0$ on $\thetadom$.  Assume
that $\KL{\q(\theta) || \pbase(\theta)}$ is well-defined and finite.

Then, for any $\epsilon > 0$ and any $M > 0$, we can find a density
$\palt(\theta)$ such that $\norm{\phi(\theta \vert \palt, p)}_p < \epsilon$ but
$\abs{\KL{q(\theta) || \palt(\theta)} - \KL{q(\theta) || \pbase(\theta)}} > M$.
\begin{proof}
See \appref{diffable_lp} for a constructive proof, the key to which is the fact
that in any $\norm{\cdot}_p$ neighborhood of zero there exist prior densities
taking values arbitrarily close to zero on sets of nonzero measure, for which
the reverse KL divergence blows up.
\end{proof}
\end{thm}
%

Recall from \secref{local_sensitivity} (and particularly \exref{beta_inf_norm}
of \appref{diffable_nonparametric}) that there exist priors that cannot be
formed from \eqref{mult_perturbation} using $\phi$ with $\norminf{\phi} <
\infty$. In light of the proof of \thmref{kl_discontinuous_main}, the limited
expressiveness of multiplicative perturbations with the $\norminf{\cdot}$ norm
looks like a feature rather than a bug. Consider the rightmost panel of
\figref{linf_examples}, which illustrates the tradeoffs between the various
norms.  The two blue and red densities are far from one another according to
reverse KL divergence since the red density takes values that are nearly zero
where the blue density has nonzero mass. The two densities are also distant in
$\norminf{\cdot}$ since it takes a large multiplicative change to turn the
nonzero blue density into the nearly zero red density. However, the two
densities are close in $\norm{\cdot}_{p}$ since the region where the red density
is nearly zero has a small measure. In order for VB approximations to be
continuous (a necessary condition for Fr{\'e}chet differentiability), one must
consider a topology on priors that is no coarser than the topology induced by
reverse KL divergence.  But since valid priors can take values close to zero, a
sacrifice in expressiveness of the neighborhood of zero must be made in order to
induce a topology that is compatible with reverse KL divergence. Multiplicative
changes and the $\norminf{\cdot}$ norm implement such a tradeoff in a natural,
easy-to-understand way.


In this sense, VB approximations based on reverse KL divergence are inherently
non-robust to priors that ablate mass nearly to zero.  No parameterization of
the space of priors will relieve this non-robustness.  Only by basing
variational approximations on divergences other than reverse KL might this
non-robustness be alleviated.

\section{Fast Computation of the Sensitivity}
\seclabel{computing_sensitivity}
A principal challenge of computing the sensitivity efficiently is the
high-dimensional nature of the parameter $\zeta$ and hence the variational
parameters $\eta$. In particular, we have seen that, in our BNP stick-breaking
model, $\zeta$ and $\eta$ both grow linearly with the number of data points $N$.
This growth leads to two major computational challenges: (1) we must solve a
high-dimensional optimization problem to extremize the VB objective, and (2) we
must solve a linear system given by the Hessian $\hessopt$. Here we show how we
can use special structure in the model to reduce to low-dimensional problems and
thereby enjoy efficient computation.

\noindent \textbf{Global and local parameters.} In both cases, the key to
reducing to a lower-dimensional problem is separating \emph{global} and
\emph{local} parameters. Global variables are common to all data points. Local
variables are unique to each data point. For instance, in a Gaussian (or other
typical) mixture model, the stick-breaking proportions $\nu$ and component
parameters $\beta$ are global, whereas the cluster assignment parameters $\z$
are local.

Let $\gamma$ denote the collection of global parameters. When we use a standard
mean-field VB parameterization, the VB distributions on $\gamma$ have their own
variational parameters, which we denote $\etaglob$. Similarly, let $\ell$ denote
the local parameters and let $\etalocal$ be the corresponding local variational
parameters.

\noindent \textbf{Reducing to optimization over the global variational
parameters.} We next show how to reduce the potentially high-dimensional
optimization problem over all of $\eta$ to optimizing over just the global
variational parameters $\etaglob$.

In all models we will consider, the conditional posterior $\p(\z \vert
\gamma,\x)$ has a tractable closed form.  Since we choose a conjugate mean-field
approximating family for $\q(\z \vert \eta)$, the optimal local variational
parameters $\etaoptlocal$ can be written as a closed-form function of the global
variational parameters $\etaglob$. For some prior parameter $\varepsilon$ (as in
\thmref{bnp_deriv}), let $\etaoptlocal(\eta_\gamma; \varepsilon)$ denote this
mapping, so that
\begin{align}\eqlabel{local_eta_optim}
\etaoptlocal(\etaglob; \varepsilon) :=
    \argmin_{\etalocal} \KL{(\eta_\gamma, \etalocal), \varepsilon}.
\end{align}
In \exref{qz_optimality} (\appref{gmm_global_local_vb}), we illustrate this
technique for a Gaussian mixture model.
Using \eqref{local_eta_optim}, we can rewrite our objective as a
function of the global parameters.  Define
\begin{align*}
\KLglobal(\etaglob, \varepsilon) :=
    \mathrm{KL}\Big((\etaglob, \etaoptlocal(\etaglob; \varepsilon)), \varepsilon\Big).
\end{align*}
The $\etaoptglob(\varepsilon)$ that minimizes $\KLglobal(\etaglob, \varepsilon)$
is the same as the corresponding sub-vector of the $\etaopt(\varepsilon)$ that
minimizes $\KL{\eta, \varepsilon}$.  

Rather than optimizing the $\KL{\eta}$ over all variational parameters, we
numerically optimize $\KLglobal$, which is a function only of the relatively
low-dimensional global parameters.  To minimize $\KLglobal(\etaglob)$ in
practice, we run the BFGS algorithm with a loose convergence tolerance followed
by the trust-region Newton conjugate gradient method to find a high-quality
optimum (the \texttt{trust-ncg} method of \texttt{scipy.optimize.minimize},
\cite{2020:scipy}; see also \citet[Chapter~7]{nocedal:2006:numerical}).  After
the optimization terminates at an optimal $\etaoptglob$, the optimal local
parameters $\etaoptlocal$ can be set in closed form to produce the entire vector
of optimal variational parameters, $\etaopt = (\etaoptglob, \etaoptlocal)$.

\subsection{Computing and inverting the Hessian} Since the dimension $\etadim$
of $\eta$ scales with $N$, we can quickly reach cases where inverting or even
instantiating a dense matrix of size $\etadim \times \etadim$ in memory would be
prohibitive. The key to efficient computation is that $\hessopt$ is not dense;
we will again exploit structure inherent in the global/local decomposition.

For generic variables $a$ and $b$, let $\hess{ab}$ denote the sub-matrix
$\evalat{\partial^2 \KL{\eta} / \partial \eta_a \eta_b^T}{\etaopt}$, the Hessian
with respect to the variational parameters governing $a$ and $b$. We decompose
the Hessian matrix $\hessopt$ into four blocks according to the global/local
decomposition:
\begin{align*}
\hessopt =
\fracat{\partial^2 \KL{\eta}}
       {\partial \eta \partial \eta^T}
       {\etaopt} ={}&
\left(
\begin{array}{cc}
   \hess{\gamma\gamma} & \hess{\gamma\ell} \\
   \hess{\ell\gamma}     & \hess{\ell\ell} \\
\end{array}
\right).
\end{align*}
Similarly, let $\crosshessian_\gamma$ be the components of $\crosshessian$
corresponding to the variational parameters $\etaglob$.  The local components,
$\crosshessian_\ell$, are zero since no local variables enter the expectation in
\eqref{bnp_vb_crosshessian} when we are perturbing the stick-breaking
distribution.
%
  %

In this notation,
\begin{align} \eqlabel{global_local_derivative_breakdown}
\fracat{d \etaopt(\t)}{d \t}{t = 0} ={}&
-\left(
\begin{array}{cc}
   \hess{\gamma\gamma} & \hess{\gamma\ell} \\
   \hess{\ell\gamma}     & \hess{\ell\ell} \\
\end{array}
\right)^{-1}
\left( \begin{array}{c} \crosshessian_\gamma \\ 0 \end{array}\right).
\end{align}
Applying the Schur complement and focusing on the global parameters (see
\appref{more_hessian} for more details), we find
\begin{align}\eqlabel{global_sens}
  \fracat{d \etaopt_\gamma(\t)}{d \t}{t = 0} &=
  - \hessopt_\gamma^{-1}\crosshessian_\gamma
  \mathwhere
  \hessopt_\gamma := \left(\hess{\gamma\gamma} -
        \hess{\gamma\ell} \hess{\ell\ell}^{-1} \hess{\ell\gamma}\right),
\end{align}
In the models we consider, $\hess{\ell\ell}$ is block diagonal, and the size of
$\hess{\gamma\gamma}$ is relatively small. Thus each term of \eqref{global_sens}
can be tractably computed, even on very large datasets. While the Schur
complement calculation is illustrative, \eqref{global_sens} is equivalent to
applying automatic differentiation to the global-only objective
$\KLglobal(\etaglob, \t)$; see \appref{more_hessian} for details.

In our BNP applications, it is not cost-effective to form and invert or
factorize $\hessopt$ in memory.  Instead, we numerically solve linear systems of
the form $\hessopt^{-1} v$ using the conjugate gradient (CG) algorithm
\citep[Chapter 5]{nocedal:2006:numerical}, which requires only Hessian-vector
products that are readily available through automatic differentiation.

\noindent \textbf{A linear approximation only in the global variational parameters}.
With the tools above, we can separate out the linear approximation in the global
parameters and then directly compute the local parameters. In particular, we
compute
\begin{align}\eqlabel{global_lin_approx}
  \etalin_\gamma(\t) := \etaopt_\gamma +
  \fracat{d \etaopt_\gamma(\t)}{d \t}{\t=0} \t ,
\end{align}
and then use $\etaoptlocal(\etaglob)$ e.g.\ in computing our quantity of
interest. By doing so, our approximation is able to retain non-linearities in
the map $\etaglob \mapsto \etaoptlocal(\etaglob)$.  We give an example for the
expected number of clusters in \appref{vb_insample_nclusters_example}.  In all
our experiments, we use \eqref{global_lin_approx} in this way.

\section{Experimental Results}
\seclabel{results}
We next evaluate our sensitivity approximations on three real data sets, each
with a different model using stick-breaking.\footnote{Code and instructions for reproducing
our experiments can be found online at
\url{https://github.com/Runjing-Liu120/BNPStickBreakingSensitivity}.}
We find that our approximations
largely agree with ground truth obtained by re-running the VB optimization, but
with the evaluation of our derivative an order of magnitude faster than
re-optimizing for a given perturbation.

    \subsection{Gaussian mixture modeling on iris data}
    \seclabel{results_iris}

We perform a clustering analysis of Fisher's iris data set
\citep{fisher:1936:iris, anderson:1936:iris}. Here each  data point (with
$N=150$ total points) represents $d=4$ measurements of a particular flower, from
one of three iris species. We use a standard Gaussian mixture model with a
conjugate Gaussian-Wishart prior for the component parameters (detailed in
\appref{app_iris}) and a mean-field VB approximation with truncation parameter
$\kmax = 15$. We consider two quantities of interest: (1) $\gclustersabbr$, the
posterior expected number of clusters among the $N$ observed data points, and
(2) $\gclusterspredabbr$, the posterior predictive expected number of clusters
in $N$ new (i.e.\ as-yet-unseen) data points. We set the base stick-breaking
prior $\pbase(\nuk)$ to be the standard $\betadist{\nuk \vert 1,\alpha}$
distribution with $\alpha = \alpha_0 = 2$. Under the base stick-breaking prior
with $\alpha_0$, the posterior expected number of clusters matches the three
iris species; see also \figref{iris_fit} in \appref{app_iris} for an
illustration.

\noindent \textbf{Sensitivity to the concentration parameter.} We approximate the
changes in the quantities of interest as $\alpha$ varies over $\alpha\in[0.1,
4.0]$, which corresponds to an {\em a priori} expected number of clusters among $N$
data points in $[1.5,15]$ (\appref{app_beta_prior}). Over this range, the shape
of a $\betadist{1,\alpha}$ density varies considerably, as shown in
\figref{beta_priors} in \appref{app_beta_prior}.

\begin{knitrout}
\definecolor{shadecolor}{rgb}{0.969, 0.969, 0.969}\color{fgcolor}\begin{figure}[!h]

{\centering \includegraphics[width=0.784\linewidth,height=0.439\linewidth]{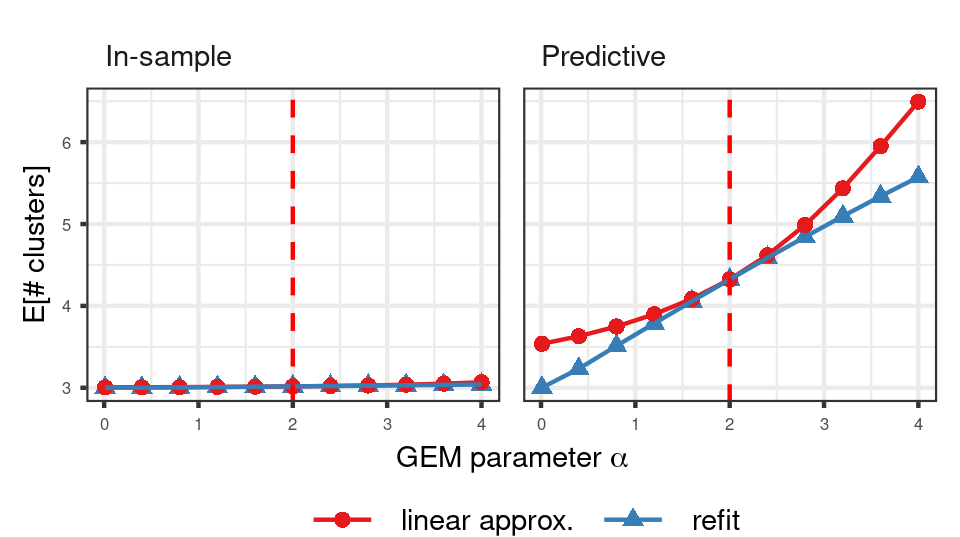}

}

\caption[The expected number of clusters in the original data set ($\gclustersabbr$, left) and in a new data set of size $N$ ($\gclusterspredabbr$, right) as $\alpha$ varies in the GMM fit of the iris data]{The expected number of clusters in the original data set ($\gclustersabbr$, left) and in a new data set of size $N$ ($\gclusterspredabbr$, right) as $\alpha$ varies in the GMM fit of the iris data. We formed the linear approximation at $\alpha_0=2$.}\label{fig:iris_alpha_sens}
\end{figure}

\end{knitrout}

\Figref{iris_alpha_sens} compares our linear approximation to ground truth on
the two quantities of interest as $\alpha$ varies. Over this range of $\alpha$,
the posterior expected number of clusters in the observed data is quite robust;
it remains nearly constant at three. The posterior predictive expected number of
clusters in $N$ new data points is less robust; it ranges roughly from
3.0 to 5.6 expected species.  Our approximation captures this
qualitative behavior. As expected, the approximation is least accurate furthest
from the $\alpha_0$, where the Taylor series is centered.

\noindent \textbf{Sensitivity to functional perturbations.} Insensitivity of the
expected number of clusters $\gclustersabbr$ to $\alpha$ does not rule out
sensitivity to other prior perturbations. We now check how our approximation
fares for the multiplicative perturbations in \eqref{mult_perturbation}. We
consider perturbations $\phi$ that are Gaussian bumps in logit stick space, with
each perturbation centered at a different location on the real line.  Each row
of \figref{iris_fsens} corresponds to a different $\phi$. Each $\phi$ is shown
in gray in the leftmost plot of its row. The middle column of
\figref{iris_fsens} shows the stick-breaking prior $\p(\nuk \vert \phi)$ induced
by the corresponding $\phi$.
The rightmost column of \figref{iris_fsens}
shows the changes produced by the $\phi$ perturbation for that row.
We see that our approximation captures the qualitative behavior of the exact
changes.

\begin{knitrout}
\definecolor{shadecolor}{rgb}{0.969, 0.969, 0.969}\color{fgcolor}\begin{figure}[!h]

{\centering \includegraphics[width=0.980\linewidth,height=0.862\linewidth]{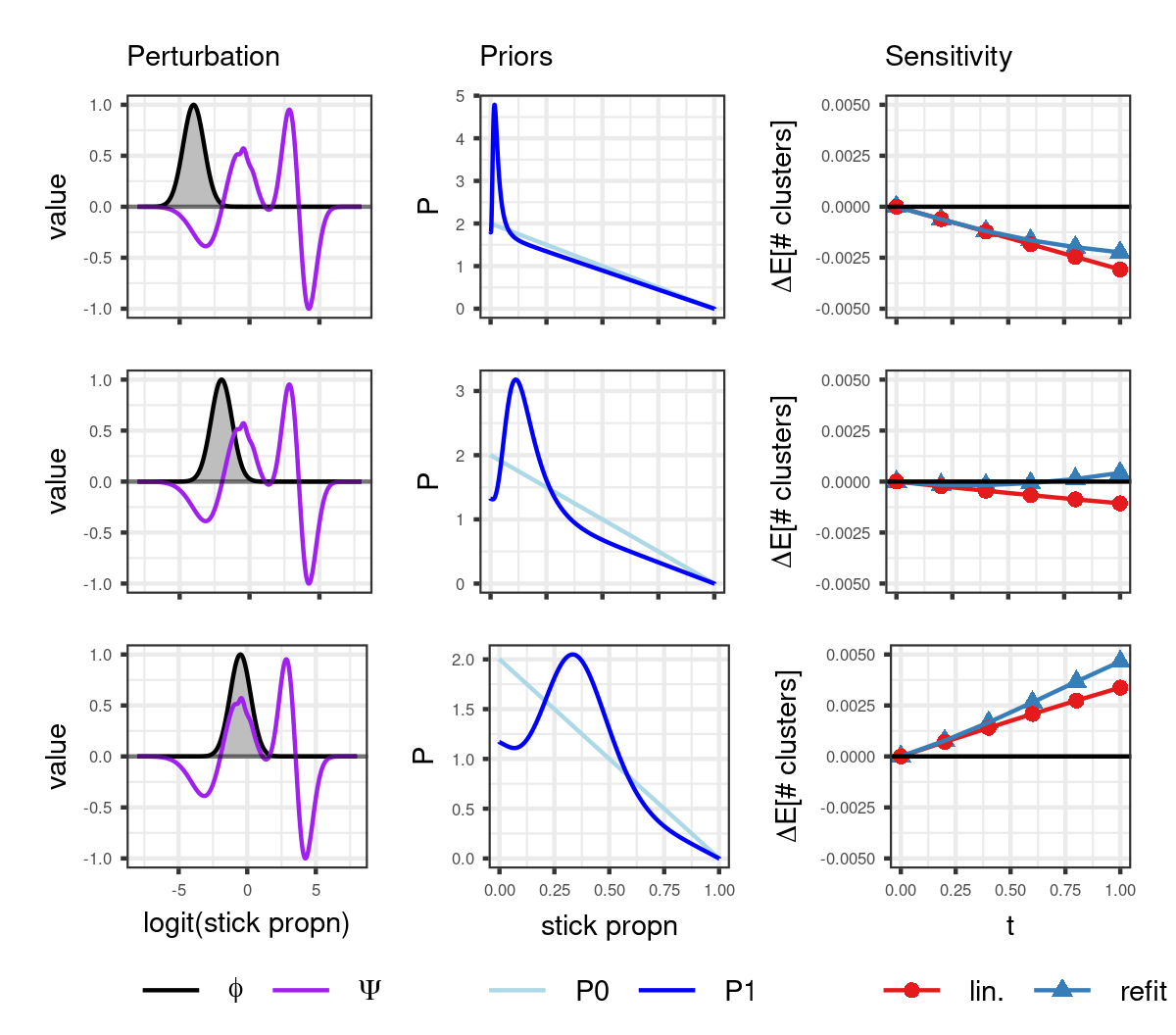}

}

\caption{Sensitivity of
        the expected number of in-sample clusters
        in the iris data set
        to three multiplicative perturbations each with $\norminf{\phi} = 1$.
        (Left) The multiplicative perturbation $\phi$ is in grey.
        The influence function $\Psi$, scaled so $\norminf{\Psi}=1$, is in purple.
        (Middle) The initial $\pbase(\nuk)$ (light blue)
        and alternative $\palt(\nuk)$ (dark blue) priors.
        (Right) The effect of the perturbation
        on the change in expected number of in-sample clusters
        for $t \in[0, 1]$.}\label{fig:iris_fsens}
\end{figure}

\end{knitrout}

We also see in this example that we can use the influence function to predict
the effect of functional changes to the stick-breaking prior. In the leftmost
column, we plot in purple the influence function in the logit
space.\footnote{\coryref{etafun_deriv_form_stick} expresses the influence
function in the stick domain $[0,1]$, but, for visualization, it is preferable
to express the influence function in the logit stick domain $\mathbb{R}$.  The
more general  \coryref{etafun_deriv_form} in \appref{diffable_nonparametric}
accomodates such transformations.}
According to \coryref{etafun_deriv_form_stick}, the sign and magnitude of the
effect of a perturbation should be determined by its integral against the
influence function.  Thus, when $\phi$ lines up with a negative part of $\infl$,
as in the first row, we expect the change to be negative.  Similarly, we expect
the perturbation of the bottom row to produce a positive change, and the middle
row, in which $\phi$ overlaps with both negative and positive parts of the
influence function, to produce a relatively small change. We see this intuition
borne out in the rightmost column.

\begin{knitrout}
\definecolor{shadecolor}{rgb}{0.969, 0.969, 0.969}\color{fgcolor}\begin{figure}[!h]

{\centering \includegraphics[width=0.980\linewidth,height=0.412\linewidth]{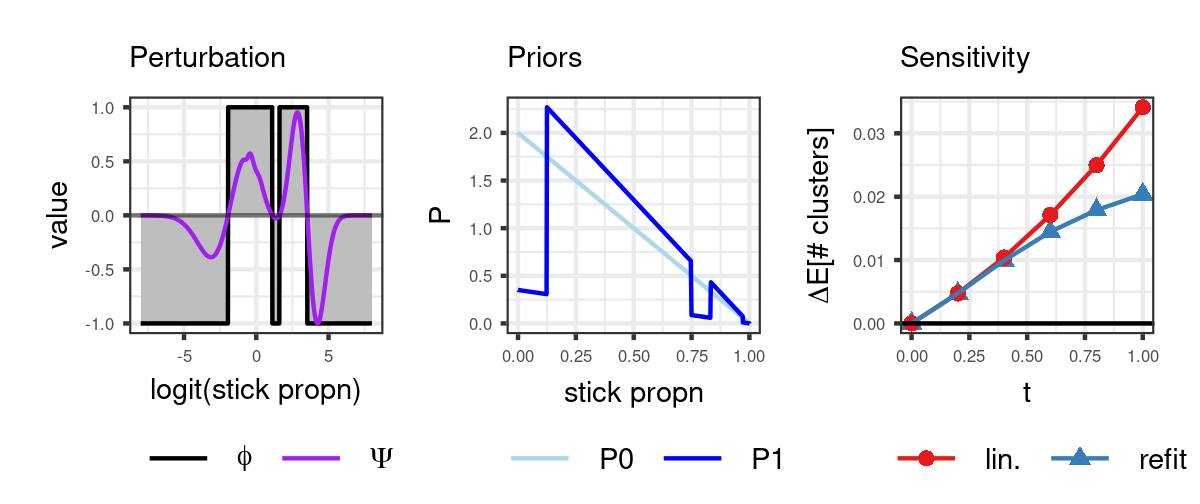}

}

\caption[Sensitivity of
        the expected number of in-sample clusters in the iris data set
        to the worst-case multiplicative perturbation with
        $\norminf{\phi} = 1$]{Sensitivity of
        the expected number of in-sample clusters in the iris data set
        to the worst-case multiplicative perturbation with
        $\norminf{\phi} = 1$.}\label{fig:iris_worstcase}
\end{figure}

\end{knitrout}

\noindent \textbf{Worst-case functional perturbation.}
Finally, \figref{iris_worstcase} shows the worst-case multiplicative
perturbation with $\norminf{\phi} = 1$, as given by
\coryref{etafun_worst_case_stick}, along with its effect on the prior and
$\gclustersabbr$. As expected, this worst-case perturbation has a much larger
effect on $\gclustersabbr$ compared to the other unit-norm perturbations in
\figref{iris_fsens}. However, even with the worst-case perturbation---which
results in an unreasonably shaped prior density---the change in $\gclustersabbr$
is still small. We conclude that $\gclustersabbr$ appears to be a robust
quantity for this model and dataset.

    \subsection{Regression mixture modeling}
    \seclabel{results_mice}

We next check our approximation on a more complex clustering task: clustering
time series, with a co-clustering matrix (and summaries thereof) as the quantity
of interest.

\noindent \textbf{Data and model.}
We use a publicly available data set of mice gene expression
\citep{shoemaker:2015:ultrasensitive}. Mice were infected with influenza virus,
and expression levels of a set of genes were assessed at 14 time points after
infection. Three measurements were taken at each time point (called biological
replicates), for a total of $\ntimepoints = 42$ measurements per gene.

The goal of the analysis is to cluster the time-course gene expression data
under the assumption that genes with similar time-course behavior may have
similar function. Clustering gene expressions is often used for exploratory
analysis and is a first step before further downstream investigation. It is
important, therefore, to ascertain the stability of the discovered clusters.

The left plot of \figref{example_genes} in \appref{app_mice} shows the
measurements of a single gene over time. We model each gene as belonging to a
latent component, where each component defines a smooth expression curve over
time. Then, observations are drawn by adding i.i.d.\ noise to the smoothed curve
along with a gene-specific offset. Following \citet{Luan:2003:clustering}, we
construct the smoothers using cubic B-splines.

Let $\x_\n\in\mathbb{R}^\ntimepoints$ be measurements of gene $\n$ at
$\ntimepoints$ time points. Let $\regmatrix$ be the $\ntimepoints \times \d$
B-spline regressor matrix, so that the $ij$-th entry of $\regmatrix$ is the
$j$-th B-spline basis vector evaluated at the $i$-th time point. The right plot
of \figref{example_genes} in \appref{app_mice} shows the B-spline basis. The
distribution of the data arising from component $k$ is
\begin{align}\eqlabel{mice_model}
\p(\x_\n | \beta_\k, \b_\n) =
\normdist{\x_\n | \regmatrix\mu_\k + \b_\n,
\tau_\k^{-1}I_{\ntimepoints \times \ntimepoints}},
\end{align}
where $\b_\n$ is a gene-specific additive offset and $I$ is the identity matrix.
We include the additive offset because we are interested in clustering gene
expressions based on their patterns over time, not their absolute level. In this
model, the component-specific parameters are $\beta_\k = (\mu_\k, \tau_\k)$, the
regression coefficients and the inverse noise variance. The component
frequencies are determined by stick-breaking according to $\nu$, and cluster
assignments $z$ are drawn as in \secref{model_bnp}.

%

Our variational approximation factorizes similarly to \eqref{vb_mf} except with
an additional factor for the additive shift. In our variational approximation,
we also make a simplification by letting $\q(\beta_\k \vert \eta) = \delta
(\beta_k \vert \eta)$, where $\delta(\cdot \vert \eta)$ denotes a point mass at
a parameterized location. See \appref{app_mice} for further details concerning
the model and variational approximation.

\noindent \textbf{Quantity of interest: the co-clustering matrix and summaries.}
In this application, we are particularly interested in which genes cluster
together, so we focus on the posterior co-clustering matrix.  Let
$\gcoclustering(\eta)\in\mathbb{R}^{\N\times\N}$ denote the matrix whose
$(i,j)$-th entry is the posterior probability that gene $i$ belongs to the same
cluster as gene $j$, given by
\begin{align*}
[\gcoclustering(\eta)]_{ij} =
\expect{\q(\z\vert\eta)}{\ind{\z_{i} = \z_{j}}}  =
\begin{cases}
\sum_{k=1}^{\kmax}\left(\expect{\q(\z_i\vert\eta)}{\z_{ik}}
\expect{\q(\z_j\vert\eta)}{\z_{jk}}\right)
& \text{for } i \not= j\\
1 & \text{for } i = j.
\end{cases}
\end{align*}
\figref{gene_initial_coclustering} shows the inferred
co-clustering matrix at $\alpha_0$.

\begin{knitrout}
\definecolor{shadecolor}{rgb}{0.969, 0.969, 0.969}\color{fgcolor}\begin{figure}[!h]

{\centering \includegraphics[width=0.588\linewidth,height=0.470\linewidth]{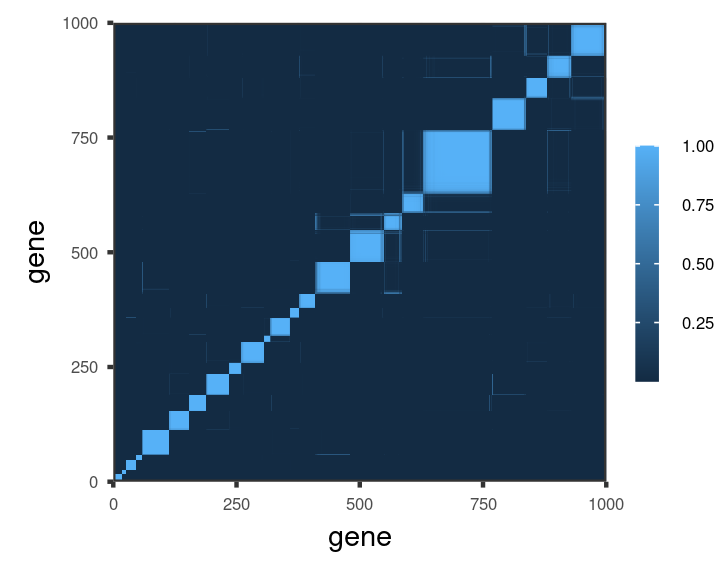} 

}

\caption[The inferred co-clustering matrix of gene expressions at $\alpha_0 = 6.$ ]{The inferred co-clustering matrix of gene expressions at $\alpha_0 = 6.$ }\label{fig:gene_initial_coclustering}
\end{figure}

\end{knitrout}

Below, we will use the influence function (\coryref{etafun_worst_case_stick}) to
try and find a perturbation that produces large changes in the co-clustering
matrix. To compute the worst-case perturbation, we must choose a univariate
summary of the $\ngenes^2$-dimensional co-clustering matrix whose derivative we
wish to extremize. We use the sum of the eigenvalues of the symmetrically
normalized graph Laplacian, as given by
\begin{align*} \laplacianevsum(\eta) = \text{Tr}\left( I - D(\eta)^{-1/2}
\gcoclustering(\eta) D(\eta)^{-1/2} \right), \end{align*}
where $D(\eta)^{-1/2}$ is the diagonal matrix with entries $d_i =
\sum_{j=1}^{\ngenes}[\gcoclustering(\eta)]_{ij}$. The quantity $\laplacianevsum$
is differentiable, and has close connection with the number of distinct
components in a graph~\citep{luxburg:2007:spectralcluster}. We expect that prior
perturbations that produce large changes in $\laplacianevsum$ will also produce large changes in the full co-clustering matrix.

\noindent \textbf{Sensitivity to the concentration parameter.}
We first evaluate the sensitivity of the co-clustering matrix $\gcoclustering$
to the choice of $\alpha$ in the stick-breaking prior.

We start at $\alpha = \alpha_0 = 6$.
We use the linear approximation to extrapolate the co-clustering matrix
under prior parameters
$\alpha = 0.1$ and $\alpha = 12$.
The {\em a priori} expected number of clusters in the original data at these values
is 2 and 50, respectively.
Despite this wide prior range, the change in the posterior
co-clustering matrix for each $\alpha$ is minuscule (\figref{gene_alpha_coclustering}). The
largest absolute changes in the co-clustering matrix is of order $10^{-2}$.
Refitting the approximate posterior at $\alpha = 0.1$ and
$\alpha = 12$ confirms the insensitivity predicted by the
linearized variational global parameters. Beyond capturing insensitivity, the
linearized parameters were also able to capture the sign and size of the
changes in the individual entries of the co-clustering matrix, even though these
changes are small.

\begin{knitrout}
\definecolor{shadecolor}{rgb}{0.969, 0.969, 0.969}\color{fgcolor}\begin{figure}[!h]

{\centering \includegraphics[width=0.980\linewidth,height=0.784\linewidth]{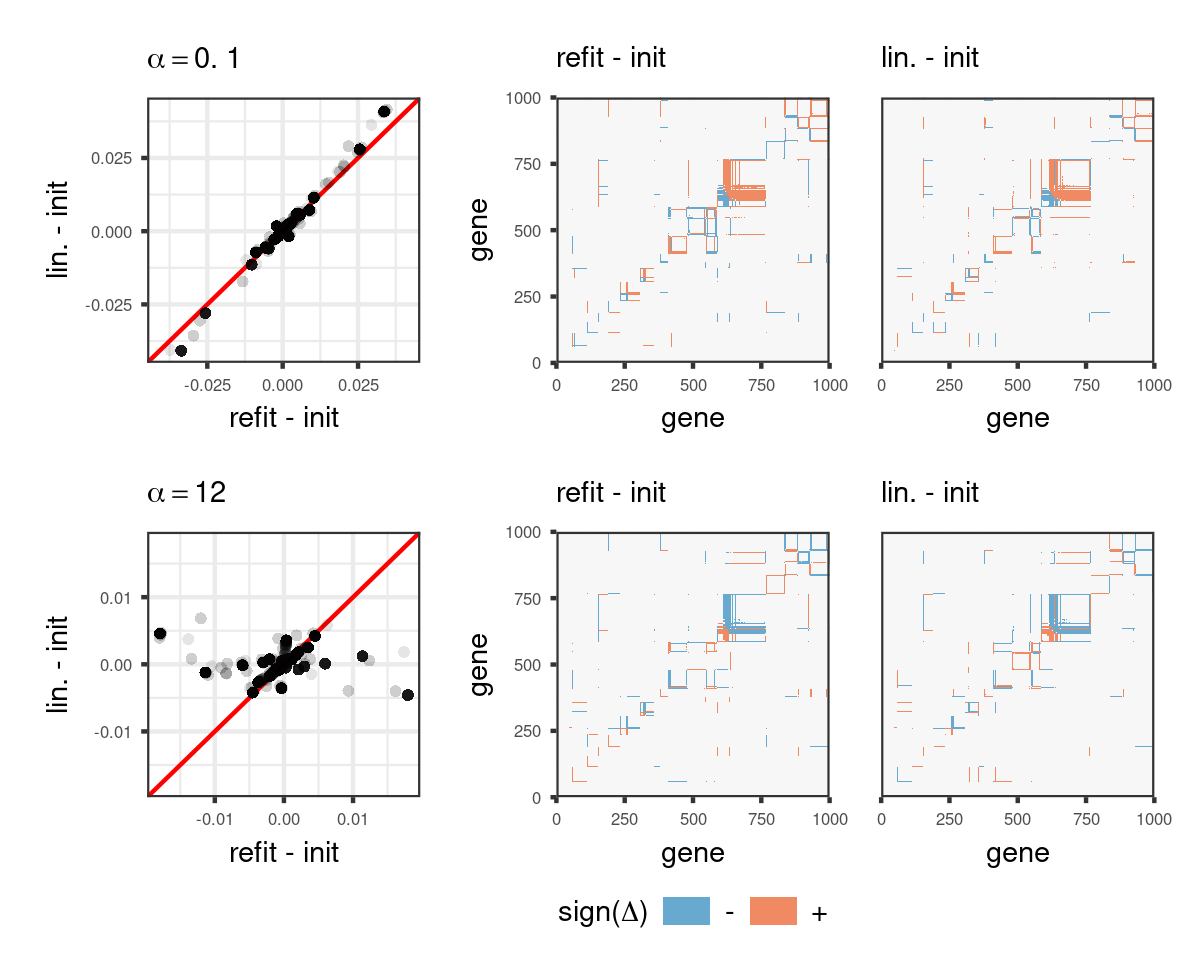} 

}

\caption[Differences in the
     co-clustering matrix at $\alpha = 0.1$ (top row)
     and $\alpha = 12$ (bottom row),
     relative to the co-clustering matrix at $\alpha_0 = 6$.
     (Left) A scatter plot of differences under the linear approximation
     against differences after refitting.
     Each point represents an entry of the co-clustering matrix.
     Note the scales of the axes:
     the largest change in an entry of the co-clustering matrix is
     $\approx 0.03$.
     (Middle) Sign changes in the co-clustering matrix observed after refitting,
     ignoring the magnitude of the change.
     (Right) Sign changes under the linearly approximated variational
     parameters.
     For visualization, changes with absolute value $< 10^{-5}$ are not colored]{Differences in the
     co-clustering matrix at $\alpha = 0.1$ (top row)
     and $\alpha = 12$ (bottom row),
     relative to the co-clustering matrix at $\alpha_0 = 6$.
     (Left) A scatter plot of differences under the linear approximation
     against differences after refitting.
     Each point represents an entry of the co-clustering matrix.
     Note the scales of the axes:
     the largest change in an entry of the co-clustering matrix is
     $\approx 0.03$.
     (Middle) Sign changes in the co-clustering matrix observed after refitting,
     ignoring the magnitude of the change.
     (Right) Sign changes under the linearly approximated variational
     parameters.
     For visualization, changes with absolute value $< 10^{-5}$ are not colored. }\label{fig:gene_alpha_coclustering}
\end{figure}

\end{knitrout}

\noindent \textbf{Sensitivity to functional perturbations.}
We now investigate sensitivity of the co-clustering matrix
to deviations from the beta prior.
In \figref{gene_fpert_coclustering}, we use the influence function for
$\laplacianevsum$ to construct a nonparametric prior perturbation that we
expect to have a large, positive effect.  The resulting prior does indeed
produce changes an order of magnitude larger than those produced by
the perturbations to $\alpha$ shown in \figref{gene_alpha_coclustering},
and our approximation
is again able to capture the qualitative changes.
The influence function is also able to
explain why $\alpha$ perturbations were unable to produce large changes in this
case: \figref{alpha_pert_logphi} shows that changing $\alpha$ (as in
\exref{beta_inf_norm}) induces large changes in the prior only where the
influence function is small.

\begin{knitrout}
\definecolor{shadecolor}{rgb}{0.969, 0.969, 0.969}\color{fgcolor}\begin{figure}[!h]

{\centering \includegraphics[width=0.980\linewidth,height=0.823\linewidth]{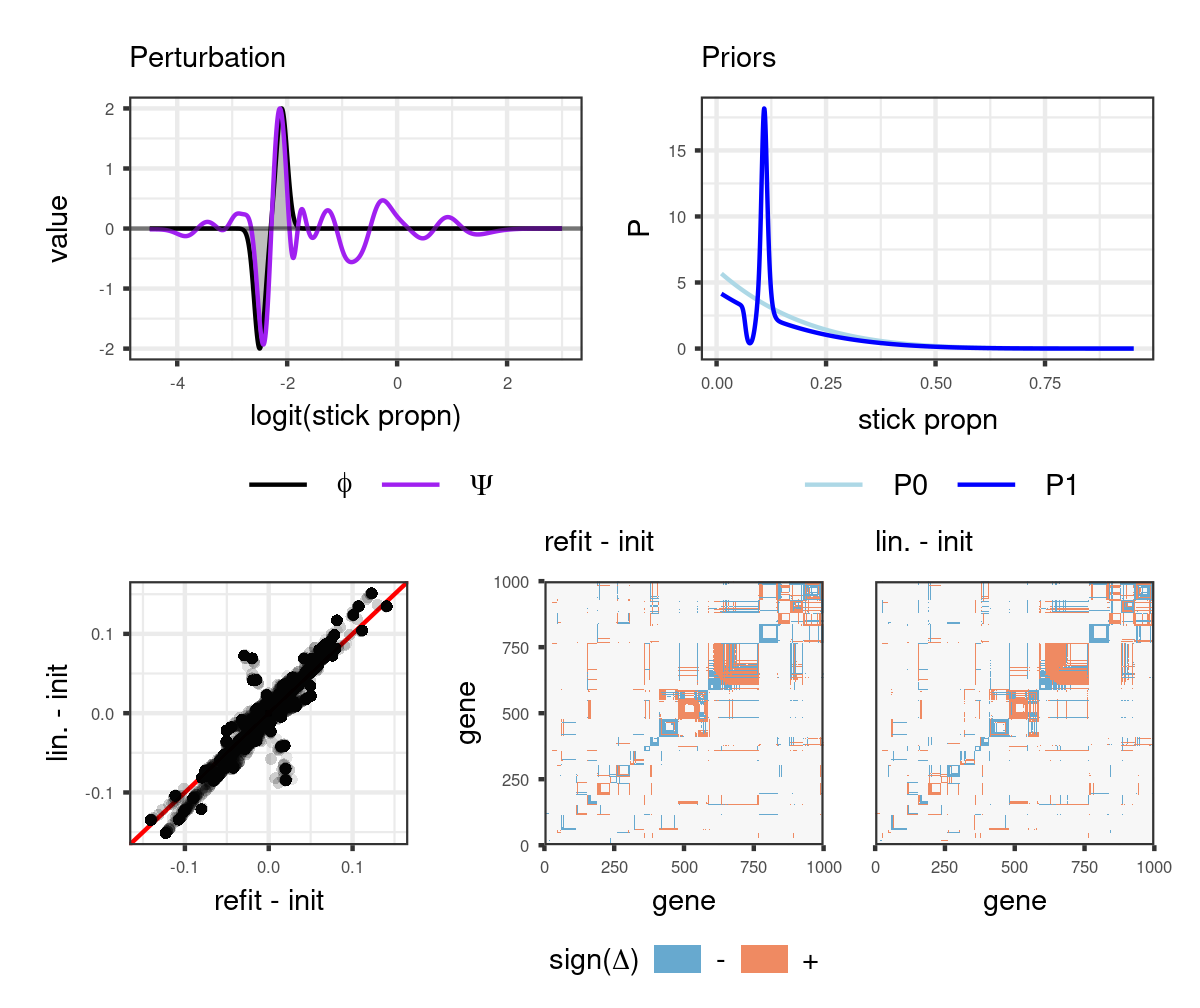} 

}

\caption[Effect on the co-clustering matrix of a multiplicative functional
     perturbation.
     (Top left) The perturbation $\phi$ is in grey,
     and the influence function is in purple.
     (Top right) The effect of this perturbation on the prior density.
     (Bottom) The effect of this perturbation on
    the co-clustering matrix.
    Note the scale of the scatter plot axes compared with
    the scatter plots in \figref{gene_alpha_coclustering}]{Effect on the co-clustering matrix of a multiplicative functional
     perturbation.
     (Top left) The perturbation $\phi$ is in grey,
     and the influence function is in purple.
     (Top right) The effect of this perturbation on the prior density.
     (Bottom) The effect of this perturbation on
    the co-clustering matrix.
    Note the scale of the scatter plot axes compared with
    the scatter plots in \figref{gene_alpha_coclustering}. }\label{fig:gene_fpert_coclustering}
\end{figure}

\end{knitrout}

\begin{knitrout}
\definecolor{shadecolor}{rgb}{0.969, 0.969, 0.969}\color{fgcolor}\begin{figure}[!h]

{\centering \includegraphics[width=0.882\linewidth,height=0.423\linewidth]{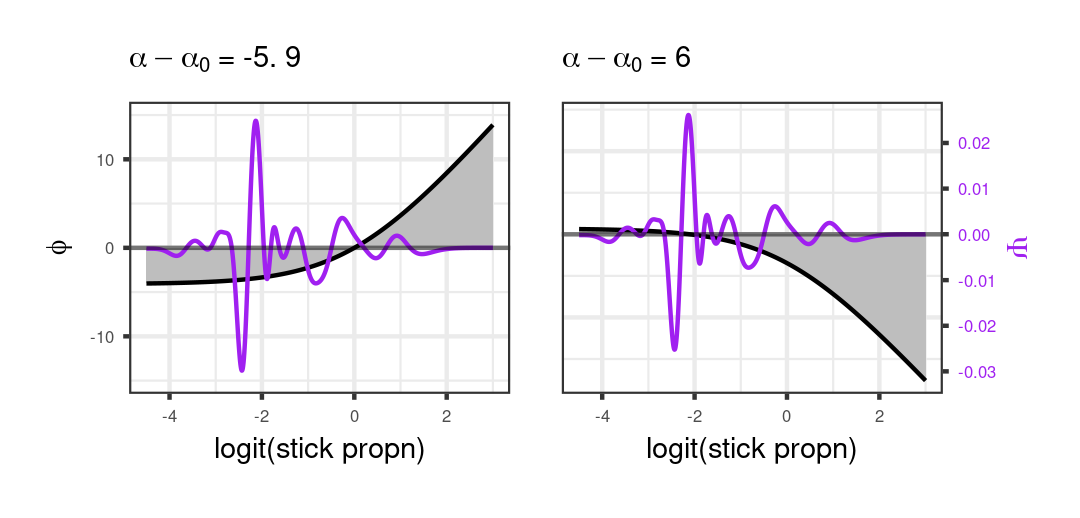} 

}

\caption[The multiplicative perturbations $\phi_\alpha(\cdot)$ that
    corresponds to decreasing (left) or increasing (right)
    the $\alpha$ parameter]{The multiplicative perturbations $\phi_\alpha(\cdot)$ that
    corresponds to decreasing (left) or increasing (right)
    the $\alpha$ parameter. }\label{fig:alpha_pert_logphi}
\end{figure}

\end{knitrout}

However, even with the (unreasonable-looking) selected functional perturbation,
the size of the differences in the co-clustering matrix remains modest. It is
unlikely that any scientific conclusions derived from the co-clustering matrix would have
changed after the functional perturbation. The co-clustering matrix appears
robust to perturbations in the stick-breaking distribution.

    \subsection{Genetic admixture modeling with fastSTRUCTURE}
    \seclabel{results_structure}

Our final analysis illustrates the use of our approximation for stick-breaking priors
beyond clustering; namely, in topic modeling.

\noindent \textbf{Data and model.}
We use a publicly available dataset that contains genotypes from $\nindiv=155$ individuals of an
endangered bird species, the Taita thrush
\citep{galbusera:2000:thrush}. Individuals were collected from four
regions in southeast Kenya (Chawia, Mbololo, Ngangao, Yale), and each individual
was genotyped at $\nloci=7$ micro-satellite loci. The four regions were once part of
a cohesive cloud forest that has been fragmented by human development. For
this endangered bird species, understanding the degree to which populations have
grown genetically distinct is important for conservation efforts: well-separated
populations with little genetic diversity are particularly at risk of
extinction.  The goal of the analysis is to infer the population of origin for specific loci and estimate the degree to which populations are admixed in each
individual.

Let $\x_{\n\l\i}\in\{1, \ldots, J_\l\}$ be the observed genotype for
individual $\n$ at locus $\l$ and chromosome $\i$. $J_\l$ is the number of
possible genotypes at locus $\l$. For example, if the measurements are all
single nucleotides (A, T, C or G) then $J_\l = 4$ for all $\l$.

A latent population is characterized by the collection $\beta_k =
(\latentpop_{\k1}, \ldots, \latentpop_{\k\nloci})$, where
$\latentpop_{\k\l}\in\Delta^{J_\l - 1}$ are the latent frequencies for the $J_l$
possible genotypes at locus $\l$. Let $\z_{\n\l\i}$ be the assignment of
observation $\x_{\n\l\i}$ to a latent population. Notice that for a given
individual $\n$, different loci (or even different chromosomes at a given locus)
may have different population assignments. The distribution of
$\x_{\n\l\i}\in\{1, \ldots, J_\l\}$ arising from population $\k$ is
$
\p(\x_{\n\l\i} \vert \latentpop_{\k}) =
\categoricaldist{\x_{\n\l\i}\vert \latentpop_{\k\l}}$.

Unlike the previous models, we now have a stick-breaking process for each
individual. Draw sticks
\begin{align*}
\nu_{\n\k} \indep \pstick(\nu_{\n\k}), \quad \n = 1, \ldots, \nindiv; \k = 1, 2, \ldots.
\end{align*}
The prior assignment probability vector $\latentadmix_{\n} =
(\latentadmix_{\n1}, \latentadmix_{\n2}, \ldots)$, now unique to each
individual, is formed by the same stick-breaking construction as before,
\begin{align*}
\latentadmix_{\n\k} = \nu_{\n\k} \prod_{\k' < \k} (1 - \nu_{\n\k'}).
\end{align*}
The population assignment $\z_{\n\l\i}$ is drawn from a multinomial
distribution
\begin{align*}
p(\z_{\n\l\i} | \latentadmix_\n) = \prod_{k=1}^{\infty} \latentadmix_{\n\k}^{\z_{\n\l\i\k}}.
\end{align*}
In this genetics application, we call $\latentadmix_{\n}$ the \textit{admixture}
of individual $\n$.

Initially we take $\pstick$ to be $\betadist{1,\alpha}$ with parameter $\alpha = \alpha_0 = 3$.
The choice of
$\alpha_0 = 3$ corresponds to roughly four distinct populations {\em a priori}, in agreement with the observation that the individuals come
from four geographic regions. Below, we will evaluate sensitivity to this prior
choice.

This model is identical to fastSTRUCTURE, a model proposed in
\citet{pritchard:2000:structure} and \citet{raj:2014:faststructure}, except that we replace
the Dirichlet prior in fastSTRUCTURE with an infinite stick-breaking process.
The result is a model similar to a hierarchical Dirichlet process for topic
modeling \citep{teh:2006:hdp}, but without the top-level Dirichlet process. In
addition, genotypes at genetic markers take the place of words in a document; in
lieu of inferring ``topics," we infer latent populations.

We use a mean-field variational approximation, and all distributions are
conditionally conjugate except for the stick-breaking proportions, which remain
logit-normal. See \appref{app_structure} for further details.

\noindent \textbf{Quantity of interest.}
The posterior quantities of interest in this application are the admixtures
$\pi_\n$. \figref{stru_init_fit} plots the inferred admixtures
$\expect{\q(\pi_\n \vert \etaopt)}{\pi_\n}$ for all individuals $\n$.

In the approximate posterior with $\alpha_0$, there appear to be three dominant
latent populations, which we arbitrarily label as populations 1, 2, and 3
(top panel of \figref{stru_func_sens}). The inferred admixture proportions generally
correspond with geographic regions: Mbololo individuals are primarily population
1, Ngangao individuals are primarily population 2, and Chawia individuals are a
mixture of populations 1, 2, and 3 (\figref{stru_init_fit} in \appref{app_structure}).

Notably, outlying admixtures among individuals from the same geographic region provide clues into the historical migration patterns of this species.
For example,
while most Mbololo individuals are dominantly population 1,
several Mbololo individuals have abnormally
large admixture proportions of population 2.
Conversely, while most Ngangao individuals are dominantly population 2,
several Ngangao individuals have abnormally large admixture
proportions of population 1. These patterns suggest that some migration has occurred
between the Mbololo and Ngangao regions.

We evaluate the sensitivity of this conclusion to possible prior perturbations.
Define the posterior quantity
\begin{align*}
\gadmix(\eta; \mathcal{N}, k) =
 \expect{\q(\pi\vert\eta)}{\frac{1}{|\mathcal{N}|}\sum_{n\in\mathcal{N}}
\pi_{\n\k}},
\end{align*}
the average admixture proportion of population $\k$ in a set of
individuals $\mathcal{N}$.

Below, we consider $\gadmix$ with three different sets of individuals:
$\mathcal{N} = \{26, ..., 31\}$, corresponding to
the outlying Mbololo individuals, labeled
``A" in \figref{stru_func_sens};
$\mathcal{N} = \{125, ..., 128\}$, corresponding to the four
outlying Ngangao individuals, labeled ``B";
and $\mathcal{N} = \{139, ..., 155\}$ corresponding to all Chawia individuals,
labeled ``C".
For individuals A, we let $\k=2$ in $\gadmix$ and
examine the robustness of the presence of population 2;
for individuals B, we use $\k = 1$;
and for individuals C, we use $\k = 3$.
The first two posterior quantities relate to the inferred migration between
the Mbololo and Ngangao regions.
In the last example, we study the robustness of having a third latent
population present, a population that primarily appears in Chawia individuals.

\noindent \textbf{Functional sensitivity.}
We construct worst-case negative perturbations
for each of the three variants of $\gadmix$, in order to see
whether the biologically interesting patterns can be made to disappear
with different prior choices.
\figref{stru_func_sens} shows the result of the worst-case perturbations
on the prior density and $\gadmix$.
After the worst-case perturbation,
the admixture proportion of population 2 in individuals A
was nearly halved.
On the other hand, the admixture of population 1 in individuals B
is more robust.
We conclude that the inferred migration from Ngangao to
Mbololo is relatively robust to the stick-breaking prior, while
conclusions about migration from Mbololo to Ngangao may be dependent on prior choices.

\begin{knitrout}
\definecolor{shadecolor}{rgb}{0.969, 0.969, 0.969}\color{fgcolor}\begin{figure}[!h]

{\centering \includegraphics[width=0.980\linewidth,height=1.098\linewidth]{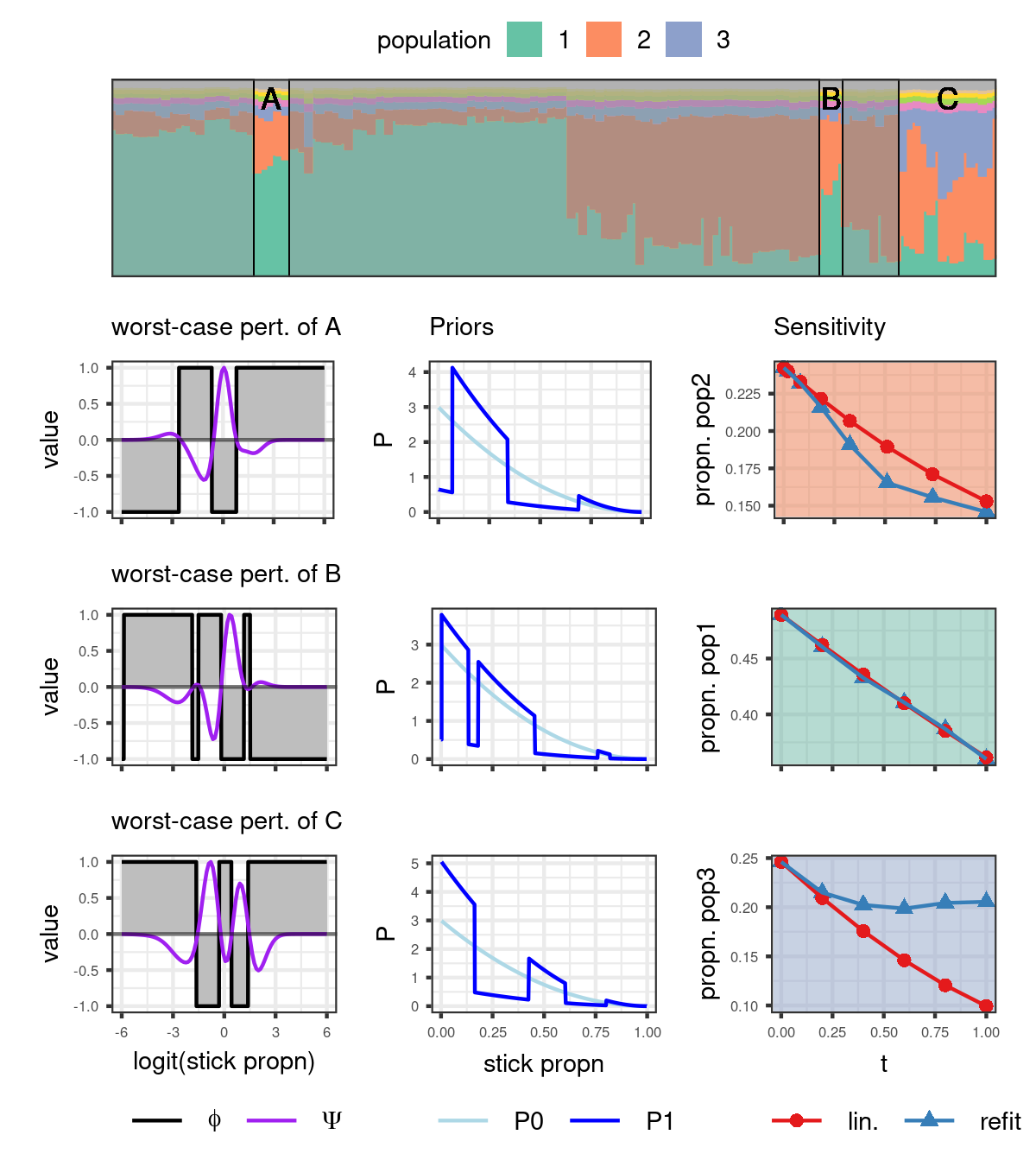} 

}

\caption[Sensitivity of inferred admixtures for several outlying individuals.
     For individuals A,
     we examine the sensitivity of the admixture proportion of population 2.
     For individuals B,
     we examine the population 1 admixture.
     For the individuals C, we examine the population 3 admixture.
     (Left column) The worst-case negative perturbation with
     $\norminf{\phi} = 1$
     in grey,
     plotted against the influence function in purple
     (scaled such that $\norminf{\psi} = 1$).
    (Middle column) The effect of the perturbation on the prior density.
    (Right column) Effects on the inferred admixture]{Sensitivity of inferred admixtures for several outlying individuals.
     For individuals A,
     we examine the sensitivity of the admixture proportion of population 2.
     For individuals B,
     we examine the population 1 admixture.
     For the individuals C, we examine the population 3 admixture.
     (Left column) The worst-case negative perturbation with
     $\norminf{\phi} = 1$
     in grey,
     plotted against the influence function in purple
     (scaled such that $\norminf{\psi} = 1$).
    (Middle column) The effect of the perturbation on the prior density.
    (Right column) Effects on the inferred admixture. }\label{fig:stru_func_sens}
\end{figure}

\end{knitrout}

In this data set and model,
the conclusions from the linear approximation did not
always agree with the conclusions from refitting the
variational approximation.
For example, the admixture proportion of population 3 in individuals C were predicted to more sensitive by our linear approximation than
were actually observed after refitting (\figref{stru_func_sens}, bottom row).

Moreover, even though the linear approximation
agreed with the refits for individuals A in
overall admixture proportion
(\figref{stru_func_sens}, second row),
the approximation does not perform uniformly well over all individuals.
\figref{stru_func_sens_admix} plots the inferred admixtures
computed using the linearized variational parameters
and the refitted variational parameters.
The admixture proportion of population 2 in individual $n = 25$
dramatically increased after refitting with the perturbed prior;
the linearized parameters failed to reproduce this change.

Even though linear approximation works less well in this example,
the influence function is still able to guide our choice of
functional perturbations at which to refit.
While the worst-case perturbations we used
may be an adversarial choice,
the influence function suggests that
we can construct a smoother perturbation
with a similar effect as the worst-case,
as we did in \secref{results_mice}.
Importantly, as we will note in the next subsection,
the influence function is cheap to compute relative to refitting.
For a further discussion of the limitations of the linear approximation,
see \appref{app_structure_results}.

\begin{knitrout}
\definecolor{shadecolor}{rgb}{0.969, 0.969, 0.969}\color{fgcolor}\begin{figure}[!h]

{\centering \includegraphics[width=0.980\linewidth,height=0.392\linewidth]{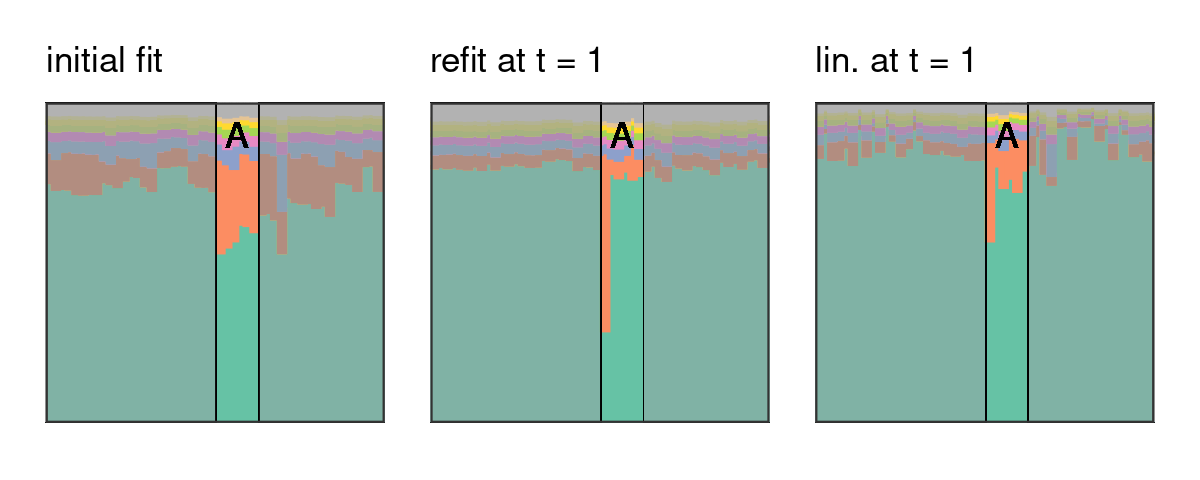} 

}

\caption{Inferred admixtures after the worst-case perturbation
     to individuals ``A" (see Figure~\ref{fig:stru_func_sens} for perturbation). }\label{fig:stru_func_sens_admix}
\end{figure}

\end{knitrout}

    \subsection{Computation time}
    \seclabel{compute_time}

The relative computational costs of the approximation and re-fitting for our
three experiments are shown in \tabref{timing_table}. The data sets we
considered in our experiments had varying degrees of complexity, and the
computational of cost of fitting the variational approximation thus also varies
accordingly. However, the cost of forming the linear approximation---the step
that requires computing and inverting the Hessian matrix---was consistently
roughly an order of magnitude faster than refitting.

Recall from \secref{computing_sensitivity} that the solution of a linear system
involving $\hessopt^{-1}$ is the computationally intensive part of the linear
approximation, and that the linear system needs to be solved only once for a
given perturbation, as described in \secref{computing_sensitivity}.  Consistent with
this observation, in all the examples, after the linear approximation is formed,
extrapolating to any new prior parameter $\alpha \ne \alpha_0$ or $\t \ne
0$ takes only fractions of a second.
For example, in the thrush data and fastSTRUCTURE model, the initial fit took
seven seconds, with subsequent refits (which we warm-started with the
initial fit) taking between five and ten seconds.  Solving a linear system to
form the linear approximation for a particular perturbation $\phi$ took less
than a second, and evaluating $\etaopt(\phi)$ was essentially free.

\begin{table}[tb]
\centering
\caption{Compute time in seconds of various quantities on each data set.
Reported times for $\etaopt(\alpha)$ and $\etalin(\alpha)$ are
median times over the set of considered $\alpha$'s.
The reported influence function time is the time required to
evaluate the influence function on a grid of 1000 points. }
\tablabel{timing_table}
\begin{tabular}{|r|r|r|r|}
    \hline
    & iris & mice  & thrush \\
    \hline
    Initial fit &
    1 &
    30 &
    7 \\
    \hline
    Hessian solve for $\alpha$ sensitivity &
    0.02 &
    3 &
    0.3 \\
    Linear approx. $\etalin(\alpha)$ &
    0.0008 &
    0.001 &
    0.0008 \\
    Refits $\etaopt(\alpha)$ &
    0.5 &
    30 &
    5 \\
    \hline
    \shortstack{ \\ The influence function \\ (at 1000 grid points)}  &
    0.09 &
    3 &
    0.6 \\
    \hline
    Hessian solve for $\phi$ &
    0.02 &
    3 &
    0.4\\
    Linear approx. $\etalin(\phi)$ &
    0.001 &
    0.001 &
    0.0008 \\
    Refit $\etalin(\phi)$ &
    0.6 &
    20 &
    10 \\
    \hline
\end{tabular}
\end{table}

\section{Conclusion}
\seclabel{conclusions}
We provide a method to approximate the effect of changing a BNP prior on a
posterior quantity of interest in a VB approximation.  Our method is generally
applicable, straightforward to implement, and computationally efficient.  In our
experiments, we show by refitting that the predictions of the approximation are
typically qualitatively accurate. Over the range of situations and quantities of
interest we considered, we discovered robustness (co-clustering in the mice
dataset), non-robustness (the predictive number of clusters in the iris
dataset), and intermediate cases where some closely related posterior quantities
were robust and others not (the population memberships of the thrush dataset).
Given such a variety of outcomes, the authors hesitate to draw any
generalizable conclusions about the robustness of DPM models, much less generic
BNP posteriors.  On the contrary, we hope that our results motivate users to
check for robustness directly, for their particular datasets and models of
interest, rather than trying to rely too heavily on intuition or general
principles. Indeed, we hope that the ease with which one can compute our prior
sensitivity measures, combined with the possibility of uncovering materially
important non-robustness, will encourage the widespread adoption of routine
prior robustness checks.

In the present work, we have focused only on the task of detecting and
characterizing non-robustness.  We have not attempted to address the critical
questions of what to do when a conclusion is non-robust, nor how to make robust
modeling choices. We hope that routine, widespread robustness checking in  a
variety of real-world problems will further inform and motivate solutions to the
important question of how to deal with, end even prevent, non-robustness in
practice.

Though the need for prior robustness checking seems particularly well-motivated
in discrete BNP problems, where the DPM prior is often chosen for computational
convenience rather than considered subjective belief, prior robustness checks
are relevant to almost all Bayesian analysis.  Despite our present focus on BNP
models and the DPM prior in particular, our methodology extends not only to
other truncated approximations of discrete BNP priors, but to any VB
approximation based on reverse KL divergence.  The best evidence for the
usefulness of our methodology of linear approximation will come from widespread
adoption and verification in many different applications and modeling
environments, and we hope that the present work is only the beginning.

\ifbool{arxiv} {
    \paragraph{Acknowledgments.}
        We are indebted to helpful discussions with Nelle Varoquaux, Matthew Stephens,
Michael C. Hughes, Eric Sudderth, and Jake Soloff, and to useful suggestions
from anonymous reviewers.  Runjing Liu is supported by the National Science
Foundation graduate research fellowship program. Ryan Giordano and Tamara
Broderick were supported in part by an NSF CAREER Award and an ONR Early Career
Grant.

} {
    \clearpage
}

\ifbool{arxiv} {
    \bibliographystyle{plainnat}
} {
    \bibliographystyle{ba}
}
\bibliography{references}

\begin{thebibliography}{55}
\newcommand{\enquote}[1]{``#1''}
\expandafter\ifx\csname natexlab\endcsname\relax\def\natexlab#1{#1}\fi
\expandafter\ifx\csname url\endcsname\relax
  \def\url#1{{\tt #1}}\fi
\expandafter\ifx\csname urlprefix\endcsname\relax\def\urlprefix{URL }\fi
\ifx\endbibitem\undefined \let\endbibitem\relax\fi

\bibitem[{Ambrogioni et~al.(2018)Ambrogioni, G{\"u}{\c{c}}l{\"u},
  G{\"u}{\c{c}}l{\"u}t{\"u}rk, Hinne, van Gerven, and
  Maris}]{ambrogioni:2018:wasserstein}
Ambrogioni, L., G{\"u}{\c{c}}l{\"u}, U., G{\"u}{\c{c}}l{\"u}t{\"u}rk, Y.,
  Hinne, M., van Gerven, M., and Maris, E. (2018).
\newblock \enquote{{Wasserstein} Variational Inference.}
\newblock {\em Advances in Neural Information Processing Systems\/}, 31:
  2473--2482.
\endbibitem

\bibitem[{Anderson(1936)}]{anderson:1936:iris}
Anderson, E. (1936).
\newblock \enquote{The species problem in iris.}
\newblock {\em Annals of the Missouri Botanical Garden\/}, 23(3): 457--509.
\endbibitem

\bibitem[{Averbukh and Smolyanov(1967)}]{averbukh:1967:theory}
Averbukh, V. and Smolyanov, O. (1967).
\newblock \enquote{The theory of differentiation in linear topological spaces.}
\newblock {\em Russian Mathematical Surveys\/}, 22(6): 201--258.
\endbibitem

\bibitem[{Barrios et~al.(2013)Barrios, Lijoi, Nieto-Barajas, and
  Pr{\"u}nster}]{barrios:2013:bnp}
Barrios, E., Lijoi, A., Nieto-Barajas, L., and Pr{\"u}nster, I. (2013).
\newblock \enquote{Modeling with normalized random measure mixture models.}
\newblock {\em Statistical Science\/}, 28(3): 313--334.
\endbibitem

\bibitem[{Basu(2000)}]{Basu:2000:robustnessBNP}
Basu, S. (2000).
\newblock {\em {B}ayesian Robustness and {B}ayesian Nonparametrics\/},
  223--240.
\newblock New York, NY: Springer New York.
\endbibitem

\bibitem[{Basu et~al.(1996)Basu, Jammalamadaka, and Liu}]{basu:1996:local}
Basu, S., Jammalamadaka, S.~R., and Liu, W. (1996).
\newblock \enquote{Local posterior robustness with parametric priors: {Maximum}
  and average sensitivity.}
\newblock In {\em Maximum Entropy and {Bayesian} Methods\/}, 97--106. Springer.
\endbibitem

\bibitem[{Baydin et~al.(2018)Baydin, Pearlmutter, Radul, and
  Siskind}]{baydin:2018:automatic}
Baydin, A., Pearlmutter, B., Radul, A., and Siskind, J. (2018).
\newblock \enquote{Automatic differentiation in machine learning: {A} survey.}
\newblock {\em Journal of Machine Learning Research\/}, 18.
\endbibitem

\bibitem[{Billingsley(1986)}]{billingsley:1986:probability}
Billingsley, P. (1986).
\newblock {\em Probability and Measure\/}.
\newblock John Wiley and Sons, second edition.
\endbibitem

\bibitem[{Bishop(2006)}]{bishop:2006:PRML}
Bishop, C.~M. (2006).
\newblock {\em Pattern Recognition and Machine Learning\/}.
\newblock Springer.
\endbibitem

\bibitem[{Blackwell and MacQueen(1973)}]{blackwell:1973:polyaurn}
Blackwell, D. and MacQueen, J.~B. (1973).
\newblock \enquote{{F}erguson distributions via {P}olya urn schemes.}
\newblock {\em The Annals of Statistics\/}, 1(2): 353 -- 355.
\endbibitem

\bibitem[{Blei and Jordan(2006)}]{blei:2006:vi_for_dp}
Blei, D. and Jordan, M.~I. (2006).
\newblock \enquote{{Variational inference for Dirichlet process mixtures}.}
\newblock {\em Bayesian Analysis\/}, 1(1): 121 -- 143.
\endbibitem

\bibitem[{Blei et~al.(2017)Blei, Kucukelbir, and
  McAuliffe}]{blei:2017:vi_review}
Blei, D., Kucukelbir, A., and McAuliffe, J. (2017).
\newblock \enquote{Variational inference: {A} review for statisticians.}
\newblock {\em Journal of the American Statistical Association\/}, 112(518):
  859–877.
\endbibitem

\bibitem[{Campbell et~al.(2019)Campbell, Huggins, How, and
  Broderick}]{campbell:2019:truncated}
Campbell, T., Huggins, J., How, J., and Broderick, T. (2019).
\newblock \enquote{Truncated random measures.}
\newblock {\em Bernoulli\/}, 25(2): 1256--1288.
\endbibitem

\bibitem[{Canale et~al.(2017)Canale, Lijoi, Nipoti, and
  Pr{\"u}nster}]{canale:2017:pitmanyor}
Canale, A., Lijoi, A., Nipoti, B., and Pr{\"u}nster, I. (2017).
\newblock \enquote{On the {Pitman--Yor} process with spike and slab base
  measure.}
\newblock {\em Biometrika\/}, 104(3): 681--697.
\endbibitem

\bibitem[{Cook(1986)}]{cook:1986:assessment}
Cook, D. (1986).
\newblock \enquote{Assessment of local influence.}
\newblock {\em Journal of the Royal Statistical Society: {S}eries {B}
  (Methodological)\/}, 48(2): 133--155.
\endbibitem

\bibitem[{Doshi et~al.(2009)Doshi, Miller, Van~Gael, and
  Teh}]{doshi:2009:ibpvariational}
Doshi, F., Miller, K., Van~Gael, J., and Teh, Y. (2009).
\newblock \enquote{Variational inference for the {Indian} buffet process.}
\newblock In {\em Artificial Intelligence and Statistics\/}, 137--144. PMLR.
\endbibitem

\bibitem[{Dudley(2018)}]{dudley:2018:real}
Dudley, R. (2018).
\newblock {\em Real Analysis and Probability\/}.
\newblock CRC Press.
\endbibitem

\bibitem[{Ferguson(1973)}]{ferguson:1973:bayesian}
Ferguson, T. (1973).
\newblock \enquote{A {B}ayesian analysis of some nonparametric problems.}
\newblock {\em The Annals of Statistics\/}, 209--230.
\endbibitem

\bibitem[{Fisher(1936)}]{fisher:1936:iris}
Fisher, R. (1936).
\newblock \enquote{The use of multiple measurements in taxonomic problems.}
\newblock {\em Annals of eugenics\/}, 7(2): 179--188.
\endbibitem

\bibitem[{Galbusera et~al.(2000)Galbusera, Lens, Schenck, Waiyaki, and
  Matthysen}]{galbusera:2000:thrush}
Galbusera, P., Lens, L., Schenck, T., Waiyaki, E., and Matthysen, E. (2000).
\newblock \enquote{Genetic variability and gene flow in the globally,
  critically-endangered Taita Thrush.}
\newblock {\em Conservation Genetics\/}, 1: 45--55.
\endbibitem

\bibitem[{Gelman et~al.(2013)Gelman, Carlin, Stern, Dunson, Vehtari, and
  Rubin}]{gelman:2013:bda}
Gelman, A., Carlin, J., Stern, H., Dunson, D., Vehtari, A., and Rubin, D.
  (2013).
\newblock {\em Bayesian Data Analysis, Third Edition\/}.
\newblock Chapman \& Hall/CRC Texts in Statistical Science. Taylor \& Francis.
\endbibitem

\bibitem[{Ghaderinezhad and Ley(2019)}]{ghaderinezhad:2019:stein}
Ghaderinezhad, F. and Ley, C. (2019).
\newblock \enquote{Quantification of the impact of priors in {Bayesian}
  statistics via {Stein’s} method.}
\newblock {\em Statistics \& Probability Letters\/}, 146: 206--212.
\endbibitem

\bibitem[{Giordano et~al.(2018)Giordano, Broderick, and
  Jordan}]{giordano:2018:covariances}
Giordano, R., Broderick, T., and Jordan, M.~I. (2018).
\newblock \enquote{Covariances, robustness and variational {B}ayes.}
\newblock {\em Journal of Machine Learning Research\/}, 19(51).
\endbibitem

\bibitem[{Gustafson(1996{\natexlab{a}})}]{gustafson:1996:marginal}
Gustafson, P. (1996{\natexlab{a}}).
\newblock \enquote{Local sensitivity of inferences to prior marginals.}
\newblock {\em Journal of the American Statistical Association\/}, 91(434):
  774--781.
\endbibitem

\bibitem[{Gustafson(1996{\natexlab{b}})}]{gustafson:1996:local}
--- (1996{\natexlab{b}}).
\newblock \enquote{Local sensitivity of posterior expectations.}
\newblock {\em Annals of Statistics\/}, 24(1): 174--195.
\endbibitem

\bibitem[{Gustafson(2000)}]{gustafson:2000:localrobustness}
--- (2000).
\newblock {\em Local Robustness in {B}ayesian Analysis\/}, 71--88.
\newblock New York, NY: Springer New York.
\endbibitem

\bibitem[{Hampel et~al.(2011)Hampel, Ronchetti, Rousseeuw, and
  Stahel}]{hampel:2011:robust}
Hampel, F., Ronchetti, E., Rousseeuw, P., and Stahel, W. (2011).
\newblock {\em Robust Statistics: {T}he Approach Based on Influence
  Functions\/}, volume 196.
\newblock John Wiley \& Sons.
\endbibitem

\bibitem[{Insua and Ruggeri(2000)}]{insua:2000:robust}
Insua, D.~R. and Ruggeri, F. (2000).
\newblock {\em Robust {B}ayesian Analysis\/}.
\newblock Springer.
\endbibitem

\bibitem[{Jaeckel(1972)}]{jaeckel:1972:infinitesimal}
Jaeckel, L. (1972).
\newblock \enquote{The Infinitesimal Jackknife, Memorandum.}
\newblock Technical report, MM 72-1215-11, Bell Lab. Murray Hill, NJ.
\endbibitem

\bibitem[{Jasra et~al.(2005)Jasra, Holmes, and
  Stephens}]{jasra:2005:mcmclabelswitch}
Jasra, A., Holmes, C., and Stephens, D. (2005).
\newblock \enquote{{{M}arkov chain {Monte Carlo} methods and the label
  switching problem in {B}ayesian mixture modeling}.}
\newblock {\em Statistical Science\/}, 20(1): 50 -- 67.
\endbibitem

\bibitem[{Krantz and Parks(2012)}]{krantz:2012:implicit}
Krantz, S. and Parks, H. (2012).
\newblock {\em The Implicit Function Theorem: {H}istory, Theory, and
  Applications\/}.
\newblock Springer Science \& Business Media.
\endbibitem

\bibitem[{Kucukelbir et~al.(2017)Kucukelbir, Tran, Ranganath, Gelman, and
  Blei}]{kucukelbir:2016:advi}
Kucukelbir, A., Tran, D., Ranganath, R., Gelman, A., and Blei, D. (2017).
\newblock \enquote{Automatic differentiation variational inference.}
\newblock {\em The Journal of Machine Learning Research\/}, 18(1): 430--474.
\endbibitem

\bibitem[{Li and Turner(2016)}]{li:2016:reyni}
Li, Y. and Turner, R. (2016).
\newblock \enquote{Variational inference with R{\'e}nyi divergence.}
\newblock {\em stat\/}, 1050: 6.
\endbibitem

\bibitem[{Lijoi et~al.(2007)Lijoi, Mena, and
  Pr{\"u}nster}]{lijoi:2007:reinforcement}
Lijoi, A., Mena, R., and Pr{\"u}nster, I. (2007).
\newblock \enquote{Controlling the reinforcement in Bayesian non-parametric
  mixture models.}
\newblock {\em Journal of the Royal Statistical Society: Series B (Statistical
  Methodology)\/}, 69(4): 715--740.
\endbibitem

\bibitem[{Liu and Wang(2016)}]{liu:2016:stein}
Liu, Q. and Wang, D. (2016).
\newblock \enquote{{Stein} variational gradient descent: {A} general purpose
  {B}ayesian inference algorithm.}
\newblock {\em Advances in Neural Information Processing Systems\/}, 29:
  2378--2386.
\endbibitem

\bibitem[{Luan and Li(2003)}]{Luan:2003:clustering}
Luan, Y. and Li, H. (2003).
\newblock \enquote{{Clustering of time-course gene expression data using a
  mixed-effects model with B-splines}.}
\newblock {\em Bioinformatics\/}, 19(4): 474--482.
\endbibitem

\bibitem[{Nielsen(1997)}]{nielsen:1997:measure}
Nielsen, O. (1997).
\newblock {\em An Introduction to Integration and Measure Theory\/}, volume~17.
\newblock Wiley-Interscience.
\endbibitem

\bibitem[{Nieto-Barajas and Prünster(2009)}]{barajas:2009:densitysens}
Nieto-Barajas, L. and Prünster, I. (2009).
\newblock \enquote{A sensitivity analysis for {B}ayesian nonparametric density
  estimators.}
\newblock {\em Statistica Sinica\/}, 19(2): 685--705.
\endbibitem

\bibitem[{Nocedal and Wright(2006)}]{nocedal:2006:numerical}
Nocedal, J. and Wright, S. (2006).
\newblock {\em Numerical Optimization\/}.
\newblock Springer Science \& Business Media.
\endbibitem

\bibitem[{Pritchard et~al.(2000)Pritchard, Stephens, and
  Donnelly}]{pritchard:2000:structure}
Pritchard, J., Stephens, M., and Donnelly, P. (2000).
\newblock \enquote{Inference of population structure using multilocus genotype
  data.}
\newblock {\em Genetics\/}, 155(2): 945--959.
\endbibitem

\bibitem[{Raj et~al.(2014)Raj, Stephens, and
  Pritchard}]{raj:2014:faststructure}
Raj, A., Stephens, M., and Pritchard, J.~K. (2014).
\newblock \enquote{fast{STRUCTURE}: {V}ariational inference of population
  structure in large {SNP} data sets.}
\newblock {\em Genetics\/}, 197(2): 573--589.
\endbibitem

\bibitem[{Ranganath et~al.(2014)Ranganath, Gerrish, and
  Blei}]{ranganath:2013:black}
Ranganath, R., Gerrish, S., and Blei, D. (2014).
\newblock \enquote{Black box variational inference.}
\newblock In {\em Artificial intelligence and statistics\/}, 814--822. PMLR.
\endbibitem

\bibitem[{Reeds(1976)}]{reeds:1976:thesis}
Reeds, J. (1976).
\newblock \enquote{On the definition of von {Mises} functionals.}
\newblock Ph.D. thesis, Statistics, Harvard University.
\endbibitem

\bibitem[{Roos et~al.(2015)Roos, Martins, Held, and
  Rue}]{roos:2015:sensitivity}
Roos, M., Martins, T., Held, L., and Rue, H. (2015).
\newblock \enquote{Sensitivity analysis for {Bayesian} hierarchical models.}
\newblock {\em Bayesian Analysis\/}, 10(2): 321--349.
\endbibitem

\bibitem[{Roychowdhury and Kulis(2015)}]{roychowdhury:2015:gammastick}
Roychowdhury, A. and Kulis, B. (2015).
\newblock \enquote{Gamma processes, stick-breaking, and variational inference.}
\newblock In {\em Artificial Intelligence and Statistics\/}, 800--808. PMLR.
\endbibitem

\bibitem[{Saha and Kurtek(2019)}]{saha:2019:geometricsens}
Saha, A. and Kurtek, S. (2019).
\newblock \enquote{Geometric sensitivity measures for {B}ayesian nonparametric
  density estimation models.}
\newblock {\em Sangkhya Series A.\/}, 81: 104--143.
\endbibitem

\bibitem[{Sethuraman(1994)}]{sethuraman:1994:constructivedp}
Sethuraman, J. (1994).
\newblock \enquote{A constructive definition of {D}irichlet priors.}
\newblock {\em Statistica Sinica\/}, 639--650.
\endbibitem

\bibitem[{Shoemaker et~al.(2015)Shoemaker, Fukuyama, Eisfeld, Zhao, Kawakami,
  Sakabe1, Maemura, Gorai, Katsura, Muramoto, Watanabe, Watanabe, Fuji,
  Matsuoka, Kitano, and Kawaoka}]{shoemaker:2015:ultrasensitive}
Shoemaker, J., Fukuyama, S., Eisfeld, A., Zhao, D., Kawakami, E., Sakabe1, S.,
  Maemura, T., Gorai, T., Katsura, H., Muramoto, Y., Watanabe, S., Watanabe,
  T., Fuji, K., Matsuoka, Y., Kitano, H., and Kawaoka, Y. (2015).
\newblock \enquote{An ultrasensitive mechanism regulates influenza
  virus-induced inflammation.}
\newblock {\em PLoS Pathogens\/}, 11(6): 1--25.
\endbibitem

\bibitem[{Sivaganesan(2000)}]{sivaganesan:2000:globallocal}
Sivaganesan, S. (2000).
\newblock \enquote{Global and local robustness approaches: {U}ses and
  limitations.}
\newblock In {\em Robust Bayesian Analysis\/}, 89--108. Springer.
\endbibitem

\bibitem[{Storey et~al.(2005)Storey, Xiao, Leek, Tompkins, and
  Davis}]{Storey:2005:significance}
Storey, J.~D., Xiao, W., Leek, J.~T., Tompkins, R.~G., and Davis, R.~W. (2005).
\newblock \enquote{{Significance analysis of time course microarray
  experiments.}}
\newblock {\em Proceedings of the National Academy of Sciences of the United
  States of America\/}, 102(36): 12837--42.
\endbibitem

\bibitem[{Teh et~al.(2006)Teh, Jordan, Beal, and Blei}]{teh:2006:hdp}
Teh, Y., Jordan, M.~I., Beal, M., and Blei, D. (2006).
\newblock \enquote{Hierarchical {D}irichlet processes.}
\newblock {\em Journal of the American Statistical Association\/}, 101(476):
  1566--1581.
\endbibitem

\bibitem[{Teh(2010)}]{Teh:2010:dp}
Teh, Y.~W. (2010).
\newblock \enquote{{D}irichlet Processes.}
\newblock In {\em Encyclopedia of Machine Learning\/}. Springer.
\endbibitem

\bibitem[{Virtanen et~al.(2020)Virtanen, Gommers, Oliphant, Haberland, Reddy,
  Cournapeau, Burovski, Peterson, Weckesser, Bright, {van der Walt}, Brett,
  Wilson, Millman, Mayorov, Nelson, Jones, Kern, Larson, Carey, Polat, Feng,
  Moore, {VanderPlas}, Laxalde, Perktold, Cimrman, Henriksen, Quintero, Harris,
  Archibald, Ribeiro, Pedregosa, {van Mulbregt}, and {SciPy 1.0
  Contributors}}]{2020:scipy}
Virtanen, P., Gommers, R., Oliphant, T., Haberland, M., Reddy, T., Cournapeau,
  D., Burovski, E., Peterson, P., Weckesser, W., Bright, J., {van der Walt},
  S., Brett, M., Wilson, J., Millman, J., Mayorov, N., Nelson, A., Jones, E.,
  Kern, R., Larson, E., Carey, C., Polat, {\.I}., Feng, Y., Moore, E.,
  {VanderPlas}, J., Laxalde, D., Perktold, J., Cimrman, R., Henriksen, I.,
  Quintero, E., Harris, C., Archibald, A., Ribeiro, A., Pedregosa, F., {van
  Mulbregt}, P., and {SciPy 1.0 Contributors} (2020).
\newblock \enquote{{{SciPy} 1.0: Fundamental algorithms for scientific
  computing in {Python}}.}
\newblock {\em Nature Methods\/}, 17: 261--272.
\endbibitem

\bibitem[{von Luxburg(2007)}]{luxburg:2007:spectralcluster}
von Luxburg, U. (2007).
\newblock \enquote{A tutorial on spectral clustering.}
\newblock {\em Statistics and Computing\/}, 17: 395--416.
\endbibitem

\bibitem[{Zeidler(1986)}]{zeidler:2013:functional}
Zeidler, E. (1986).
\newblock {\em Nonlinear Functional Analysis and Its Applications {I}: {F}ixed
  point theorems\/}.
\newblock Springer Verlag New York, Inc.
\endbibitem

\end{thebibliography}


\begin{thebibliography}{16}
\newcommand{\enquote}[1]{``#1''}
\expandafter\ifx\csname natexlab\endcsname\relax\def\natexlab#1{#1}\fi
\expandafter\ifx\csname url\endcsname\relax
  \def\url#1{{\tt #1}}\fi
\expandafter\ifx\csname urlprefix\endcsname\relax\def\urlprefix{URL }\fi
\ifx\endbibitem\undefined \let\endbibitem\relax\fi

\bibitem[{Averbukh and Smolyanov(1967)}]{averbukh:1967:theory}
Averbukh, V. and Smolyanov, O. (1967).
\newblock \enquote{The theory of differentiation in linear topological spaces.}
\newblock {\em Russian Mathematical Surveys\/}, 22(6): 201--258.
\endbibitem

\bibitem[{Billingsley(1986)}]{billingsley:1986:probability}
Billingsley, P. (1986).
\newblock {\em Probability and Measure\/}.
\newblock John Wiley and Sons, second edition.
\endbibitem

\bibitem[{Bishop(2006)}]{bishop:2006:PRML}
Bishop, C.~M. (2006).
\newblock {\em Pattern Recognition and Machine Learning\/}.
\newblock Springer.
\endbibitem

\bibitem[{Blackwell and MacQueen()}]{blackwell:1973:polyaurn}
Blackwell, D. and MacQueen, J.~B. (????).
\newblock \enquote{{F}erguson distributions via {P}olya urn schemes.}
\endbibitem

\bibitem[{Blei et~al.(2017)Blei, Kucukelbir, and
  McAuliffe}]{blei:2017:vi_review}
Blei, D.~M., Kucukelbir, A., and McAuliffe, J.~D. (2017).
\newblock \enquote{Variational inference: {A} review for statisticians.}
\newblock {\em Journal of the American Statistical Association\/}, 112(518):
  859–877.
\endbibitem

\bibitem[{Dudley(2018)}]{dudley:2018:real}
Dudley, R. (2018).
\newblock {\em Real Analysis and Probability\/}.
\newblock CRC Press.
\endbibitem

\bibitem[{Gustafson(1996)}]{gustafson:1996:local}
Gustafson, P. (1996).
\newblock \enquote{Local sensitivity of posterior expectations.}
\newblock {\em Annals of Statistics\/}, 24(1): 174--195.
\endbibitem

\bibitem[{Krantz and Parks(2012)}]{krantz:2012:implicit}
Krantz, S. and Parks, H. (2012).
\newblock {\em The Implicit Function Theorem: {H}istory, Theory, and
  Applications\/}.
\newblock Springer Science \& Business Media.
\endbibitem

\bibitem[{Kucukelbir et~al.(2016)Kucukelbir, Tran, Ranganath, Gelman, and
  Blei}]{kucukelbir:2016:advi}
Kucukelbir, A., Tran, D., Ranganath, R., Gelman, A., and Blei, D. (2016).
\newblock \enquote{Automatic differentiation variational inference.}
\newblock {\em arXiv preprint arXiv:1603.00788\/}.
\endbibitem

\bibitem[{Luan and Li(2003)}]{Luan:2003:clustering}
Luan, Y. and Li, H. (2003).
\newblock \enquote{{Clustering of time-course gene expression data using a
  mixed-effects model with B-splines}.}
\newblock {\em Bioinformatics\/}, 19(4): 474--482.
\endbibitem

\bibitem[{Nielsen(1997)}]{nielsen:1997:measure}
Nielsen, O. (1997).
\newblock {\em An Introduction to Integration and Measure Theory\/}, volume~17.
\newblock Wiley-Interscience.
\endbibitem

\bibitem[{Reeds(1976)}]{reeds:1976:thesis}
Reeds, J. (1976).
\newblock \enquote{On the definition of von {Mises} functionals.}
\newblock Ph.D. thesis, Statistics, Harvard University.
\endbibitem

\bibitem[{Shoemaker et~al.(2015)Shoemaker, Fukuyama, Eisfeld
  et~al.}]{shoemaker:2015:ultrasensitive}
Shoemaker, J.~E., Fukuyama, S., Eisfeld, A.~J., et~al. (2015).
\newblock \enquote{An ultrasensitive mechanism regulates influenza
  virus-induced inflammation.}
\newblock {\em PLoS Pathogens\/}, 11(6): 1--25.
\endbibitem

\bibitem[{Storey et~al.(2005)Storey, Xiao, Leek, Tompkins, and
  Davis}]{Storey:2005:significance}
Storey, J.~D., Xiao, W., Leek, J.~T., Tompkins, R.~G., and Davis, R.~W. (2005).
\newblock \enquote{{Significance analysis of time course microarray
  experiments.}}
\newblock {\em Proceedings of the National Academy of Sciences of the United
  States of America\/}, 102(36): 12837--42.
\endbibitem

\bibitem[{Teh(2010)}]{Teh:2010:dp}
Teh, Y.~W. (2010).
\newblock \enquote{{D}irichlet Processes.}
\newblock In {\em Encyclopedia of Machine Learning\/}. Springer.
\endbibitem

\bibitem[{Zeidler(1986)}]{zeidler:2013:functional}
Zeidler, E. (1986).
\newblock {\em Nonlinear Functional Analysis and Its Applications {I}: {F}ixed
  point theorems\/}.
\newblock Springer Verlag New York, Inc.
\endbibitem

\end{thebibliography}

\notbool{arxiv} {
    \begin{acks}[Acknowledgments]
        We are indebted to helpful discussions with Nelle Varoquaux, Matthew Stephens,
Michael C. Hughes, Eric Sudderth, and Jake Soloff, and to useful suggestions
from anonymous reviewers.  Runjing Liu is supported by the National Science
Foundation graduate research fellowship program. Ryan Giordano and Tamara
Broderick were supported in part by an NSF CAREER Award and an ONR Early Career
Grant.

    \end{acks}
}

\clearpage
\begin{appendix}

\notbool{arxiv}{
    \begin{frontmatter}
        \title{Evaluating Sensitivity to the Stick-Breaking Prior in Bayesian Nonparametrics: Supplementary Materials}

        \runtitle{Evaluating Sensitivity to the Stick-Breaking Prior in BNP: Supplement}
        \authors{}
    \end{frontmatter}
} {
    \appendixpage
}

\section{General differentiability results}
\applabel{diffable_intro}
Our goal is to approximate the dependence of the optimal VB parameters on the
prior using a Taylor series, which requires that the optimal VB parameters must
be continuously differentiable as a function of the prior specification. In this
section we state general conditions under which VB optima based on reverse KL
divergence are differentiable functions of both parametric and nonparametric
prior perturbations. We will state our conditions and results in terms of a
generic VB approximation and prior perturbation, which we articulate in
\defref{prior_t}.

\begin{defn}\deflabel{prior_t}
For some parameter $\theta \in \thetadom \subseteq \mathbb{R}^{\thetadim}$, let
$\p(\theta \vert \t)$ denote a class of probability densities relative to
a sigma-finite measure $\mu$, defined for $\t$ in an open set $\ball_\t
\subseteq \mathbb{R}$ containing $0$.  Let $\q(\theta \vert \eta)$ be a
family of approximating densities, also defined relative to $\mu$.

Let the variational objective factorize as
\begin{align}
\KL{\eta, \t} :={}&
    \KL{\eta} -
    \expect{\q(\theta \vert \eta)}
       {\left(\log \p(\theta \vert \t) - \log \p(\theta \vert \t=0)\right)}           \eqlabel{perturbed_objective}\\
\etaopt(\t) :={}& \argmin_{\eta \in \etadom} \KL{\eta, \t}.
    \eqlabel{perturbed_optimum}
\end{align}
Let $\etaopt$ with no argument refer to $\etaopt(0)$, the minimizer
of $\KL{\eta}$.
\end{defn}

In general, we expect $\theta$ of \defref{prior_t} to be some subset of the
model parameters whose prior is being perturbed. The decomposition in
\eqref{perturbed_objective} is always possible for VB approximations based on
reverse KL divergence, in the sense that one could always take $\theta$ to be
all model parameters and $\KL{\eta} = 0$. We decompose the objective in this way
in order to state strict regularity assumptions only on the part of the reverse
KL divergence that is being perturbed.  Indeed, we will require little from the
$\KL{\eta}$ part of the decomposition other than that it can be differentiated
and optimized.

By identifying $\t$ with some hyperparameter (e.g. the concentration parameter,
as in \exref{alpha_perturbation} below), we can use \defref{prior_t} to study
parametric perturbations.  Furthermore, by parameterizing a path through the
space of general densities, \defref{prior_t} will allow us to study
nonparametric perturbations (e.g. \exref{gem_mult_perturbation} below and the
detailed analysis of \apprangeref{diffable_nonparametric}{diffable_lp}).  We
can thus study VB prior robustness in general by studying problems of the
form in \defref{prior_t}.

\begin{ex}\exlabel{alpha_perturbation}
For the BNP model with the $\pstick$ prior on the stick breaks, take $\theta = (\nu_1, \ldots,
\nu_{\kmax-1})$, and take $\mu$ to be the Lebesgue measure on $[0,1]^{\kmax-1}$.
Let $\alpha_0$ be some initial value of the concentration parameter, and
let $\t$ be $\alpha - \alpha_0$, so that deviations of $\t$ away from
$0$ represent deviations of $\alpha$ away from $\alpha_0$.

Expanding the reverse KL divergence in \eqref{kl_def}, we see that the prior
$\p(\nuk \vert \alpha)$ enters the VB objective in a term of the form
$\sum_{\k=1}^\infty \expect{\q(\nuk \vert \eta)}{\log \p(\nuk \vert \alpha)}$.
Adding and subtracting the this term evaluated at $\alpha_0$ gives
\begin{align*}
\KL{\eta, \alpha} = \KL{\eta, \alpha_0}
-\sum_{\k=1}^{\kmax - 1}
            \left(
                \expect{\q(\nuk \vert \eta)}{\log \p(\nuk \vert \alpha)} -
                \expect{\q(\nuk \vert \eta)}{\log \p(\nuk \vert \alpha_0)}
             \right).
\end{align*}
Plugging in the definition of $\p(\nuk \vert \alpha)$, recognizing that the
normalizing constant does not depend on $\nuk$ and so can be neglected in the
optimization, letting $\KL{\eta} := \KL{\eta, \alpha_0}$, and substituting $\t =
\alpha - \alpha_0$ gives
\begin{align*}
\KL{\eta, \t} = \KL{\eta, \alpha_0}
-\t \sum_{\k=1}^{\kmax - 1}
    \expect{\q(\nuk \vert \eta)}{\log (1 - \nuk)}.
\end{align*}
\end{ex}

\begin{ex}\exlabel{gem_mult_perturbation}
As in \exref{alpha_perturbation}, take $\theta = (\nu_1, \ldots, \nu_{\kmax-1})$
and $\mu$ to be the Lebesgue measure on $[0,1]^{\kmax-1}$. Let $\pbase(\nuk) :=
\betadist{\nuk \vert 1, \alpha_0}$, and let $\palt(\nuk)$ be a density, not
in the beta family, that shifts mass towards zero:
\begin{align*}
\palt(\nuk) :=
    \frac{\exp(-\nuk)\pbase(\nuk)}{\int \exp(-\nuk')\pbase(\nuk') d\nuk'}.
\end{align*}
For $\t \in [0,1]$ define the multiplicatively perturbed prior
\begin{align*}
\p(\nuk \vert \t) :=
    \frac{\palt(\nuk)^{\t} \pbase(\nuk)^{1-\t}}
         {\int \palt(\nuk')^{\t} \pbase(\nuk')^{1-\t} d\nuk'}.
\end{align*}
When $\t = 0$, $\p(\nuk \vert \t) = \pbase(\nuk)$, when $\t = 1$,
$\p(\nuk \vert \t)  = \palt(\nuk)$.  For $\t \in (0,1)$
$\p(\nuk \vert \t)$ varies smoothly between $\pbase$ and $\palt$.

As in \exref{alpha_perturbation}, up to constants not depending on
$\nuk$ we can write
\begin{align*}
\log \p(\nuk \vert \t) - \log \p(\nuk \vert \t=0) ={}&
    -\t \log \pbase(\nuk) + \t \log \palt(\nuk) + \const
\\={}& -\t \nuk + \const \Rightarrow
\\
\KL{\eta, \t} ={}& \KL{\eta} -\t \expect{\q(\nuk\vert\eta)}{\nuk} + \const.
\end{align*}
Different choices for $\palt(\nuk)$ would give different additive perturbations
to the reverse KL divergence.
\end{ex}
%

    \subsection{Parametric prior perturbations}
    \applabel{diffable_parametric}
    We now state conditions under which $\t \mapsto \etaopt(\t)$, as defined by
\defref{prior_t}, is continuously differentiable.  Our key theoretical tool will
be the implicit function theorem \citep[see, e.g.,][]{krantz:2012:implicit}, applied
to the first-order conditions for the VB optimization problem.

Our results can be expressed in terms of unnormalized densities, which can
simplify some computation.  To that end, let $\qtil$ and $\ptil$ refer to
potentially unnormalized (but normalizable) versions of the respectively
corresponding $\q$ and $\p$ given in \defref{prior_t}, so that
\begin{align*}
\q(\theta \vert \eta) :={}
    \frac{\qtil(\theta \vert \eta)}
    {\int \qtil(\theta' \vert \eta) \mu(d\theta')} \mathand
\p(\theta \vert \t) :={}
    \frac{\ptil(\theta \vert \t)}
    {\int \ptil(\theta' \vert \t) \mu(d\theta')}.
\end{align*}

In \assuref{kl_opt_ok}, stated in \secref{local_sensitivity} above, we require
some mild regularity conditions for the ``initial problem,'' $\KL{\eta}$. As we
discuss in \appref{diffable_concentration}, \assuref{kl_opt_ok} states
conditions that are typically satisfied when $\KL{\eta}$ can be optimized
numerically using unconstrained optimization.

Next, we will require some differentiability conditions for the perturbation and
the variational approximation.

%
\begin{assu}\assulabel{exchange_order}
Assume that the map $\eta \mapsto \log \qtil(\theta \vert \eta)$ is twice
continuously differentiable, and that the map $\t \mapsto \log  \ptil(\theta
\vert \t)$ is continuously differentiable.

Further, assume that we can exchange the order of integration and
differentiation in the expressions $\int \qtil(\theta \vert \eta) \log
\ptil(\theta \vert \t) \mu(d\theta)$ and $\int \qtil(\theta \vert \eta)
\mu(d\theta)$ at $\eta = \etaopt$ and $\t = 0$ for the derivatives $\partial /
\partial \eta$, $\partial^2 / \partial \eta^2$, and $\partial^2 / \partial \eta
\partial \t$.
\end{assu}

In certain cases, one can verify \assuref{exchange_order} directly, such as when
$\expect{\q(\theta \vert \eta)}{\log \ptil(\theta \vert \t)}$ has a closed form.
For more general situations, the following assumptions allow us to satisfy
\assuref{exchange_order} using the dominated convergence theorem \citep[Theorem
16.8]{billingsley:1986:probability}.

%
\begin{assu}\assulabel{exchange_order_f}
Let $f(\theta, \eta, \t)$ be a function taking values in $\mathbb{R}$. Assume
that the partial derivatives $\partial / \partial \eta$, $\partial^2 / \partial
\eta^2$, and $\partial^2 / \partial \eta \partial \t$ of $f$ exist, are
continuous functions of $\eta$ and $\t$, and are $\mu$-measureable functions of
$\theta$ on some open set $\ball_\eta \times \ball_\t$.

Let $M(\theta) > 0$ be a measurable function with $\int M(\theta) \mu(d\theta) <
\infty$.  Assume that, for all $\eta, \t \in \ball_\eta \times \ball_\t$,
$M(\theta)$ is $\mu$-almost everywhere greater than each of the following
functions: $\abs{f(\theta, \eta, \t)}$, $\norm{\partial f(\theta, \eta, \t) /
\partial \eta}_2$, $\norm{\partial^2 f(\theta, \eta, \t) / \partial \eta
\partial \eta^T}_2$, and $\norm{\partial^2 f(\theta, \eta, \t) / \partial \eta
\partial \t}_2$.
\end{assu}
%

%
\begin{assu}\assulabel{exchange_order_dom}
(Sufficient conditions for \assuref{exchange_order}.)
Let \assuref{exchange_order_f} hold with the function $f(\theta, \eta, \t) =
\qtil(\theta \vert \eta) \log \ptil(\theta \vert \t)$ as well as with $f(\theta,
\eta, \t) = \qtil(\theta \vert \eta)$.
\end{assu}

By the dominated convergence theorem, \assuref{exchange_order_dom} implies
\assuref{exchange_order} (see \lemref{exchange_order} in \appref{proofs} for a
proof). The advantage of \assuref{exchange_order_dom} over
\assuref{exchange_order} is that the conditions of \assuref{exchange_order_dom}
can typically be verified even when the expectation $\expect{\q(\theta \vert
\eta)}{\log \ptil(\theta \vert \t)}$ does not have a closed form.  In
\appref{diffable_concentration}, we will discuss how different choices of
variational approximations for the stick lengths lend themselves to either
\assuref{exchange_order} of \assuref{exchange_order_dom}.  Furthermore,
\assuref{exchange_order_f} will be essential to analyzing nonparametric
perturbations in \appref{diffable_nonparametric}.

We are now in a position to define the quantities that occur in the derivative
and state our main result.

\begin{defn}\deflabel{deriv_quantities}
Under the conditions of \defref{prior_t}, when \assuref{kl_opt_ok,
exchange_order} hold, define
\begin{align*}
\hessopt :={}& \fracat{\partial^2 \KL{\eta}}
                      {\partial \eta \partial \eta^T}
                      {\etaopt} \mathand \\
\lqgradbar{\theta \vert \eta} :={}&
    \lqgrad{\theta \vert \eta} -
    \expect{\q(\theta \vert \eta)}{\lqgrad{\theta \vert \eta}}.
\end{align*}


Further, define

\begin{align*}
\crosshessian :={}&
    \fracat{\partial
            \expect{\q(\theta \vert \eta)}
                   {\fracat{\partial \log \ptil(\theta \vert \t)}
                           {\partial \t}{\t=0} }
            }
        {\partial \eta}{\eta = \etaopt}
={}
    \expect{\q(\theta \vert \etaopt)}{
          \lqgradbar{\theta \vert \etaopt}
          \fracat{\partial \log \ptil(\theta \vert \t)}
                 {\partial \t}{\t=0}},
\end{align*}
where the final equality follows from differentiating under the integral using
\assuref{exchange_order} (see \lemref{logq_continuous} in \appref{proofs} for
more details).
\end{defn}

\begin{thm}\thmlabel{etat_deriv}
Under the conditions of \defref{prior_t, deriv_quantities}, let
\assuref{kl_opt_ok, exchange_order} hold.   Then the map $\t \mapsto
\etaopt(\t)$ is continuously differentiable at $\t=0$ with derivative
\begin{align}\eqlabel{vb_eta_sens}
\fracat{d \etaopt(\t)}{d \t}{0} ={}&
    - \hessopt^{-1} \crosshessian.
\end{align}
(For a proof, see \appref{proofs} \proofref{etat_deriv}.)
\end{thm}


    \subsection{Differentiability of BNP models with respect to $\alpha$}
    \applabel{diffable_concentration}
    In this section, we return to the BNP problem and prove carefully that the map
$\alpha \mapsto \etaopt(\alpha)$ satisfies \assuref{kl_opt_ok, exchange_order},
and so the conditions of \thmref{etat_deriv}.  As in \exref{alpha_perturbation},
we will take $\mu$ to be the Lebesgue measure on $[0,1]^{\kmax - 1}$.

Recall from \secref{model_vb} that we take $\q(\nuk \vert \eta)$ to be a normal
density on the logit-transformed sticks, $\lnu_\k$.  For the duration of
this section, write $\q(\lnuk \vert \eta) = \normdist{\lnuk \vert \mu_\k,
\sigma^2_\k}$, so that the subvector of $\eta$ parameterizing $\q(\lnuk \vert
\eta)$ is $\etanuk = (\mu_\k, \sigma_\k)$.
By the formula for transformation of probability densities,
\begin{align*}
\q(\nuk \vert \etanuk) =
    \normdist{\log\left(\frac{\nu_\k}{1 - \nu_\k} \right)
        \Big\vert  \mu_\k, \sigma^2_\k}
    \frac{1}{\nuk (1 - \nuk)},
\end{align*}
where we have used the fact that $\fracat{d \lnu_\k}{ d\nuk}{\nuk} =
\frac{1}{\nuk (1 - \nuk)}$.  Similarly, for any function $f(\nuk)$ of the stick
lengths, we can transform the expectations as $\expect{\q(\nuk \vert
\etanuk)}{f(\nuk)} = \expect{\q(\lnuk \vert \etanuk)}{f\left(
\frac{\exp(\lnuk)}{1 + \exp(\lnuk)}  \right))}$, using the fact that
$\nuk = \frac{\exp(\lnuk)}{1 + \exp(\lnuk)}$.

Differentiability of $\KL{\eta}$ (\assuitemref{kl_opt_ok}{kl_diffable}) is
immediately satisfied for the $\eta$ that parameterize $\q(\beta \vert \eta)$
and $\q(\z \vert \eta)$ by our use of conjugate approximating families and
standard parameterizations.  The stick length density, $\q(\nuk \vert \etanuk)$
is not a standard exponential family
\footnote{In this section, we continue to take $\mu$ to be the Lebesgue measure
on $[0,1]$ as in \exref{alpha_perturbation}.  We could have equivalently taken
$\mu$ to be the Lebesgue measure on $\mathbb{R}$ and analyzed $\p(\lnuk \vert
\alpha)$ instead of $\p(\nuk \vert \alpha)$.  Had we done so, the log Jacobian
term $\log (\nuk(1 - \nuk))$ now appearing in the entropy would have instead
appeared in the $\log \ptil(\lnuk \vert \alpha)$ term, and so been part of
\assuref{exchange_order} rather than \assuitemref{kl_opt_ok}{kl_diffable}.
Nevertheless, the needed assumptions would be substantively the same. For
essentially this reason, the choice of dominating measure in \defref{prior_t}
does not matter.}
, so we must show that the entropy $\expect{\q(\nuk \vert \etanuk)}{\log \q(\nuk
\vert \etanuk)}$ is  twice continuously differentiable. The entropy is given up
to a constant by
\begin{align*}
\MoveEqLeft
\expect{\q(\nuk \vert \etanuk)}{\log \q(\nuk \vert \etanuk)}
\\={}&
    \expect{\q(\nuk \vert \etanuk)}
           {\log \normdist{\log\left(\frac{\nu_\k}{1 - \nu_\k} \right)
               \Big\vert  \mu_\k, \sigma^2_\k}} +
    \expect{\q(\nuk \vert \etanuk)}
           {\log \left(\nuk (1 - \nuk)\right)}
\\={}&
   \expect{\q(\lnuk \vert \etanuk)}
          {\log \normdist{\lnuk \Big\vert  \mu_\k, \sigma^2_\k}} +
   \expect{\q(\lnuk \vert \etanuk)}{\lnuk}
\\={}&
    \frac{1}{2} \log \sigma^2_\k + \mu_\k + \const,
\end{align*}
which is twice continuously differentiable by inspection.
Indeed, \assuitemref{kl_opt_ok}{kl_diffable} is typically satisfied in VB
problems; when it is not, many black-box optimization methods also do not apply.

Non-singularity of the Hessian matrix $\hessopt$
(\assuitemref{kl_opt_ok}{kl_hess}) is satisfied whenever $\etaopt$ is at a local
optimum of $\KL{\eta}$.  In practice, we compute $\etaopt$ and (approximately)
check \assuitemref{kl_opt_ok}{kl_hess} numerically as part of computing the
sensitivity $\hessopt^{-1} \crosshessian$.  As with
\assuitemref{kl_opt_ok}{kl_diffable}, if \assuitemref{kl_opt_ok}{kl_hess} is
violated, then the user will probably have difficulty optimizing $\KL{\eta}$.

\assuitemref{kl_opt_ok}{kl_opt_interior} essentially requires that $\KL{\eta}$
be well-defined in an $\mathbb{R}^\etadim$ neighborhood of $\etaopt$, and can
require some care in choosing the parameterization $\eta$.  As an example of a
parameterization that would violate \assuitemref{kl_opt_ok}{kl_opt_interior},
consider parametrizing $\q(\z_{\n} \vert \eta)$ by the $\kmax$ expectations
$m_\k := \expect{\q(\z_{\n} \vert \eta)}{\z_{\n\k}}$.  The set $(m_1, \ldots,
m_\kmax)$ completely specify $\q(\z_{\n} \vert \eta)$, but violate
\assuitemref{kl_opt_ok}{kl_opt_interior}, since any valid parameterization
satisfies $\sum_{\k=1}^\kmax m_\k = 1$, and so no open ball in
$\mathbb{R}^\etadim$ can be contained in $\etadom$.  However,
\assuitemref{kl_opt_ok}{kl_opt_interior} is satisfied we use an {\em
unconstrained parameterization} for $\q(\zeta \vert \eta)$.   Unconstrained
parameterizations of variational distributions allow the use of unconstrained
optimization for variational inference and are a good practice when available
\citep{kucukelbir:2016:advi}.  For details on our parameterizations, see
the corresponding appendices.

Verifying \assuref{exchange_order} is the principal technical challenge of
satisfying the conditions of \thmref{etat_deriv}. Recall from
\exref{alpha_perturbation} that $\log \ptil(\nuk \vert \t) = t \log (1 - \nuk)$,
so we need to establish \assuref{exchange_order} for
\begin{align*}
-\expect{\q(\nuk \vert \etanuk)}{t \log (1 - \nuk)} =
\expect{\q(\lnuk \vert \etanuk)}
      {t \log (1 + \exp(\lnuk))}.
\end{align*}
Since the preceding equality holds for all $\t$ and $\etanuk$, it suffices to
establish that we can exchange the order of integration and differentiation for
the right hand side.  Since the normal density has a term of the form
$\exp(-\const \lnuk^2)$, and since $\log (1 + \exp(\lnuk)) \exp(-\abs{\lnuk})  <
\infty$ for all $\lnuk \in \mathbb{R}$ as long as the variational variance is
finite, one can show that the conditions of \assuref{exchange_order_dom} are
satisfied within $\ball_\eta \times \ball_\t$.  (See
\lemref{normal_q_is_regular} in \appref{proofs} for a proof.)
Note that
derivatives with respect to any components of $\eta$ other than $\etanuk$ are
zero and so \assuref{exchange_order} is trivially satisfied.

\Assuref{exchange_order_dom} implies \assuref{exchange_order}.  Since both
\assuref{kl_opt_ok, exchange_order} are satisfied, \thmref{etat_deriv} applies,
and the map $\alpha \mapsto \etaopt(\alpha)$ is continuously differentiable.

We end this section by observing that the only real technical challenge was
showing that the assumptions were satisfied for the logit-normal densities
$\q(\nuk \vert \etanuk)$.  Had we instead used the conjugate beta density
parameterized by its natural parameters, then both \assuref{kl_opt_ok} and
\assuref{exchange_order} would follow immediately by standard properties of the
Beta distribution.  In particular, the expectation $\expect{\q(\nuk \vert
\etanuk)}{t \log (1 - \nuk)}$ needed for \assuref{kl_opt_ok} is simply $\t$
times the Beta distribution's moment parameter, which is known to be an
infinitely-differentiable function of the natural parameters.

    \subsection{Nonparametric prior perturbations}
    \applabel{diffable_nonparametric}
    We now show how, by parameterizing a
path between two arbitrary densities, we can apply \thmref{etat_deriv} to
nonparametric perturbations of the prior density.
Again let us return to the abstract setting of \defref{prior_t}. Let us fix an
initial prior density, $\pbase(\theta)$, at which we have computed a VB
approximation, and suppose we wish to ask what the variational optimum would
have been had we used some alternative prior density, $\palt(\theta)$.  Let us
write $\etaopt(\pbase)$ and $\etaopt(\palt)$ for these two approximations,
respectively, so we are interested in quantifying the change $\g(\etaopt(\palt)) -
\g(\etaopt(\pbase))$. If this change is large, we say that our quantity of
interest is not robust to replacing $\pbase$ with $\palt$.

To approximately assess robustness using the local sensitivity approach, we must
somehow define a continuous path from $\pbase(\theta)$ to $\palt(\theta)$
parameterized, say, by $\t \in [0, 1]$. One way to do so is to define a
multiplicative path
\begin{align}
\log \ptil(\theta \vert \t) ={}&
    (1 - \t)\log \pbase(\theta) + \t \log \palt(\theta).
        \eqlabel{mult_pert_simple}
\end{align}
Under \eqref{mult_pert_simple}, when $\t=0$, $\p(\theta \vert \t) =
\pbase(\theta)$, when $\t=1$, $\p(\theta \vert \t, \pbase, \palt) =
\palt(\theta)$, and $\t \in (0,1)$ smoothly parameterizes a path between the
two.  If we can verify that \thmref{etat_deriv} applies to the perturbation
given in \eqref{mult_pert_simple}, then, just as in the parametric case, we can
form the Taylor series approximation,
\begin{align*}
\etaopt(\palt) \approx
    \etaopt(\pbase) + \fracat{d \etaopt(\t)}{d\t}{\t=0} (1 - 0).
\end{align*}

Our first task is then to state conditions under which \thmref{etat_deriv}
applies to \eqref{mult_pert_simple}.  In \eqref{mult_pert_simple} we have
assumed that $\palt$ is a density, but it will be more convenient to observe
that, when $\palt \ll \pbase$, we can re-write
\begin{align*}
\log \ptil(\theta \vert \t) ={}&
    \log \pbase(\theta) +
        \t \log \frac{\palttil(\theta)}{\pbasetil(\theta)} +
        \const. & \constdesc{\theta}
\end{align*}
Defining the generic function $\phi(\theta) := \log
\frac{\palttil(\theta)}{\pbasetil(\theta)}$ motivates consideration of
perturbations of the form $\log \ptil(\theta \vert \t) = \pbase(\theta) + \t
\phi(\theta)$, where $\phi(\theta)$ is some generic measurable function. We can
then ask what $\phi$ give rise to valid densities as well as differentiable maps
$\t \mapsto \etaopt(\t)$.

\begin{defn}\deflabel{prior_nl_pert}
Let $\mu$ denote a measure on the Borel sets of $\thetadom$ and fix
$\pbase(\theta)$, a density with respect to $\mu$.  Assume that
$\mu\left(\{\theta: \pbase(\theta) = 0 \}\right) = 0$, so that (in a slight
abuse of notation) $\mu \ll \pbase$.  For any measurable $\phi: \thetadom
\mapsto \mathbb{R}$ for which the expressions are well-defined, let
\begin{align*}
\ptil(\theta \vert \phi) :={}& \pbase(\theta)\exp(\phi(\theta)).
\end{align*}
As usual, when $0 < \int \ptil(\theta \vert \phi) \mu(d\theta) < \infty$, we let
$\p(\theta \vert \phi)$ be the normalized version of $\ptil(\theta \vert \phi)$.
Further, define the norm $\norminf{\phi} := \esssup_{\theta \sim \mu}
\abs{\phi(\theta)}$, and let $\ball_\phi(\delta) := \left\{ \phi: \norminf{\phi} <
\delta \right\}$.
\end{defn}
%

The class of perturbations defined in \defref{prior_nl_pert} are one of the
family of ``nonlinear'' functional perturbations given by
\citet{gustafson:1996:local}, though we deviate from
\citet{gustafson:1996:local} by allowing $\phi$ to take on negative values. The
following result, which motivates the use of the $\norminf{\cdot}$ norm to
measure the ``size'' of a perturbation $\phi$, is only a minor modification of
the corresponding result from \citet{gustafson:1996:local} to allow negative
perturbations.

\begin{lem}\lemlabel{pert_invariance}
(\citet{gustafson:1996:local})
Fix the quantities given in \defref{prior_nl_pert}.  For a fixed probability
measure $\palt \ll \mu$ with density $\palt(\theta)$ with respect to $\mu$, let
$\phi(\theta \vert \palt) := \log \palt(\theta) / \pbase(\theta)$.  Then $\palt
\mapsto \norminf{\phi(\cdot \vert \palt)}$ is a norm, does not depend on $\mu$,
and is invariant to invertible transformations of $\theta$.

Furthermore, for any $\phi$ with $\norminf{\phi} < \infty$, the quantity
$\ptil(\theta \vert \phi)$ gives rise to a valid prior, in the sense that
$\ptil(\theta \vert \phi) \ge 0$ $\mu$-almost everywhere, and
$0 < \int \ptil(\theta \vert \phi) \mu(d\theta) < \infty$.
\seeproof{pert_invariance}
\end{lem}

The set of priors $\left\{\p(\theta \vert \phi) : \phi \in
\ball_\phi(\delta)\right\}$ live in a multiplicative band around the original
prior, $\pbase$, as shown in \figref{linf_examples}. Although
\lemref{pert_invariance} proves that every $\phi$ with $\norminf{\phi}$ is a
valid prior, the converse is not true, and the Beta prior perturbation of
\exref{alpha_perturbation} is a counterexample.

\begin{ex}\exlabel{beta_inf_norm}
Take $\mu$ to be the Lebesgue measure on $[0,1]$, let $\pbase(\theta) =
\betadist{\theta \vert 1, \alpha_0}$ and $\palt(\theta) = \betadist{\theta \vert
1, \alpha_1}$ for $\alpha_0 \ne \alpha_1$.  Taking
$\phi(\theta) = (\alpha_1 - \alpha_0) \log(1 - \theta)$ parameterizes
a path from $\pbase$ to $\palt$ as in \eqref{mult_pert_simple}, and
\begin{align*}
\norminf{\phi} =
    \abs{\alpha_1 - \alpha_0} \sup_{\theta \in [0,1]} \abs{\log(1 - \theta)} =
    \infty.
\end{align*}
Therefore, in general, there exist valid priors that cannot be expressed by
\defref{prior_nl_pert} with $\phi$ with $\norminf{\phi} < \infty$.
\end{ex}

We now show that, when $\norminf{\phi} < \infty$, we can apply
\thmref{etat_deriv}.  We still require the following assumption on the VB
density, which is strictly weaker than \assuref{exchange_order_dom}.

\begin{assu}\assulabel{exchange_order_q}
Assume that \assuref{exchange_order_f} applies with the function $f(\theta,
\eta, \t) = \q(\theta \vert \eta)$ (no $\t$ dependence).
\end{assu}


\begin{cor}\corylabel{etafun_deriv_form}
Fix the quantities given in \defref{prior_nl_pert}, and let \assuref{kl_opt_ok,
exchange_order_q} hold. Let $g(\eta): \etadom \mapsto \mathbb{R}$ denote a
continuously differentiable real-valued function of interest.  Define the
``influence function'' $\infl: \thetadom \mapsto \mathbb{R}$:
\begin{align}\eqlabel{infl_defn}
\infl(\theta) :={}&
    - \fracat{d g(\eta)}{ d \eta^T}{\etaopt} \hessopt^{-1}
        \lqgradbar{\theta \vert \etaopt}
        \q(\theta \vert \etaopt).
\end{align}
Then, if $\norminf{\phi} < \infty$, the map $\t \mapsto g(\etaopt(\t \phi))$ is
continuously differentiable at $\t=0$ with derivative
\begin{align}\eqlabel{vb_eta_infl_sens}
\fracat{d g(\etaopt(\t \phi))}{d \t}{0} ={}&
    \int \infl(\theta) \phi(\theta) \mu(d\theta).
\end{align}
\begin{proof}
It suffices to show that \assuref{exchange_order_q} implies
\assuref{exchange_order} for the perturbation given in \defref{prior_nl_pert}
when $\norminf{\phi} < \infty$.  Observe that $\log \ptil(\theta \vert \t) = \t
\phi(\theta)$, so, for any $f(\theta, \eta, \t)$ that satisfies the conditions
of \assuref{exchange_order_f},
%
%
$\phi(\theta) f(\theta, \eta, \t) \le \norminf{\phi} M(\theta)$.
%
%
Therefore \assuref{exchange_order_f} is satisfied by $\phi(\theta) f(\theta,
\eta, \t)$ as well.  It follows that \assuref{exchange_order_q} $\Rightarrow$
\assuref{exchange_order_dom} $\Rightarrow$ \assuref{exchange_order}.
The form of the influence function is then given by gathering terms in
\eqref{vb_eta_sens}.
\end{proof}
\end{cor}


The influence function can be a useful summary of the effect of making generic
changes to the prior density, as we will show in the experiments of
\secref{results}.  For visualization, it can be useful to reduce the dimension
of the domain of the influence function, as we discuss in the following example.

\begin{ex}\exlabel{infl_univariate}
In the BNP example, we are perturbing each of the sticks, so we take $\theta \in
[0,1]^{\kmax - 1}$.  Formally, $\phi: [0,1]^{\kmax - 1} \mapsto \mathbb{R}$ can
express different perturbations for the density of each of the $\kmax - 1$
sticks.  However, when we describe ``changing the stick breaking density,'' we
mean changing each stick's prior density in the same way.

To represent perturbing all the sticks simultaneously, take some univariate
perturbation $\phi_{u}: [0,1] \mapsto \mathbb{R}$, and set $\phi(\nu_1, \ldots,
\nu_{\kmax - 1}) = \sum_{\k=1}^{\kmax - 1} \phi_{u}(\nuk)$. By linearity of the
derivative \coryref{etafun_deriv_form},
\begin{align*}
\fracat{d g(\etaopt(\t \phi))}{d \t}{0} ={}&
    \int \infl(\theta) \left(
        \sum_{\k=1}^{\kmax - 1} \phi_{u}(\nuk) \right)
    d\nu_1 \ldots d \nu_{\kmax - 1}.
\end{align*}
By definition, $\expect{\q(\theta \vert \etaopt)}{\lqgradbar{\theta \vert
\etaopt}} = 0$, so $\int \infl(\theta) \mu(d\theta) = 0$.  By the mean-field
assumption, $\infl(\nu_1, \ldots, \nu_{\kmax - 1}) = \prod_{\k=1}^{\kmax - 1}
\infl_\k(\nuk)$, where $\infl_\k(\nuk)$ is derived from \eqref{infl_defn} but
using $\theta = \nuk$.  Letting $\nu_0 \in [0,1]$ denote the variable of
integration and plugging in the preceding observations gives
\begin{align*}
\int \infl(\theta) \phi(\theta) \mu(d\theta) =
    \int_0^1 \left(\sum_{\k=1}^{\kmax - 1} \infl_k(\nu_0) \right)
        \phi_{u}(\nu_0) d \nu_0.
\end{align*}
Thus we can say that the influence function for perturbing all the stick
breaking densities simultaneously is given by the sum of the
individual sticks' influence functions, which maps $[0,1] \mapsto \mathbb{R}$.
\end{ex}

    \subsection{Worst-case prior perturbations and Fr{\'e}chet differentiability}
    \applabel{diffable_worst_case}
    As we saw in \coryref{etafun_deriv_form}, the derivative of perturbations given
by \defref{prior_nl_pert} takes the form of an integral of the influence
function against the perturbation.  It is natural to use the influence function
to {\em explore} the space of priors, e.g., to find alternative priors with
large influence but small $\norminf{\phi}$.  Consider as an example the
following corollary, which is the VB analogue of \citet[Result
11]{gustafson:1996:local}.


\begin{cor}\corylabel{etafun_worst_case}
The ``worst-case'' derivative in $\ball_\phi(\delta)$ is given by
\begin{align*}
\sup_{\phi \in \ball_\phi(\delta)}
    \fracat{d g(\etaopt(\t \phi))}{d \t}{0} =
        \delta \int \abs{\infl(\theta)} \mu(d\theta),
\end{align*}
which is achieved at the perturbation
$\phi^*(\theta) = \delta \, \mathrm{sign}\left(\infl(\theta)\right)$.
\begin{proof}
The result follows immediately from applying H{\"o}lder's inequality
\citep[][Theorem 5.1.2 and subsequent discussion]{dudley:2018:real}
to \eqref{vb_eta_infl_sens}.
\end{proof}
\end{cor}

%

As discussed in \secref{influence_function} above, we also wish to show that the
map $\phi \mapsto \g(\etaopt(\phi))$ is  continuously Fr{\'e}chet differentiable
as a map from $L_\infty$ to $\mathbb{R}^\etadim$.

\begin{thm}\thmlabel{eta_phi_deriv}
Let \assuref{kl_opt_ok, exchange_order_q} hold. Then the map $\phi \mapsto
\etaopt(\phi)$ is well-defined and continuously Fr{\'e}chet differentiable in a
neighborhood of $0$ as a map from $\linf$ to $\mathbb{R}^\etadim$, with the
derivative given in \coryref{etafun_deriv_form}.

(For a proof, see \appref{proofs} \proofref{eta_phi_deriv}.)

\end{thm}


    \subsection{Other nonparametric prior perturbations}
    \applabel{diffable_lp}
One might ask whether one could consider paths through the space of priors other
than multiplicative, such as additive perturbations.  In this section, we
briefly consider a broader class of nonlinear perturbations investigated by
\citet{gustafson:1996:local}, of which additive and multiplicative perturbations
are special cases, and show that, within this class, only multiplicative
perturbations lead to Fr{\'e}chet differentiable VB optima.

As in \appref{diffable_nonparametric}, suppose we have an initial prior $\pbase$
and an alternative $\palt$, and that we wish to parameterize a continuous path
between them.  Deviating from multiplicative perturbations, for some $p \in [1,
\infty)$, let
\begin{align}\eqlabel{p_pert_simple}
\ptil(\theta \vert \tp) :=
    \left((1 - \tp)\pbase(\theta)^{1/p} +
    \tp \frac{1}{p}\palt(\theta)^{1/p} \right)^{p}.
\end{align}
As with \eqref{mult_pert_simple}, $\p(\theta \vert \tp = 0) = \pbase(\theta)$,
$\p(\theta \vert \tp = 1) = \palt(\theta)$, and $\p(\theta \vert \tp)$ moves
continuously between the two in $\tp \in (0, 1)$.  When $p = 1$,
$\ptil(\theta \vert \tp)$ defines an ``additive perturbation,'' and
the limit as $p \rightarrow \infty$ gives the multiplicative perturbation
of \eqref{mult_pert_simple}.

\citet[Result 2]{gustafson:1996:local} states a result analogous to
\lemref{pert_invariance} for \eqref{p_pert_simple}, where the
$\norminf{\cdot}$ norm is replaced by
\begin{align}\eqlabel{phi_lp_norm}
\phi(\theta \vert \palt, p) :={}
    \palt(\theta)^{1/p} - \pbase(\theta)^{1/p} \mathand
\norm{\phi}_p :={} \left(\int \abs{\phi(\theta)}^p \right)^{1/p}.
\end{align}
We refer the reader to \citet{gustafson:1996:local} for
details.%
\footnote{\citet{gustafson:1996:local} in fact considers only pointwise
positive perturbations $\phi(\theta \vert \palt, p) > 0$, $\mu$-almost
everywhere.  It is not hard to extend \lemref{pert_invariance} \citep[Result
2]{gustafson:1996:local}
to permit negative perturbations, except for the fact
that $\norm{\cdot}_p$-neighborhoods of the zero function will always contain
pointwise negative ``priors.''
We allow for negative $\phi(\theta \vert \palt, p)$ because otherwise
$\norm{\cdot}_p$ leads to counter intuitive notions of the ``size'' of prior
perturbations, as we discuss in \appref{positive_pert}, and because standard
results in functional analysis used in the proof of \thmref{eta_phi_deriv}
require open neighborhoods.

However, we must acknowledge that the main result of this section,
\thmref{kl_discontinuous_main} below, relies on the possibility that
$\phi(\theta \vert \palt, p)$ can be negative.   In light of this, one might
reasonably wonder whether we should in fact restrict to positive perturbations
in an attempt to avoid the consequences of \thmref{kl_discontinuous_main}. In
the view of the authors, restricting to pointwise positive perturbations is a
somewhat artificial solution to a fundamental disconnect between the
$\norm{\cdot}_p$ norm and reverse KL divergence which we dicuss at the end of
the present section. We believe that the disconnect is resolved more
transparently and naturally through the use of the $\norminf{\cdot}$ norm and
multiplicative perturbations which are allowed to be negative.}
For our present discussion, what matters is that the use of the perturbation
in \eqref{p_pert_simple} strongly motivates the use of the norm
$\norm{\phi(\theta \vert \palt, p)}_p$ when forming, for example, worst-case
perturbations as in \coryref{etafun_worst_case}.

Though the $\norm{\phi(\theta \vert \palt, p)}_p$ norm does not appear to cause
major difficulties for the full Bayesian posterior,
\footnote{Other than the fact that there exist pointwise negative priors
induced by $\phi(\theta \vert \palt, p)$ in every neighborhood of the
zero function.}
the $\norm{\phi(\theta \vert \palt, p)}_p$ norm is not compatible with KL
divergence, in the sense that reverse KL divergence is {\em discontinuous} in
this norm. Prior changes that are arbitrarily small according to
$\norm{\phi(\theta \vert \palt, p)}_p$ can induce arbitrarily large changes in
the reverse KL divergence, and so (in general) arbitrarily large changes in its
optimum.  The precise result is stated in \thmref{kl_discontinuous_main} of
\secref{influence_function} above; we now provide the proof.

%

\proofof{\thmref{kl_discontinuous_main}}\prooflabel{kl_discontinuous_main}
For the duration of the proof, we will use the shorthand that a density
applied to a set represents the integral of the density over the set.
For example, for a set $S$, $\p(S) = \int_S \p(\theta)\mu(d\theta)$.

The proof will be constructive, based on an alternative $\palt(\theta)$ formed
by driving $\pbase(\theta)$ to zero in a small interval.  By making the interval
narrow, we can make $\norm{\phi(\theta \vert \palt, p)}_p$ small, but by making
the $\palt(\theta)$ sufficiently close to zero, we can make the reverse KL
divergence difference large irrespective of how narrow the interval is.

First, observe that
\begin{align*}
\KL{q(\theta) || \palt(\theta)} -
\KL{q(\theta) || \pbase(\theta)} ={}&
\expect{\q(\theta)}{\log \frac{\palt(\theta)}{\pbase(\theta)}}.
\end{align*}

For any set $S$ with $\pbase(S) = \epsilon$, define
\begin{align*}
\palt(\theta \vert S, \delta) :=
    \frac{\delta^{\ind{\theta \in S}}}{1 + \epsilon(1 - \delta)} \pbase(\theta).
\end{align*}
Then $\palt(\theta \vert S, \delta)$ is a valid density, and
\begin{align*}
\KL{q(\theta) || \palt(\theta)} - \KL{q(\theta) || \pbase(\theta)}
    ={}& \q(S) \log \delta - \log\left( 1 + \epsilon(1 - \delta) \right).
\end{align*}
By \eqref{phi_lp_norm},
\begin{align*}
\phi(\theta \vert \palt, p) ={}&
    \pbase(\theta)^{1/p} \left(
        \frac{\left(\delta^{1/p}\right)^{\ind{\theta \in S}}}
             {\left( 1 + \epsilon(1 - \delta) \right)^{1/p}} - 1 \right) \mathand\\
\norm{\phi(\theta \vert \palt, p)}_p^p ={}&
\epsilon \left(
   \frac{\left(\delta^{1/p}\right)}
        {\left( 1 + \epsilon(1 - \delta) \right)^{1/p}} - 1 \right) +
(1 - \epsilon) \left(
   \frac{1}
        {\left( 1 + \epsilon(1 - \delta) \right)^{1/p}} - 1 \right).
\end{align*}

Since $\mu$ is absolutely continuous with respect to the Lebesgue measure, there
exists a sequence $\epsilon_n \rightarrow 0$ with $\epsilon_n > 0$ and a
sequence of corresponding sets $S_n$ such that $\pbase(S_n) = \epsilon_n$. (See
\lemref{continuity_partition} for a proof of this fact, which is a
straightforward consequence of \citet[Proposition 15.5]{nielsen:1997:measure}
and the continuity of the Lebesgue measure.) Since $\q(\theta) > 0$ on
$\thetadom$, $\q(S_n) > 0$ for all $n$.  Since $\KL{\q(\theta) ||
\pbase(\theta)}$ is finite, we must have $\lim_{n \rightarrow} \q(S_n) = 0$.

Take $\delta_n  = \exp(-1 / (\q(S_n)^2))$, and take $\palt(\theta) =
\palt(\theta \vert S_n, \delta_n)$.  Then $\epsilon_n (1 - \delta_n) \rightarrow
0$, and $\q(S_n)\log \delta_n = -1 / \q(S_n)$, so
\begin{align*}
\abs{\KL{q(\theta) || \palt(\theta \vert S_n, \delta_n)} -
    \KL{q(\theta) || \pbase(\theta)}} \rightarrow{}& \infty, \quad \textrm{but}\\
\norm{\phi(\theta \vert \palt(\cdot \vert S_n, \delta_n), p)}_p^p
    \rightarrow{}& 0.
\end{align*}
Thus, for sufficiently large $n$, the conclusion follows.
\qed
%

\section{Detailed Proofs}\applabel{proofs}
In this section, we provide detailed proofs for results stated above.

A standard consequence of the dominated convergence theorem is the ability to
exchange integration and differentiation.  Since we will use this result
frequently, we state it here in our own notation as \thmref{dct}.

\begin{thm}\thmlabel{dct}
\citep[Theorem 16.8]{billingsley:1986:probability}
Let $\mu$ be sigma-finite measure on $\thetadom$, and let $S_\t \subseteq
\mathbb{R}$.  Let $f:\thetadom \times S_\t \mapsto \mathbb{R}$.

If there exists a function $M(\theta)$ with $\int M(\theta) \mu(d\theta) <
\infty$ such that $\abs{f(\theta, \t)} \le M(\theta)$, $\mu$-almost surely,
for all $\t \in S_\t$, then the map $\t \mapsto \int f(\theta, \t)
\mu(d\theta)$ is continuous.

Further, suppose that the derivative $\fracat{\partial f(\theta, \t)}{\partial
\t}{\t}$ exist $\mu$-almost surely for $\t \in S_\t$.  If there exists
an $M'(\theta)$ such that $\int M'(\theta) \mu(d\theta) < \infty$ and
$\abs{\fracat{\partial f(\theta, \t)}{\partial \t}{\t}} \le M'(\theta)$,
$\mu$-almost surely and for all $\t \in S_\t$, then
\begin{align*}
\fracat{\partial \int f(\theta, \t) \mu(d\theta)}{\partial \t}{\t} =
     \int \fracat{\partial f(\theta, \t)}{\partial \t}{\t} \mu(d\theta).
\end{align*}
\end{thm}

%
\begin{lem}\lemlabel{exchange_order}
Under \assuref{exchange_order_f}, at any $\eta, \t \in \ball_\eta \times
\ball_\t$, we can exchange the order of integration and differentiation in $\int
f(\theta, \eta, \t) \mu(d\theta)$ for the derivatives $\partial / \partial
\eta$, $\partial^2 / \partial \eta^2$, and $\partial^2 / \partial \eta \partial
\t$.
\begin{proof}

Let $\eta_d$ denote the $d-$th entry of the vector $\eta$.  Then
\begin{align*}
\abs{\partial f(\theta, \eta, \t) / \partial \eta_d} \le{}&
  \norm{\partial f(\theta, \eta, \t) / \partial \eta}_2
\textrm{,}\\
\abs{\partial^2 f(\theta, \eta, \t) / \partial \eta_d \partial \t} \le{}&
    \norm{\partial f(\theta, \eta, \t) / \partial \eta \partial \t}_2
\textrm{, and}\\
\abs{\partial^2 f(\theta, \eta, \t) /
       \partial \eta_{d_1} \partial \eta_{d_2}} \le{}&
     \norm{\partial f(\theta, \eta, \t) / \partial \eta \partial\eta^T}_2.
\end{align*}
The conclusion follows by repeatedly applying \thmref{dct} to the components
of the derivatives.

\end{proof}
\end{lem}
%


\begin{lem}\lemlabel{logq_continuous}

Under \assuref{exchange_order}, the map $\eta, \t \mapsto \expect{\q(\theta
\vert \eta)}{\log \ptil(\theta \vert \t)}$ has continuous partial derivatives
$\partial / \partial \eta$, $\partial^2 / \partial \eta^2$, and $\partial^2 /
\partial \eta \partial \t$ at all $\eta, \t \in \ball_\eta \times \ball_\t$.
Furthermore,

\begin{align}
\fracat{\partial \expect{\q(\theta \vert \eta)}
              {\log \ptil(\theta \vert \t)}}{\partial \eta}{\eta}
={}&
\expect{\q(\theta \vert \eta)}
       {\lqgradbar{\theta \vert \eta}
       \log \ptil(\theta \vert \t)}
       \eqlabel{q_sens_is_cov}\\
\fracat{\partial^2 \expect{\q(\theta \vert \eta)}
      {\log\ptil(\theta \vert \t)}}{\partial \eta \partial \t}{\eta, \t}
={}&
\expect{\q(\theta \vert \eta)}
       {\lqgradbar{\theta \vert \eta}
       \fracat{\partial \log\ptil(\theta \vert \t)}{\partial \t}{\t}}
\eqlabel{q_sens_psi_grad_is_cov}.
\end{align}
\begin{proof}
We can write
\begin{align*}
R(a, b) :={} \frac{a}{b} \quad \Rightarrow
\expect{\q(\theta \vert \eta)}{\psi(\theta, \t)} ={}
R\left(\int \qtil(\theta \vert \eta) \psi(\theta, \t) \mu(d\theta),
  \int \qtil(\theta \vert \eta) \mu(d\theta)\right).
\end{align*}
If necessary, we can shrink $\ball_\eta$ so that the denominator $\int
\qtil(\theta \vert \eta) \mu(d\theta)$ is bounded below by a positive constant
for all $\eta \in \ball_\eta$.  With the denominator strongly positive, $R(a,b)$
is is a continuously differentiable function to all orders for all $\t, \eta \in
\ball_\t \times \ball_\eta$.  The desired results follow from
\assuref{exchange_order} by the chain rule.

\end{proof}
\end{lem}

\vspace{1em}

\prooflabel{etat_deriv}
\proofof{\thmref{etat_deriv}}
By \assuitemref{kl_opt_ok}{kl_diffable} and \lemref{logq_continuous}, $\eta
\mapsto \KL{\eta, \t}$ is continuously differentiable for all $\eta, \t \in
\ball_\eta \times \ball_\t$.  So, for all $\t \in \ball_\t$, the optimal
$\etaopt(\t)$ satisfies the first order condition:
\begin{align}\eqlabel{vb_first_order_condition}
\fracat{\partial \KL{\eta, \t}}{ \partial \eta}{\etaopt(\t), \t} ={}
\fracat{\partial \KL{\eta}}{ \partial \eta}{\etaopt(\t)} +
\fracat{\partial
    \expect{\q(\theta \vert \eta)}{\log \ptil(\theta \vert \t)}}
    {\partial \eta}
    {\etaopt(\t)}
={} 0
\end{align}

We wish to apply the implicit function theorem \citep[Theorem
3.3.1]{krantz:2012:implicit} to the estimating equation defined by
\eqref{vb_first_order_condition}. Again, by \assuitemref{kl_opt_ok}{kl_diffable}
and \lemref{logq_continuous}, the estimating equation given is continuously
differentiable in both $\eta$ and $\t$. The Jacobian of the estimating equation
is nonsingular by \assuitemref{kl_opt_ok}{kl_hess}, and valid in an open ball by
\assuitemref{kl_opt_ok}{kl_opt_interior}. Finally, the form of the derivative is
given by \citet[Theorem 3.3.1]{krantz:2012:implicit}, together with
\eqref{q_sens_psi_grad_is_cov} of \lemref{logq_continuous}.

For convenience, \tabref{kranz_notation} shows the correspondence between our
notation and that of \citet[Theorem 3.3.1]{krantz:2012:implicit}.

\begin{center}
\begin{tabular}{|c|c|}
\hline Krantz \& Parks notation & Our notation \\\hline
$\Phi(x)$                       & $\KL{\eta, \t}$ \\\hline
$Q$                             & $1$ \\\hline
$M$                             & $\etadim$ \\\hline
$U$                             & $\ball_\eta \times \ball_\t$ \\\hline
$W$                             & $\ball_\t$ \\\hline
$x_1,\ldots,x_Q$                & $\t$ \\\hline
$x_{Q+1},\ldots,x_N$            & $\eta$ \\\hline
$f_1(x_a), \ldots,f_M(x_a)$     & $\etaopt(\t)$ \\\hline
%
\end{tabular}\tablabel{kranz_notation}
\end{center}

\qed
%

\vspace{1em}

\prooflabel{pert_invariance}\proofof{\lemref{pert_invariance}}
Let $\mu$ and $\mu'$ denote two mutually absolutely continuous candidate
dominating measures for $\pbase$, with respective densities (Radon-Nikodym
derivatives) $\pbase(\theta)$ and $\pbase'(\theta)$.  Let the respective
densities of the measure $\p$ be denoted $\palt(\theta)$ and $\palt'(\theta)$ as
well.  Let $R(\theta) = \fracat{d\mu}{d\mu'}{\theta}$ denote the Radon-Nikodym
derivative of $\mu$ with respect to $\mu'$, and note that $\pbase'(\theta) =
R(\theta) \pbase(\theta)$ and $\palt'(\theta) = R(\theta) \palt(\theta)$.

We have that the perturbations for $\mu$ and $\mu'$ are given respectively by
\begin{align*}
\phi(\theta \vert \beta, \palt) ={}&
  \log \palt(\theta) - \log \pbase(\theta) + \log \beta \\
\phi'(\theta \vert \beta, \palt') ={}&
    \log \palt'(\theta) - \log \pbase'(\theta) + \log \beta
\\={}&
\log \palt(\theta) - \log R(\theta)
    - \log \pbase(\theta) + \log R(\theta)+ \log \beta
\\={}&
\phi(\theta \vert \beta, \palt).
\end{align*}
It follows that $\norminf{\phi(\cdot \vert \beta, \palt)} = \norminf{\phi'(\cdot
\vert \beta, \palt')}$.

Next, let $\tau := \tau(\theta)$ be an invertible transformation with Jacobian
$J(\theta) := \mathrm{det}\left(\fracat{d\tau}{d\theta^T}{\theta}\right)$. For
the dominating measure $\mu$, let $\pbase(\theta)$ and $\palt(\theta)$ denote
the densities of $\theta$ and $\pbase'(\tau)$ and $\palt'(\tau)$ denote the
densities of $\tau$.  The desired result follows by the exact same formal
argument as for the change of measure, except with $J(\theta) \mu(d\theta)$
and $\mu(d\tau)$ taking the place of $R(\theta) \mu(d\theta)$ and
$\mu'(d\theta)$, respectively.

We now prove that $\phi$ gives rise to valid priors when $\norminf{\phi} <
\infty$. Since the exponential function is positive, for any $\phi(\theta)$,
\begin{align*}
\ptil(\theta \vert \phi) = \pbase(\theta) \exp(\phi(\theta)) > 0,
\end{align*}
$\mu$-almost everywhere.  Furthermore, since $\int \pbase(\theta)
\lambda(d\theta) = 1$,
\begin{align*}
\exp(-\norminf{\phi}) \le{}
\int \pbase(\theta) \exp\left(\phi(\theta)\right) \mu(d\theta)
\le{}
\exp(\norminf{\phi}).
\end{align*}
so that $0 < \int \ptil(\theta \vert \phi) \mu(d\theta) < \infty$
whenever $\norminf{\phi} < \infty$.
\qed

\vspace{1em}

\begin{lem}\lemlabel{exchange_order_q_suffices}
Under \defref{prior_nl_pert},
\assuref{exchange_order_q} implies \assuref{exchange_order} when
$\norminf{\phi} < \infty$.

\begin{proof}
Since $\qtil(\theta \vert \eta) \phi(\theta) \le \qtil(\theta \vert \eta)
\delta$, and $\qtil(\theta \vert \eta)$ satisfies \assuref{exchange_order_f}
with some $M(\theta)$ by \assuref{exchange_order_q}, we can satisfy
\assuref{exchange_order_f} for $\qtil(\theta \vert \eta) \phi(\theta)$ with
$\max\{1, \delta\} M(\theta)$.  Finally, \lemref{exchange_order} implies that
\assuref{exchange_order} is satisfied.
\end{proof}
\end{lem}

\begin{lem}\lemlabel{objective_is_frechet}

Under \assuref{exchange_order_q}, the map $\eta, \phi \mapsto \partial
\expect{\q(\theta \vert \eta)}{\phi(\theta)} / \partial \eta$ is  continuously
Fr{\'e}chet differentiable as a map from $\mathbb{R}^\etadim \times \linf
\mapsto \mathbb{R}^\etadim$.
\begin{proof}
The map $\eta, \phi \mapsto  \partial \expect{\q(\theta \vert
\eta)}{\phi(\theta)} / \partial \eta$ is a map from the Banach space
$\mathbb{R}^\etadim \times \linf$ into the Banach space $\mathbb{R}$. Let us
take the L2 norm $\norm{\cdot}_2$ on $\mathbb{R}^{\etadim}$ and $\mathbb{R}$.
Let $\ball$ denote the ball $\ball_\eta \times \{ \phi: \norminf{\phi} <
\delta\}$ for some $\delta > 0$.  Let $\linop$ denote a linear operator from
$\ball$ to $\mathbb{R}^\etadim$, and define the dual norm
\begin{align*}
\norm{\linop}^* :=
    \sup_{\Delta \eta: \norm{\eta}_2 \le 1}
    \sup_{\Delta \phi: \norminf{\phi} \le 1}
     \norm{\linop(\Delta \eta, \Delta \phi)}_2.
\end{align*}
Formally, $\Delta \eta$ and $\Delta \phi$ are members of $\mathbb{R}^\etadim$
and $\linf$ respectively, but in the preceding display they can be thought of as
directions on which the linear operator $\linop$ operates.

Observe that the directional derivatives are linear operators, and so
$\norm{\cdot}^*$ defines a norm on the space of linear operators. We will prove
Fr{\'e}chet differentiability using the fact that a functional is Fr{\'e}chet
differentiable if its directional derivatives are continuous in $\norm{\cdot}^*$
as a function of the location at which they are evaluated
(\citet[Proposition 4.8(c)]{zeidler:2013:functional}, \citet[Corollary
1.4]{averbukh:1967:theory} and \citep[Appendix A]{reeds:1976:thesis}). Further,
it suffices by \citet[Proposition 4.14(c)]{zeidler:2013:functional} to show that
the partial derivatives with respect to $\eta$ and $\phi$ are
continuously Fr{\'e}chet to
show that the joint map is continuously Fr{\'e}chet differentiable.


First, consider the partial derivative with respect to $\eta$.   Observe that,
by \lemref{exchange_order_q_suffices}, \lemref{logq_continuous} applies with
$\log \ptil(\theta \vert \t) = \phi(\theta)$ (no $\t$ dependence). Consequently the
map $\eta \mapsto \expect{\q(\theta \vert \eta)}{\phi(\theta)}$ is twice
continuously differentiable.   The linear operator corresponding to the
directional derivative in the $\Delta \eta$ direction is given by
\begin{align*}
\linop_\eta(\Delta \eta, \Delta \phi) =
    \fracat{\partial^2 \expect{\q(\theta \vert \eta)}{\phi(\theta)}}
           {\partial \eta \partial\eta^T}{\eta} \Delta \eta,
\end{align*}
with no dependence on $\Delta \phi$.  Define for the moment the the $\etadim
\times \etadim$ matrix $\mathscr{H}(\eta, \phi) := \partial^2 \expect{\q(\theta
\vert \eta)}{\phi(\theta)} / \partial \eta \partial \eta^T$.  Then the dual norm
of the derivative is simply the operator norm of $\mathscr{H}$, i.e.,
$\norm{\linop_\eta}^* = \norm{\mathscr{H}(\eta, \phi)}_{op}$. Thus we must show
that $\norm{\mathscr{H}(\eta, \phi)}_{op}$ is continuous in $\eta, \phi$.  For
any $\eta', \phi'$ and $\eta'', \phi''$ in $\ball_\eta \times
\ball_\phi(\delta)$,
\begin{align*}
\MoveEqLeft
\norm{\mathscr{H}(\eta', \phi') - \mathscr{H}(\eta'', \phi'')}_{op} \\
&\le
\norm{\mathscr{H}(\eta', \phi') - \mathscr{H}(\eta', \phi'')}_{op} +
\norm{\mathscr{H}(\eta', \phi'') - \mathscr{H}(\eta'', \phi'')}_{op}.
\end{align*}
For the first term in the preceding display, for all $\eta'$,
\begin{align*}
\norm{\mathscr{H}(\eta', \phi') - \mathscr{H}(\eta', \phi'')}_{op}
    \le{}&
    \fracat{\partial^2 \expect{\q(\theta \vert \eta)}{1}}
           {\partial \eta \partial\eta^T}{\eta'} \norminf{\phi' - \phi''}
           \Rightarrow\\
\lim_{\phi' \rightarrow \phi''}
\norm{\mathscr{H}(\eta', \phi') - \mathscr{H}(\eta', \phi'')}_{op} ={}& 0.
\end{align*}
For the second term, by \lemref{logq_continuous}, for all $\phi''$,
\begin{align*}
\lim_{\eta' \rightarrow \eta''}
    \norm{\mathscr{H}(\eta', \phi'') - \mathscr{H}(\eta'', \phi'')}_{op} = 0.
\end{align*}
It follows that $\norm{\mathscr{H}(\eta, \phi)}_{op}$ is continuous in $\eta,
\phi$, and so the partial derivative with respect to $\eta$ is
a continuous Fr{\'e}chet derivative.


Next, we consider the partial derivative with respect to $\phi$.  By \eqref{q_sens_is_cov}, we can write
\begin{align*}
\fracat{\partial \expect{\q(\theta \vert \eta)}{\phi(\theta)}}
       {\partial \eta}{\eta}
={}
\expect{\q(\theta \vert \eta)}
       {\lqgradbar{\theta \vert \eta} \phi(\theta)}.
\end{align*}
Since this expression is linear in $\phi$, the linear operator for the partial
derivative with respect to $\phi$ is given by
\begin{align*}
\linop_\phi(\Delta \eta, \Delta \phi) =
    \expect{\q(\theta \vert \eta)}
           {\lqgradbar{\theta \vert \eta} \Delta \phi(\theta)},
\end{align*}
with no dependence on $\Delta \eta$.

In order to be a valid partial derivaive, we must verify that $\linop_\phi$ is a
bounded linear operator.  Boundedness follows from H{\"o}lder's inequality and
\assuref{exchange_order_q} since
\begin{align*}
\sup_{\Delta\phi: \norminf{\Delta\phi} \le 1}
    \norm{\linop_\phi(\Delta \eta, \Delta \phi)}_2
\le{}&
\expect{\q(\theta \vert \eta)}
       {\norm{\lqgradbar{\theta \vert \eta}}_1} \norminf{\Delta \phi}
\\\le{}&
\sqrt{\etadim}\expect{\q(\theta \vert \eta)}
       {\norm{\lqgradbar{\theta \vert \eta}}_2}
\\\le{}&
\sqrt{\etadim} \int M(\theta) \mu(d\theta) < \infty.
\end{align*}

Similarly, the dual norm of the $\phi$ partial derivative is given by
\begin{align*}
\norm{\linop_\phi}^* ={}& \expect{\q(\theta \vert \eta)}
       {\norm{\lqgradbar{\theta \vert \eta}}_1}.
\end{align*}
We thus need to show that $\eta \mapsto \expect{\q(\theta \vert \eta)}
{\norm{\lqgradbar{\theta \vert \eta}}_1}$ is a continuous function of $\eta$
(there is no $\phi$ dependence).  To show this, observe that
\begin{align}
\expect{\q(\theta \vert \eta)}
       {\norm{\lqgradbar{\theta \vert \eta}}_1} ={}&
\frac{\int \qtil(\theta \vert \eta)
           \norm{\lqgradbar{\theta \vert \eta}}_1 \mu(d\theta)}
     {\int \qtil(\theta \vert \eta) \mu(d\theta)}.
    \eqlabel{phi_partial_dual}
\end{align}
By \assuref{exchange_order_q}, we have that there exists a finitely integrable
envelope function $M(\theta)$ such that, for all $\eta \in \ball_\eta$,
\begin{align*}
\qtil(\theta \vert \eta) \le{}& M(\theta) \mathand \\
\qtil(\theta \vert \eta)
           \norm{\lqgradbar{\theta \vert \eta}}_1
    \le{}&
\sqrt{\etadim} \qtil(\theta \vert \eta)
           \norm{\lqgradbar{\theta \vert \eta}}_2
           \le{} M(\theta).
\end{align*}
Therefore, by the dominated convergence theorem, we can exchange limits and
integrals in the numerator and denominator of \eqref{phi_partial_dual}.  It
follows that, for any $\eta'$ and $\eta''$ in $\ball_\eta$,
\begin{align*}
\lim_{\eta' \rightarrow \eta''}
\abs{\int \qtil(\theta \vert \eta') \mu(d\theta) -
     \int \qtil(\theta \vert \eta'') \mu(d\theta)}
\le{}&
\lim_{\eta' \rightarrow \eta''}
\int  \abs{\qtil(\theta \vert \eta') - \qtil(\theta \vert \eta'')}
\mu(d\theta)
\\={}&
\int \lim_{\eta' \rightarrow \eta''}  \abs{
\qtil(\theta \vert \eta') - \qtil(\theta \vert \eta'')
}
={} 0.
\end{align*}
Thus the numerator of \eqref{phi_partial_dual} is continuous in $\eta$. The
denominator of \eqref{phi_partial_dual} is also continuous by an analogous
argument.  Since the denominator of \eqref{phi_partial_dual} is bounded away
from zero, $\expect{\q(\theta \vert \eta)} {\norm{\lqgradbar{\theta \vert
\eta}}_1}$ is a continuous composition of continuous functions, and itself
continuous.  It follows that the $\phi$ partial derivative is
continuously Fr{\'e}chet differentiable.

Since its partial derivatives are continuous, it follows by \citet[Proposition
4.14(c)]{zeidler:2013:functional} that the joint map $\eta, \phi \mapsto
\partial \expect{\q(\theta \vert \eta)}{\phi(\theta)} / \partial \eta$ is
continously Fr{\'e}chet differentiable.

\end{proof}
\end{lem}


\prooflabel{eta_phi_deriv}\proofof{\thmref{eta_phi_deriv}}
Recall that, by \lemref{exchange_order_q_suffices}, \lemref{logq_continuous}
applies with $\log \ptil(\theta \vert \t) = \phi(\theta)$ (no $\t$ dependence).
Therefore, as in the proof of \thmref{etat_deriv}, for any $\phi \in
\ball_\phi(\delta)$, $\etaopt(\phi)$ satisfies the first-order condition
\begin{align}\eqlabel{vb_first_order_condition_phi}
\fracat{\partial \KL{\eta}}{ \partial \eta}{\etaopt(\phi)} +
\fracat{\partial
    \expect{\q(\theta \vert \eta)}{\phi(\theta)}}
    {\partial \eta}
    {\etaopt(\phi)}
={} 0.
\end{align}

As in the proof of \thmref{etat_deriv}, we wish to employ an implicit
function theorem, but this time for general Banach spaces.  We will
use \citet[Theorem 4.B]{zeidler:2013:functional}.

First, \citet[Chapter 4 Condition 21b]{zeidler:2013:functional} holds since
$\hessopt$ is invertible by \assuitemref{kl_opt_ok}{kl_hess}.   So we satisfy
conditions (i), (ii), and (iii) of \citet[Theorem
4.B(c)]{zeidler:2013:functional}, giving that the function $\etaopt(\phi)$
exists.

Moreover, by \assuitemref{kl_opt_ok}{kl_diffable} and
\lemref{objective_is_frechet}, the estimating equation
\eqref{vb_first_order_condition_phi} is continuously Fr{\'e}chet differentiable
($C^1$ in the notation of Zeidler) in a neighborhood of $\etaopt, 0$.
It follows from \citet[Theorem 4.B(d)]{zeidler:2013:functional}, $\etaopt(\phi)$
is also continuously Fr{\'e}chet differentiable.
\qed
%


\begin{lem}\lemlabel{continuity_partition}
If $\mu$ is absolutely continuous with respect to the Lebesgue measure,  and
$\pbase$ is a probability measure with a density relative to $\mu$, then there
exists a sequence $\epsilon_n \rightarrow 0$ with $\epsilon_n > 0$ and a
sequence of corresponding sets $S_n$ such that $\pbase(S_n) = \epsilon_n$.
\begin{proof}
Let $\epsilon'_n = n^{-1}$.  Since $\pbase \ll \mu \ll \lambda$ (where $\lambda$
is the Lebesgue measure), by applying \citet[Proposition
15.5]{nielsen:1997:measure}, for each $n$ there exists a $\delta'_n$ such that,
for any measureable set $A$ with $\mu(A) < \delta'_n$, $\pbase(A) <
\epsilon'_n$.  Again applying \citet[Proposition 15.5]{nielsen:1997:measure},
there similarly exists a $\delta_n$ such that for any measureable set $A$ with
$\lambda(A) < \delta_n$, $\mu(A) < \delta'_n \Rightarrow \pbase(A) <
\epsilon'_n$.

For each $n$, partition $\thetadom$ into a countable number of sets $A_{m}$ such
that $\sum_{m} \lambda(A_{m}) = 1$ and $\lambda(A_{m}) < \delta_n$. (This is
possible by dividing $\thetadom$ into sufficiently small rectangles, for
example.)  Then $\pbase(A_{m}) < \epsilon'_n$ for all $m$.  Since $\pbase$ is a
probability measure, $\sum_m \pbase(A_{m}) = 1$, so there must exist at least $1 /
\epsilon'_n$ indices $m'$ such that $\pbase(A_{m'}) > 0$. Take any such $m'$ and
let $\epsilon_n = \pbase(A_{m'})$ and $S_n = A_{m'}$.
\end{proof}
\end{lem}

\begin{lem}\lemlabel{normal_q_is_regular}
Let $\mu$ denote the Lebesgue measure on $\mathbb{R}$. Let $\sigma$ denote the
standard deviation of a normal distribution, let $\eta$ denote a continuously
differentiable function of the normal distribution's natural parameters, let
$\normdist{\theta \vert \eta}$ denote a normal density with respect to $\mu$,
and let $\ball_\eta$ denote an open ball in $\mathbb{R}^2$ such that $0 < \sigma <
\infty$ for all $\eta \in \ball_\eta$.
Let $\ball_\t$ denote an open ball in $\mathbb{R}$, and let $\psi(\theta, \t)$
be a function such that $\theta \mapsto \psi(\theta, \t)$ is $\mu$-measurable
for all $\t \in \ball_\t$.

If there exists a constant $C > 0$ such that
\begin{align*}
\sup_{\t \in \ball_\t} \abs{\psi(\theta, \t)}
    \le C \exp(\abs{\theta})
\mathand
\sup_{\t \in \ball_\t}
    \abs{\frac{\partial \psi(\theta, \t)}{\partial \t}}
    \le C \exp(\abs{\theta}),
\end{align*}
then one can exchange the order of expectation and differentation in the
expression $\expect{\normdist{\theta \vert \eta}}{\psi(\theta, \t)}$ for the
derivatives $\partial / \partial \eta$, $\partial^2 / \partial \eta \partial
\eta$, and $\partial^2 / \partial \eta  \partial \t$, eavluated at any $\t \in
\ball_\t$ and $\eta \in \ball_\eta$.

\begin{proof}
%
For the moment, let $\eta$ denote the exponential family natural parameters of
the normal distribution. By properties of the exponential family,
\begin{align*}
\fracat{\partial \log \qtil(\theta \vert \eta)}
       {\partial \eta}{\eta}
      ={} (\theta, \theta^2)^T \mathand&
\fracat{\partial^2 \log \qtil(\theta \vert \eta)}
      {\partial \eta \partial \eta^T}{\eta}
      ={} 0_{2\times2} \Rightarrow\\
\norm{
    \fracat{\partial \log \qtil(\theta \vert \eta)}
           {\partial \eta}{\eta}
}_2^2 ={} \theta^2 + \theta^4 \mathand&
\norm{
    \fracat{\partial^2 \log \qtil(\theta \vert \eta)}
          {\partial \eta \partial \eta^T}{\eta}
}_2 ={} 0.
\end{align*}
Let $\ballclosed_\eta$ denote the closure of $\ball_\eta$, and let
\begin{align*}
\eta^* := \argmax_{\eta \in \ballclosed_\eta}
    \expect{\q(\theta \vert \eta)}{\exp(\abs{\theta})}.
\end{align*}
By standard properties of the normal and the boundedness of $\sigma(\eta)$, the
right hand side of the preceding display is finite.
Then
\begin{align*}
\int \q(\theta \vert \eta) \psi(\theta, \t) \mu(d \theta) \le{}&
    \left( \sup_{\theta} \sup_{\t \in \ball_\t}
        \abs{\psi(\theta, \t)} \exp(-\abs{\theta}) \right)
    \int \q(\theta \vert \eta) \exp(\theta) \mu(d \theta)
\\\le{}&
    \const
    \expect{\q(\theta \vert \eta^*)}{\exp(\abs{\theta})}.
    \quad\constdesc{\eta, \t}
\end{align*}
Therefore, for \assuref{exchange_order_f}, we can take $M(\theta)
\propto \q(\theta \vert \eta^*) \exp(\abs{\theta})$. The other terms follow
similarly, since each multiplier of $\q(\theta \vert \eta)$ is dominated by
$\exp(-\abs{\theta})$.  The final $M(\theta)$ simply takes the largest
of the five constants.

Finally, if $\tilde{\eta}$ is a twice-continuously differentiable function of
the natural parameters $\eta$ (e.g the mean and variance), then the derivatives
with respect to $\tilde{\eta}$ are equal to the derivatives with respect to
$\eta$ times bounded (on $\ball_\eta$) functions of $\eta$ that do not depend on
$\theta$. Thus a constant multiple of $M(\theta)$ will bound the new
derivatives.
\end{proof}
\end{lem}

\section{Positive Perturbations Are Counterintuitive}\applabel{positive_pert}

The following example illustrates how, by requiring perturbations to be positive,
one can induce counterintuitive notions of the ``size'' of a perturbation that
ablates prior mass.

\begin{ex}\exlabel{positive_pert_large}
\SimPositivePertFig

Take $\mu$ to be the Lebesgue measure on $[0,1]$. Let $\pbase(\theta) = \ind{0
\le \theta \le 1}$.  For some $\delta > 0$ and $0 < \epsilon \ll 1$, let
\begin{align*}
\palt(\theta) :={}&
    \left(\frac{1-\delta \epsilon}{1 - \epsilon} \right)
        \ind{\epsilon \le \theta \le 1} +
    \delta \ind{0 \le \theta \le \epsilon}.
\end{align*}

We can use \eqref{p_pert_simple} to give $\ptil(\theta \vert \tp=1) = \palt(\theta)$ by using, for any $\alpha > 0$,
\begin{align*}
\phi(\theta) ={}&
    \left( \alpha\left(\frac{1-\delta \epsilon}{1-\epsilon} \right)^{1/p}
        - 1
    \right)
        \ind{\epsilon \le \theta \le 1} +
    \left(\alpha \delta^{1/p} - 1 \right) \ind{0 \le \theta \le \epsilon}.
\end{align*}
It follows that
\begin{align*}
\norm{\phi}_p ={}&
    \left( \alpha\left(\frac{1-\delta \epsilon}{1-\epsilon} \right)^{1/p} - 1
    \right) (1- \epsilon) +
    \left(\alpha \delta^{1/p} - 1 \right) \epsilon.
\end{align*}
For $\phi$ to be positive, we require
\begin{align*}
\alpha^p \ge \frac{1 - \epsilon}{1 - \delta \epsilon}
    \mathtxt{and}
\alpha^p \ge \frac{1}{\delta}.
\end{align*}

First, let us consider adding a small amount of prior mass, taking $\delta = 2 -
\epsilon$; let the corresponding perturbation be $\phi^+$.  For $\delta > 1$,
then we achieve $\phi \ge 0$ by taking $\alpha^p = \frac{1 - \epsilon}{1 -
\delta \epsilon}$.  Using the fact that $\epsilon \ll 1$ and keeping only
leading-order terms,
\begin{align*}
\frac{1-\epsilon}{1 - \delta \epsilon} \approx{}&
    (1- \epsilon)(1 + \delta \epsilon)
\\\approx{}& 1 + (\delta - 1) \epsilon
\\\approx{}& 1 + \epsilon,
\end{align*}
so
\begin{align*}
\norm{\phi^+}_p  ={}&
    \left(\alpha \delta^{1/p} - 1 \right) \epsilon
\\\approx{}&
    \left(
        \left( \left(1 + \epsilon\right) \left(2 - \epsilon \right)\right)^{1/p}
        - 1 \right) \epsilon
\\\approx{}&
\left( 2^{1/p} - 1 \right) \epsilon.
\end{align*}

Next, consider removing the same amount of mass with the symmetric change
$\delta = \epsilon$, letting $\phi^-$ be the corresponding perturbation. Then we
can ensure that $\phi(\theta) \ge 0$ with $\alpha^p \ge \epsilon^{-1}$, and
$\epsilon \ll 1$ gives
\begin{align*}
\frac{1-\delta\epsilon}{1 - \epsilon} \approx{}& 1- \epsilon,
\end{align*}
and
\begin{align*}
\norm{\phi^-}_p  ={}&
    \left( \alpha\left(\frac{1-\delta \epsilon}{1-\epsilon} \right)^{1/p} - 1
    \right) (1- \epsilon)
\\\approx{}&
\left(\left(\frac{1- \epsilon}{\epsilon}  \right)^{1/p} - 1\right)(1 - \epsilon)
\\\approx{}&
    \left( \frac{1}{\epsilon}\right)^{1/p}.
\end{align*}

Since $\epsilon$ is small, $\norm{\phi^-}_p \approx \left(
\frac{1}{\epsilon}\right)^{1/p} \gg \norm{\phi^+}_p \approx \left( 2^{1/p} - 1
\right) \epsilon$, despite the two perturbations respectively removing and
adding the same amount of arbitrarily small probability mass.

\end{ex}

\section{Computational details}

\subsection{The optimal local parameters}
\applabel{gmm_global_local_vb}
In all models we consider,
the optimal local variational parameters $\etaoptlocal$ can be written
as a closed-form function of the global variational parameters $\etaglob$.
Let $\etaoptlocal(\eta_\gamma; \t)$ denote this mapping; that is,
\begin{align*}
  \etaoptlocal(\etaglob; \t) := \argmin_{\etalocal} \KL{(\eta_\gamma, \etalocal), \t}.
\end{align*}

The next example details this mapping for the Gaussian mixture model.

\begin{ex}[Optimalility of $\etalocal$ in a GMM]\exlabel{qz_optimality}
Recall that under the truncated variational approximation,
the cluster assignment $\z_\n$ is a discrete random variable
over $\kmax$ categories.

Let $\eta_{\z_\n}$ be the categorical parameters in its
exponential family natural parameterization.
That is, we let $\eta_{\z_\n} = (\rho_{\n1}, \rho_{\n2}, ..., \rho_{\n(\kmax-1)})$
be an unconstrained vector in $\mathbb{R}^{\kmax-1}$;
in this parameterization, the assignment probabilities are
\begin{align*}
  p_{\n\k} := \expect{\q(\z_\n \vert \etaz)}{\z_{\n\k}} =
  \frac{\exp(\rho_{\n\k})}{1 + \sum_{\k'=1}^{\kmax-1}\exp(\rho_{\n\k})}
\end{align*}
We use the exponential family parameterization because
we require the optimal variational parameters $\etaopt$
to be interior to $\etadom$ in \thmref{etat_deriv}.
In the mean parameterization,
$\sum_{\k=1}^\kmax p_{\n\k} = 1$, so the
optimal mean parameters $\hat p_{\n}$ cannot be
interior to $\Delta^{\kmax - 1}$.
On the other hand, $\eta_{\z_\n}$ as defined
is unconstrained in $\mathbb{R}^{\kmax - 1}$.

Fixing $\q(\beta\vert\etabeta)$ and $\q(\nu\vert\etanu)$,
the optimal $\etaopt_{\z_\n}$ must satisfy
\begin{align*}
& \q(\z_\n | \etaopt_{\z_\n}) \propto \exp\left(\tilde \rho_{\n\k}\right)\\
& \mathwhere \tilde \rho_{\n\k} := \expect{\q(\beta, \nu \vert \eta)}
       {\log\p(\x_n \vert \beta_\k) + \log \pi_\k}.
\end{align*}
See \citet{bishop:2006:PRML} and \citet{blei:2017:vi_review} for details.
To satisfy this optimality condition,
we set the optimal $\etaopt_{\z_\n}$ to be
\begin{align*}
\etaopt_{\z_\n} = \left(\log\frac{\tilde\rho_{\n1}}{\tilde\rho_{\n\kmax}},
\log\frac{\tilde\rho_{\n2}}{\tilde\rho_{\n\kmax}}, \ldots,
\log\frac{\tilde\rho_{\n(\kmax-1)}}{\tilde\rho_{\n\kmax}}\right).
\end{align*}
Thus, as long as the expectations $\tilde\rho_{\n\k}$ can be provided
as a closed-form function of
$(\etabeta, \etanu)$, the optimal $\etaopt_{\z_\n}$ can be also be set in closed-form as
a function of $(\etabeta, \etanu)$.
\end{ex}

\subsection{More details on computing and inverting the Hessian}
\applabel{more_hessian}
We fill in more details for the efficient computation of the Hessian outlined in
\secref{computing_sensitivity}.

We start from our formula in \eqref{global_local_derivative_breakdown},
\begin{align*}
\fracat{d \etaopt(\t)}{d \t}{t = 0} ={}&
-\left(
\begin{array}{cc}
   \hess{\gamma\gamma} & \hess{\gamma\ell} \\
   \hess{\ell\gamma}     & \hess{\ell\ell} \\
\end{array}
\right)^{-1}
\left( \begin{array}{c} \crosshessian_\gamma \\ 0 \end{array}\right),
\end{align*}
and an application of the Schur complement gives
\begin{align*}
\fracat{d \etaopt(\t)}{d \t}{t = 0} ={}&
-\left(\begin{array}{c}
I_{\gamma\gamma} \\
\hess{\ell\ell}^{-1} \hess{\ell\gamma}
\end{array}\right)
\left(\hess{\gamma\gamma} -
      \hess{\gamma\ell} \hess{\ell\ell}^{-1} \hess{\ell\gamma}\right)^{-1} \crosshessian_\gamma,
\end{align*}
where $I_{\gamma\gamma}$ is the identity matrix with
the same dimension as $\eta_\gamma$.
Specifically, observe that the sensitivity of the global parameters
is given by
\begin{align*}
  \fracat{d \etaopt_\gamma(\t)}{d \t}{t = 0} &=
  - \hessopt_\gamma^{-1}\crosshessian_\gamma
  \mathwhere
  \hessopt_\gamma := \left(\hess{\gamma\gamma} -
        \hess{\gamma\ell} \hess{\ell\ell}^{-1} \hess{\ell\gamma}\right),
\end{align*}
In the models we consider, $\hess{\ell\ell}$ is sparse, and the size of
$\hess{\gamma\gamma}$ does not grow with $\N$. Thus, each term of
$\hessopt_\gamma$ can be tractably computed, stored in  memory, and inverted,
even on very large datasets.

One can derive the exact same identity using the optimality of
$\etaoptlocal(\eta_\gamma)$.  By applying the chain rule, one can
verify that
\begin{align}\eqlabel{global_kl_hess}
\hessopt_{\gamma} &=
    \frac{\partial^2}{\partial\eta_\gamma\partial\eta_\gamma^T}
    \KLglobal(\etaopt_\gamma, 0).
\end{align}
In practice, we evaluate $\hessopt_\gamma$ using automatic differentiation and
\eqref{global_kl_hess} rather than the Schur complement.

\subsection{Expressing $\g$ using global parameters only}
\applabel{vb_insample_nclusters_example}

Given a posterior quantity $\g$,
we again take advantage of the fact that the optimal
local parameters can be found in closed form given global parameters.
In general, $\g$ will be a function of the entire vector of variational parameters.
However, in the same way that $\KLglobal$ implicitly sets the local parameters at their optimum
and is a function of only global parameters and the prior parameter $\t$,
we can construct an analogous mapping for $\g$,
\begin{align}\eqlabel{g_as_global}
(\t, \etaglob) \mapsto g\Big(\big(\etaglob, \etaoptz(\etaglob, \t))\Big).
\end{align}

We illustrate this mapping
when our quantity of interest is the in-sample expected posterior number of clusters.

\begin{ex}\exlabel{vb_insample_nclusters_globallocal}
Let
$\gclustersabbr(\eta)$ denote the variational approximation to
$\expect{\p(\z\vert\x)}{\nclusters(\z)}$.   Using the fact that
$\p(\z_\n\vert \beta, \nu, \x) = \q(\z_\n \vert \etaopt_{\z_\n})$
is available in closed form, we can then take
\begin{align*}
\gclustersabbr(\etaopt) :={}&
    \expect{\q(\beta, \nu \vert\etaopt)}{
        \expect{\p(\z \vert \beta, \nu, \x)}{\nclusters(\z)}
    }
\\\approx{}&
    \expect{\p(\beta, \nu \vert \x)}{
        \expect{\p(\z \vert \beta, \nu, \x)}{\nclusters(\z)}
    }
    = \expect{\p(\z\vert\x)}{\nclusters(\z)} \Rightarrow \\
\gclustersabbr(\eta) ={}&
    \sumkm \left(1 -  \prod_{\n=1}^\N
        \left(1 - \expect{\q(\beta, \nu \vert \eta_\beta, \eta_\nu)}
                    {\expect{\p(\z_{\n} \vert \beta, \nu, \x)}{\z_{\n\k}}}
                    \right)\right).
\end{align*}
In this way, $\gclustersabbr(\eta)$ depends only on $\eta_\beta$ and $\eta_\nu$,
which are much lower-dimensional than $\eta_\z$, and retains nonlinearities in
the map
\begin{align*}
\eta_\beta, \eta_\nu \mapsto \expect{\q(\beta, \nu \vert \eta_\beta,
\eta_\nu)} {\expect{\p(\z_{\n} \vert \beta, \nu, \x)}{\z_{\n\k}}}.
\end{align*}
\end{ex}

The mapping \eqref{g_as_global} can be constructed for any posterior quantity $\g$.
Therefore, linearizing the global parameters using \eqref{global_sens, global_lin_approx} is sufficient:
we do not need to invert the full Hessian
and linearize the entire set of variational parameters, global and local.

\subsection{Evaluating stick expectations}\applabel{gh_quadrature}
We describe how to compute expectations with repsect to the stick-breaking
proportion $\nu_\k$. Let $f: \mathbb{R}\mapsto\mathbb{R}$ be a smooth function,
and we are interested in expectations of the form
\begin{align*}
  \expect{\q(\nuk \vert \eta)}{f(\nuk)}.
\end{align*}
For example, $f$ might be $f(\nu_\k) = \log \p(\nu_\k)$, whose
expectation appears in the $\mathrm{KL}$ divergence.

Recall that we chose the distribution on the logit-transformed
stick-breaking proportions $\lnu_\k$ to be normally distributed.
Let $\lnumean_\k$ and $\lnusd_\k$ be the location and scale, respectively,
of the Gaussian distribution on $\lnu_\k$.
Also let $\s$ be the sigmoid function, so that $\nu_\k = \s(\lnu_\k)$.

To compute expectations of a smooth function
$f(\nuk)$, the law of the unconscious statistician states that
\begin{align*}
  \expect{\q(\nuk \vert \eta)}{f(\nuk)} ={}&
  \expect{\q(\lnu_\k \vert \eta)}
         {f\circ \s\left(\lnu_\k\right)}.
\end{align*}
By choosing $\q(\lnu_\k \vert \eta)$ to be Gaussian,
the right-hand side of is a Gaussian integral,
which we approximate
using GH quadrature with $\ngh$ knots,
located at $\xi_g$, weighted by $\omega_g$:
\begin{align}\eqlabel{gh_integral}
\expect{\q(\lnu_\k \vert \eta)}
       {f\circ \s\left(\lnu_\k\right)}
\approx{}&
    \sum_{g=1}^{\ngh} \omega_g f\circ \s \left(\lnusd_\k \xi_{g} + \lnumean_\k\right)
\end{align}
Using GH quadrature to approximate the expectation
is similar to the ``reparameterization trick,'' only using
GH points rather than standard normal draws.

\subsection{Unconstrained variational parameterizations}
\applabel{app_vb_unconstrained}
Recall from \thmref{etat_deriv} that we require the optimal variational parameters
$\etaopt$ to be in the interior of its domain. One way to achieve this
is to use only \textit{unconstrained} parameterizations for the
component distributions of $\q$. One such parameterization was
presented in \exref{qz_optimality}, where we let
$\eta_{\z_\n}$, which parameterize the cluster assignments, be allowed to
take any value in $\mathbb{R}^{\kmax - 1}$; the assignment probabilities
$m_{\n}\in\mathbb{R}^{\kmax}$, which are constrained to sum to one,
are then formed with an appropriate transform of the unconstrained parameters
$\eta_{\z_\n}$

Other variables require careful parameterization as well.
For instance, instead of parameterizing the normal distribution on
logit-sticks $\lnu_\k$ using a mean and variance, we let $\eta_{\nu_\k}\in\mathbb{R}^2$
be the mean and \textit{log} variance. The variance is constrained to be positive;
the log-variance is unconstrained on the real line.
In general, a real-valued parameter $\mu_i$ which must be constrained
$a < \mu_i < b$ can be transformed to its unconstrained parameterization
by letting
\begin{align*}
  \eta_i = \log(\mu_i - a) - \log(b - \mu_i).
\end{align*}

In the variational approximation to the GMM model,
we let the component variables $\beta_\k$ be Normal-Wishart.
In this case, the scale matrix of the Normal-Wishart, $W_\k\in\mathbb{R}^{d\times d}$,
is constrained be positive definite.
Because $W$ is symmetric, we only neeed $d(d + 1) / 2$ parameters to represent it.
To form an unconstrained parameterization, we factorize $W$ using the Cholesky decomposition,
\begin{align*}
W = L^T L,
\end{align*}
where $L$ is a lower-triangular matrix, with positive diagonal entries.
The unconstrained parameterization of $W$ is then taken to be
the strictly lower-diagonal entries of $L$,
along with the $\log$ of the diagonal entries of $L$.

\section{Additional Experimental Details}
\applabel{app_results}

First, we review the Beta prior on the stick-breaking proportions,
which are common to each model we considered.
Then, we give some additional modeling details for each experiment.


\newcommand{\BetaPriorsEx}{

\begin{knitrout}
\definecolor{shadecolor}{rgb}{0.969, 0.969, 0.969}\color{fgcolor}\begin{figure}[!h]

{\centering \includegraphics[width=0.980\linewidth,height=0.392\linewidth]{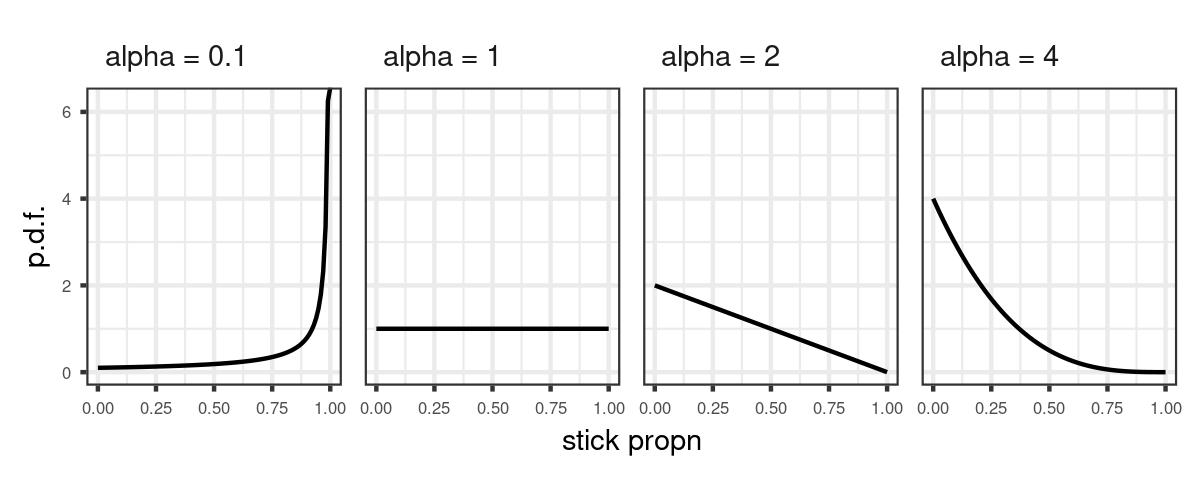} 

}

\caption[Probability density functions of $\text{Beta}(1, \alpha)$ distributions, under various $\alpha$ considered for the iris data set]{Probability density functions of $\text{Beta}(1, \alpha)$ distributions, under various $\alpha$ considered for the iris data set.}\label{fig:beta_priors}
\end{figure}

\end{knitrout}
}

\newcommand{\IrisInitFit}{

\begin{knitrout}
\definecolor{shadecolor}{rgb}{0.969, 0.969, 0.969}\color{fgcolor}\begin{figure}[!h]

{\centering \includegraphics[width=0.588\linewidth,height=0.470\linewidth]{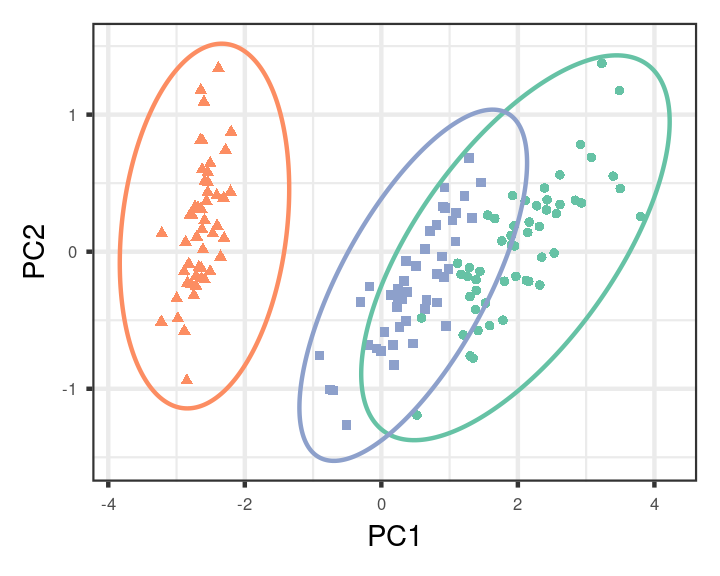} 

}

\caption[The iris data in principal component space and
                      GMM fit at $\alpha = 2$]{The iris data in principal component space and
                      GMM fit at $\alpha = 2$. Colors denote inferred memberships and
                      ellipses represent estimated covariances. }\label{fig:iris_fit}
\end{figure}

\end{knitrout}
}


\newcommand{\MiceExampleGenes}{

\begin{knitrout}
\definecolor{shadecolor}{rgb}{0.969, 0.969, 0.969}\color{fgcolor}\begin{figure}[!h]

{\centering \includegraphics[width=0.980\linewidth,height=0.392\linewidth]{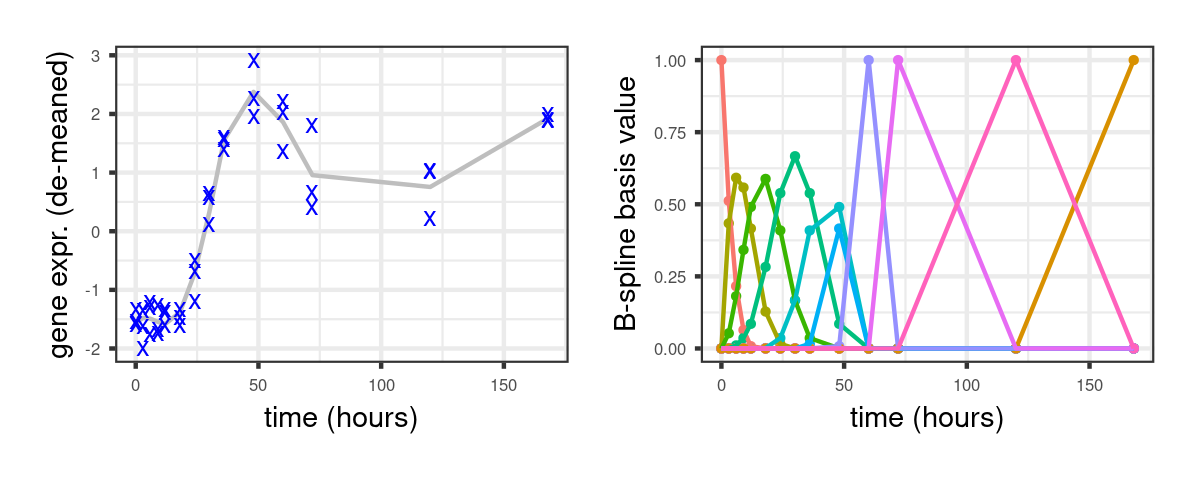} 

}

\caption[(Left) An example gene and its expression measured at 14 unique time points
    with three biological replicates at each time point.
     (Right) The cubic B-spline basis with 7 degrees of freedom,
    along with three indicator functions for the last three time points,
    $\timeindx = 72, 120, 168$]{(Left) An example gene and its expression measured at 14 unique time points
    with three biological replicates at each time point.
     (Right) The cubic B-spline basis with 7 degrees of freedom,
    along with three indicator functions for the last three time points,
    $\timeindx = 72, 120, 168$.}\label{fig:example_genes}
\end{figure}

\end{knitrout}
}

\newcommand{\MiceSmoothers}{

\begin{knitrout}
\definecolor{shadecolor}{rgb}{0.969, 0.969, 0.969}\color{fgcolor}\begin{figure}[!h]

{\centering \includegraphics[width=0.980\linewidth,height=0.627\linewidth]{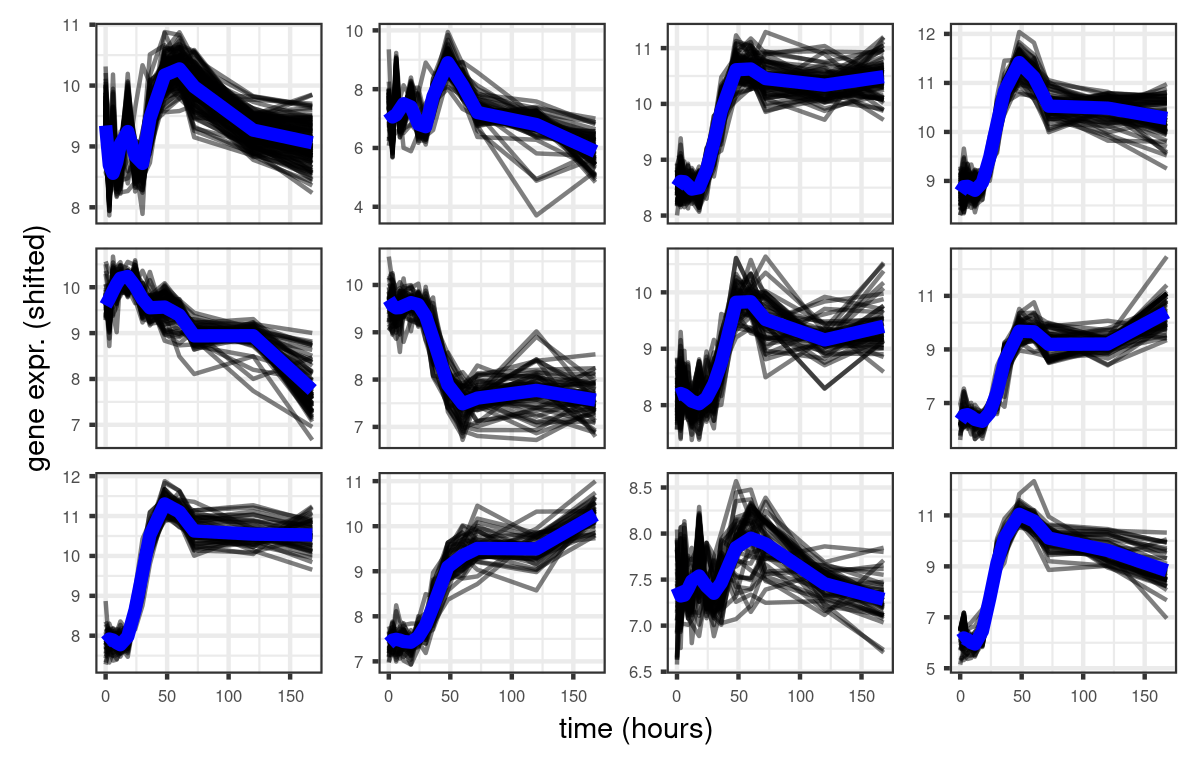} 

}

\caption[Inferred clusters in the mice gene expression dataset.
    Shown are the twelve most occupied clusters.
    In blue, the inferred cluster centroid.
    In grey, gene expressions averaged over replicates and
    shifted by their inferred intercepts]{Inferred clusters in the mice gene expression dataset.
    Shown are the twelve most occupied clusters.
    In blue, the inferred cluster centroid.
    In grey, gene expressions averaged over replicates and
    shifted by their inferred intercepts. }\label{fig:gene_centroids}
\end{figure}

\end{knitrout}
}


\newcommand{\StructureInitialFit}{

\begin{knitrout}
\definecolor{shadecolor}{rgb}{0.969, 0.969, 0.969}\color{fgcolor}\begin{figure}[!h]

{\centering \includegraphics[width=0.980\linewidth,height=0.470\linewidth]{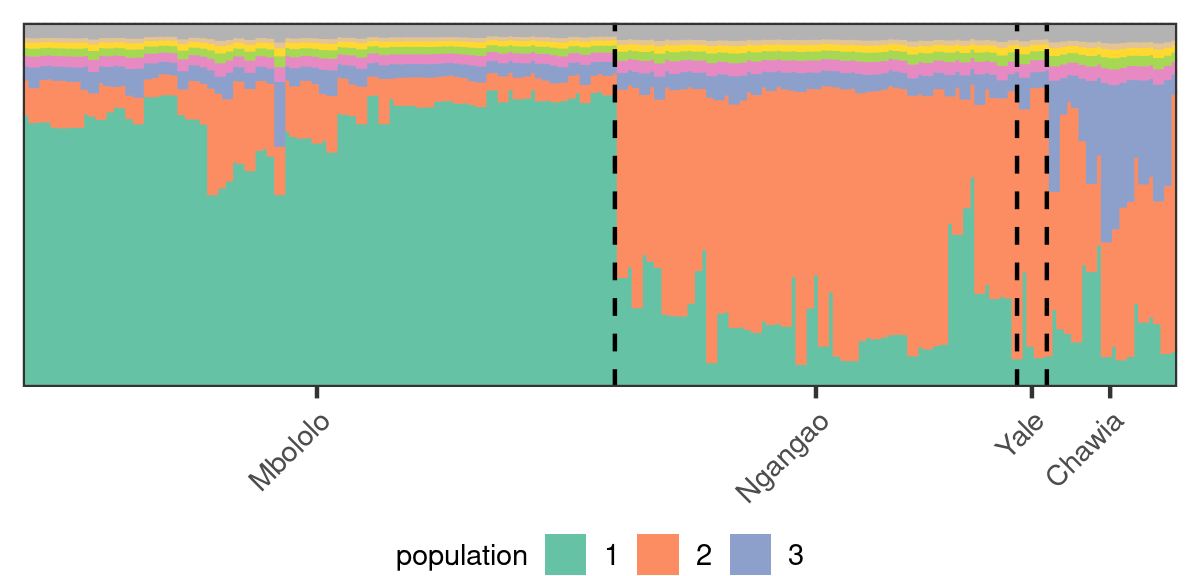} 

}

\caption[The inferred individual admixtures at $\alpha_0 = 3$.
    Each vertical strip is an individual and each color
    a latent population.
    Lengths of colored segments represent the inferred admixture proportions.
    Individuals are ordered by the geographic region from which they were sampled
    (Mbololo, Ngangao, Yale, and Chawia).
    In the text, we refer to the green, orange, and purple latent populations
    as population 1, 2, and 3, respectively]{The inferred individual admixtures at $\alpha_0 = 3$.
    Each vertical strip is an individual and each color
    a latent population.
    Lengths of colored segments represent the inferred admixture proportions.
    Individuals are ordered by the geographic region from which they were sampled
    (Mbololo, Ngangao, Yale, and Chawia).
    In the text, we refer to the green, orange, and purple latent populations
    as population 1, 2, and 3, respectively. }\label{fig:stru_init_fit}
\end{figure}

\end{knitrout}
}

\newcommand{\StructureLimitationsA}{

\begin{knitrout}
\definecolor{shadecolor}{rgb}{0.969, 0.969, 0.969}\color{fgcolor}\begin{figure}[!h]

{\centering \includegraphics[width=0.980\linewidth,height=0.666\linewidth]{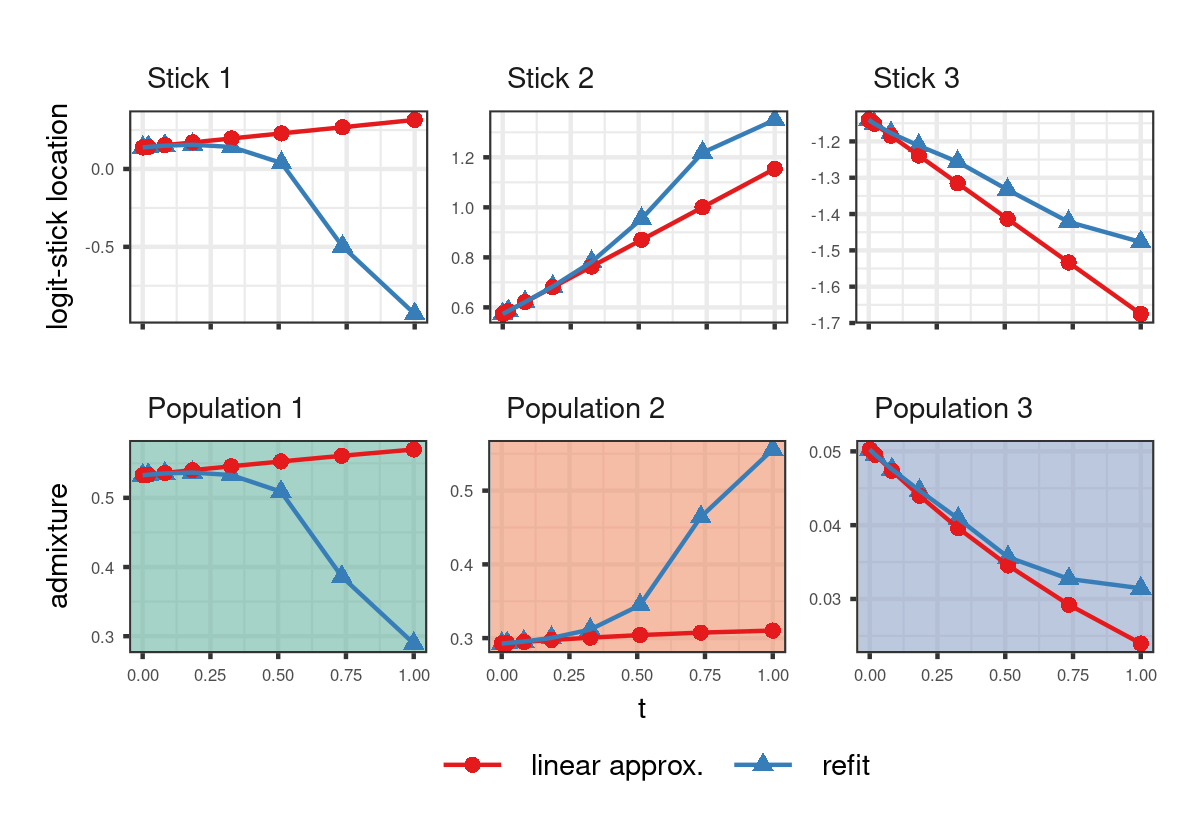} 

}

\caption[An individual $(\n = 26)$ for which
    the linearly approximated variational parameters
    poorly captured the
    change in admixture observed after refitting
    as $\t \rightarrow 1$.
    (Top row) the change in location parameter of the normally
    distributed logit-sticks, for the first three sticks.
    The response here is a variational parameter, so
    the approximation (red) is necessarily linear with respect to $\t$.
    (Bottom row) the change in the inferred admixtures for
    populations 1, 2, and 3]{An individual $(\n = 26)$ for which
    the linearly approximated variational parameters
    poorly captured the
    change in admixture observed after refitting
    as $\t \rightarrow 1$.
    (Top row) the change in location parameter of the normally
    distributed logit-sticks, for the first three sticks.
    The response here is a variational parameter, so
    the approximation (red) is necessarily linear with respect to $\t$.
    (Bottom row) the change in the inferred admixtures for
    populations 1, 2, and 3. }\label{fig:stru_lin_bad_example}
\end{figure}

\end{knitrout}
}

\newcommand{\StructureLimitationsB}{

\begin{knitrout}
\definecolor{shadecolor}{rgb}{0.969, 0.969, 0.969}\color{fgcolor}\begin{figure}[!h]

{\centering \includegraphics[width=0.980\linewidth,height=0.666\linewidth]{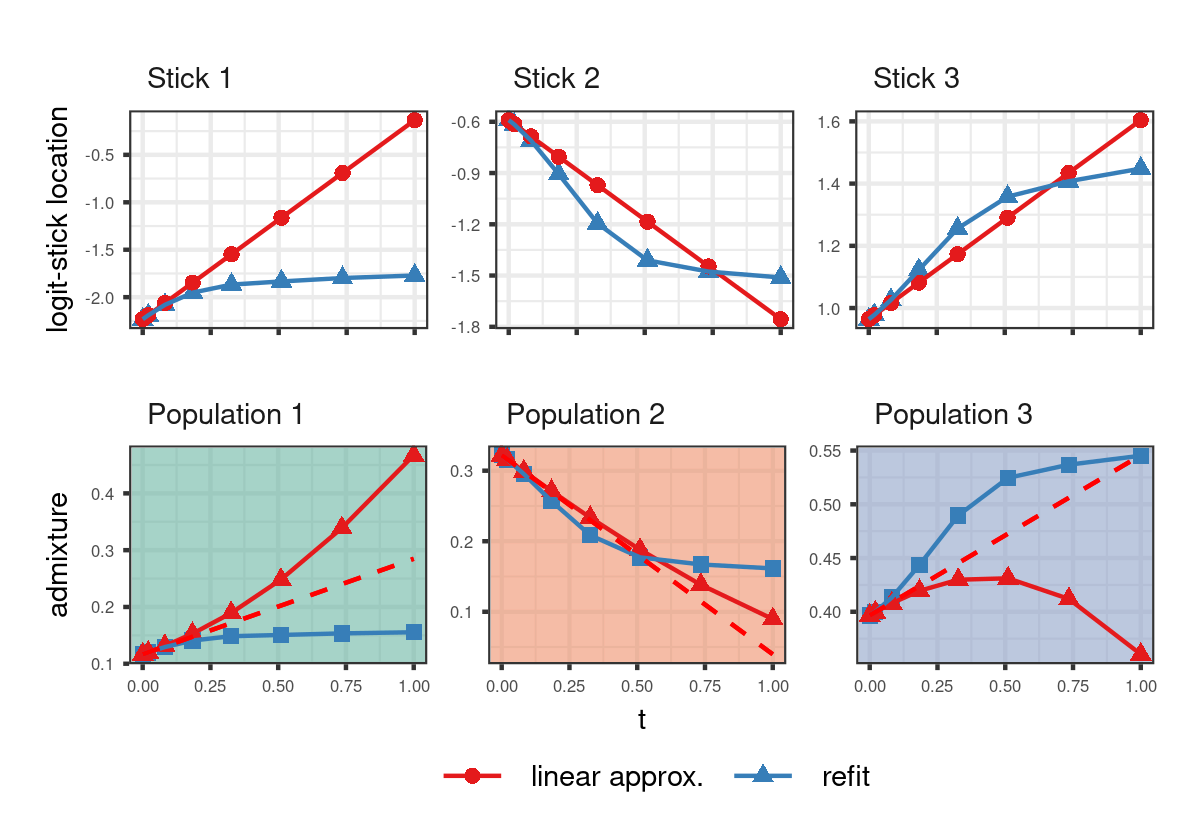} 

}

\caption[An example where
    linearizing the posterior quantity itself outperforms
    linearizing the variational parameters only.
    Shown are logit-stick location parameters (top row) and
    inferred admixtures (bottom row)
    for individual $n = 74$ and populations $k = 1, 2$ and $3$.
    Dashed red is the approximation $\glin(\t)$ formed by linearizing the
    inferred admixture $\expect{\q}{\pi_{\n\k}}$ with respect to prior
    parameter $t$.
    On the admixture proportion of population 3,
    $\glin(\t)$ outperforms $\g(\etalin(\t))$ (solid red)]{An example where
    linearizing the posterior quantity itself outperforms
    linearizing the variational parameters only.
    Shown are logit-stick location parameters (top row) and
    inferred admixtures (bottom row)
    for individual $n = 74$ and populations $k = 1, 2$ and $3$.
    Dashed red is the approximation $\glin(\t)$ formed by linearizing the
    inferred admixture $\expect{\q}{\pi_{\n\k}}$ with respect to prior
    parameter $t$.
    On the admixture proportion of population 3,
    $\glin(\t)$ outperforms $\g(\etalin(\t))$ (solid red). }\label{fig:stru_fully_lin_example}
\end{figure}

\end{knitrout}
}

\newcommand{\StructureNClusters}{

\begin{knitrout}
\definecolor{shadecolor}{rgb}{0.969, 0.969, 0.969}\color{fgcolor}\begin{figure}[!h]

{\centering \includegraphics[width=0.980\linewidth,height=0.431\linewidth]{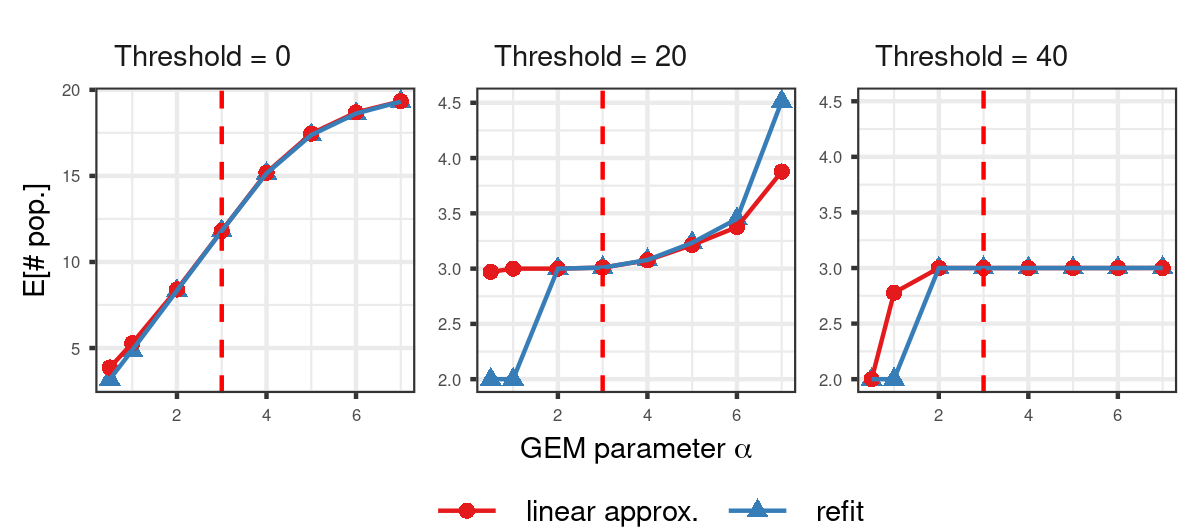} 

}

\caption[The expected number of (thresholded)
    populations in the thrush data as $\alpha$ varies.
    We computed the linear approximation at $\alpha_0 = 3$, and
    we compare the
    results under the linearly approximated variational parameters with the
    results observed after refitting.
    Thresholds at $\tau = 20$ and $\tau = 40$ corresponding to
    approximately
    $2\%$ and $4\%$
    of the total number of loci in the data set, respectively]{The expected number of (thresholded)
    populations in the thrush data as $\alpha$ varies.
    We computed the linear approximation at $\alpha_0 = 3$, and
    we compare the
    results under the linearly approximated variational parameters with the
    results observed after refitting.
    Thresholds at $\tau = 20$ and $\tau = 40$ corresponding to
    approximately
    $2\%$ and $4\%$
    of the total number of loci in the data set, respectively. }\label{fig:stru_alpha_nclusters}
\end{figure}

\end{knitrout}
}

\newcommand{\StructureClusterWeights}{

\begin{knitrout}
\definecolor{shadecolor}{rgb}{0.969, 0.969, 0.969}\color{fgcolor}\begin{figure}[!h]

{\centering \includegraphics[width=0.980\linewidth,height=0.588\linewidth]{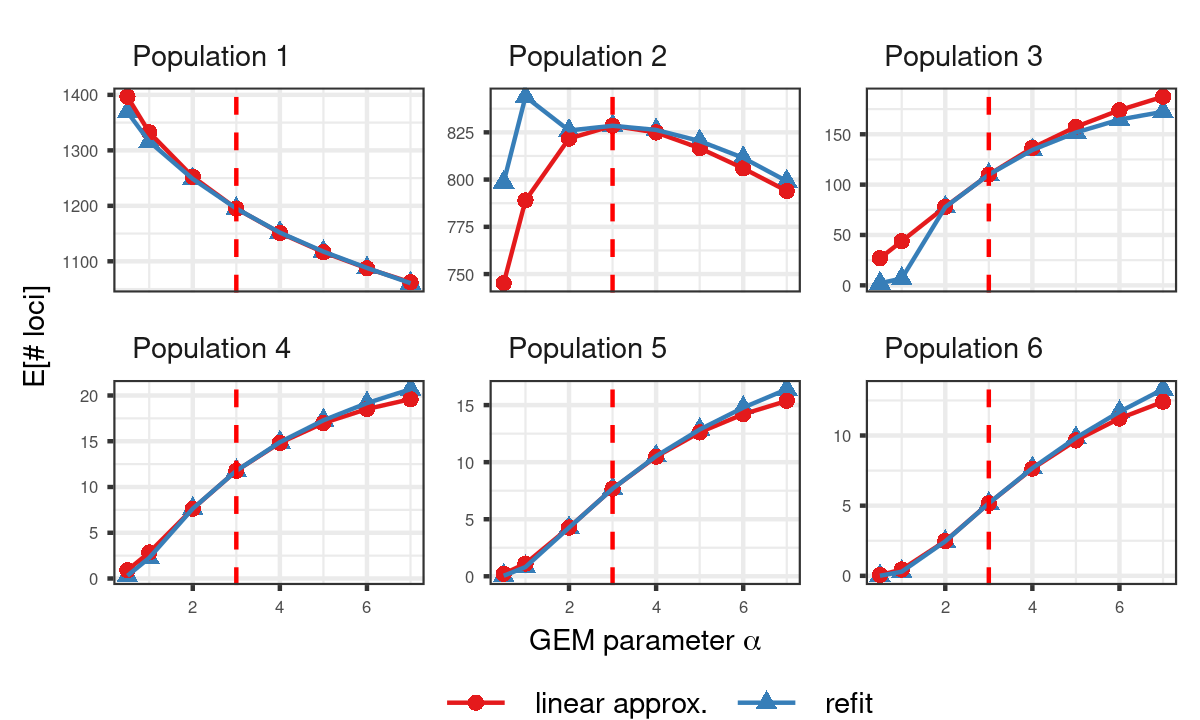} 

}

\caption[The expected number of loci per population as $\alpha$ varies]{The expected number of loci per population as $\alpha$ varies. }\label{fig:stru_alpha_cluster_weights}
\end{figure}

\end{knitrout}
}

    \subsection{The Beta prior}
    \applabel{app_beta_prior}
    In the iris experiment, we considered $\alpha\in[0.1, 4.0]$.
Over this range, the shape of
the $\betadist{1, \alpha}$ stick-breaking density varies considerably, as shown in
\figref{beta_priors}.

\BetaPriorsEx

To help us understand the effect of the concentration parameter
$\alpha$ we often use the following fact. Under the $\gem$ prior, the {\em a priori} expected
number of distinct clusters in a dataset of size $N$ is given by
\begin{align}\eqlabel{prior_num_clusters}
\expect{\p(\z \vert \pi)\p(\pi \vert \alpha)}{\nclusters(\z)} =
\sum_{\n = 1}^\N \frac{\alpha}{\alpha + \n - 1}.
\end{align}
See \citet{blackwell:1973:polyaurn} and \citet[Equation 11]{Teh:2010:dp}.

    \subsection{Gaussian mixture modeling on iris data}
    \applabel{app_iris}
    The observations are vectors $\x_\n \in \mathbb{R}^\d$, and we model each
component with a multivariate Gaussian. In this model, $\beta_\k = (\mu_k,
\Lambda_\k)$, where $\mu_\k \in \mathbb{R}^\d$, $\Lambda_\k$ is a $\d\times\d$
positive definite information matrix, and
\begin{align*}
\p(\x_\n \vert \beta_\k) ={}& \normdist{\x_n \vert \mu_\k, \Lambda_\k^{-1}} \\
\log\p(\x_\n \vert \beta_\k) ={}&
    -\frac{1}{2}(\x_n - \mu_k)^T \Lambda_\k (\x_n - \mu_k)
    + \frac{1}{2} \log |\Lambda_\k| + \const.\\
    & \constdesc{\beta_\k}
\end{align*}
We let $\pbetaprior(\beta_\k)$ be the conjugate prior, which in this case is normal-Wishart:
\begin{align*}
  \pbetaprior(\beta_\k) &= \normalwishart{\beta_\k \vert \tau_0, n_0, p_0, V_0}\\
  \log\pbetaprior(\beta_\k) &=
      -\frac{\tau_0}{2}(\mu_\k - \mu_0)^T \Lambda_\k (\mu_\k - \mu_0)\\
      &{} + \frac{n_0 - p_0 - 1}{2} \log |\Lambda_\k| -
      \frac{1}{2} \textrm{Tr}(V_0 \Lambda_\k) + \const,
\end{align*}
where $(\tau_0, n_0, p_0, V_0)$ are fixed prior parameters.

In this model, the conditionally conjugate variational distribution on $\beta_\k$ is
normal-Wishart, which we denote as
$\q(\beta_\k \vert \eta) = \normalwishart{\beta_k \vert \eta_{\beta_\k}}$,
with $\eta_{\beta_\k}$ the normal-Wishart parameters.
The conditionally conjugate variational distribution on $\z$ are multinomial.

\Figref{iris_fit} shows the
inferred clustering for $\alpha_0 = 2$,
which recovers that there are three iris species.

\IrisInitFit

    \subsection{Regression mixture modeling}
    \applabel{app_mice}
    \noindent \textbf{The data.}
The data come from a publicly available data set of mice gene expression
\citep{shoemaker:2015:ultrasensitive}.
Our analysis focuses on mice treated with the ``A/California/04/2009'' strain.
We normalize the data as described in
\citet{shoemaker:2015:ultrasensitive} and then apply the differential
analysis tool EDGE \citep{Storey:2005:significance} to rank the genes from most to least significantly differentially expressed.
We run our analysis on the top $\ngenes = 1000$ genes.

The left plot of \figref{example_genes}
shows the measurements of a single gene over time.
We model each gene as belonging to a latent component,
where each component defines a smooth expression curve over time.
Then, observations are drawn by adding i.i.d.\ noise to the smoothed
curve along with a gene-specific offset.

\MiceExampleGenes

\noindent \textbf{The B-spline basis.}
Notice from \figref{example_genes}, which shows an example time-course for a single gene,
that the time points are unevenly spaced, with more frequent observations at the beginning.
Following \citet{Luan:2003:clustering} we use cubic B-splines to smooth the time course expression data.
Specifically, we model the first 11 time points using
cubic B-splines with 7 degrees of freedom.
For the last three time points, $\timeindx = 72, 120, 168$ hours,
we use indicator functions.
That is, if $\tilde \regmatrix$ is the design
matrix where each column is a
B-spline basis vector evaluated at the $\ntimepoints$ measurement times,
we append to $\tilde \regmatrix$ three additional columns:
in these columns, entries are 1
if $\timeindx = 72, 120,$ or 168, receptively, and 0 otherwise.
The resulting matrix is the full design matrix $\regmatrix$.
We use indicators for the last three time points for numerical stability;
without the indicator columns,
the matrix $\tilde \regmatrix^T \tilde \regmatrix$ is nearly singular
because the later time points are more spread out.
The left column of \figref{example_genes} shows our basis functions.

\noindent \textbf{The generative model.}
\eqref{mice_model} gives the per-component conditional likelihood.
We use a normal prior for the shifts $\b_\n$,
a multivariate normal prior for the coefficients $\mu_\k$,
and a gamma prior for the inverse variance $\tau_\k$.
The prior on the mixture weights $\pi$ are constructed using the stick-breaking
construction in the main-text, and the cluster assignments $\z_\n$
are drawn from a multinomial with wieghts $\pi$, as usual.

\noindent \textbf{The variational approximation.}
The variational approximation, factorizes as
\begin{align*}
\q(\zeta \vert \eta) =
    \left( \prod_{\k=1}^{\kmax - 1} \q(\nuk \vert \eta) \right)
    \left( \prod_{\k=1}^{\kmax} \q(\beta_\k \vert \eta) \right)
    \left( \prod_{\n=1}^{\N} \q(\z_{\n} \vert \eta)
    \q(\b_{\n} \vert \z_{\n}, \eta)\right).
\end{align*}
Note that the variational distribution for $\b_\n$ conditions on $\z$.
We set $\q(\b_{\n} \vert \z_{\n} = k, \eta)$ to be Gaussian
with variational parameters dependent on $\k$.
For simplicity in this application,
we let $\q(\beta_\k \vert \eta) = \delta (\beta_k \vert \eta)$,
where $\delta(\cdot \vert \eta)$ denotes a point mass at a parameterized location.

As discussed in \exref{qz_optimality},
the optimal distribution $\q(\z_\n\vert\eta)$ is multinomial whose parameters
can be set in closed form as a function of the global variational parameters only.
We allow the distribution of $\b_\n$ to depend on $\z_{\n\k}$ so that
the its optimal distribution can also be set in closed form as a function of
global parameters.

The optimal distribution $q(\b_\n\vert \z_{\n\k} = 1, \eta)$ is Gaussian,
\begin{align*}
q(\b_\n\vert \z_{\n\k} = 1, \eta) = \normdist{\b_\n \vert \hat\mu_{\b_{\n\k}}, \hat\sigma^2_{\b_{\n\k}}}.
\end{align*}
To define the optimal parameters $\hat\mu_{\b_{\n\k}}, \hat\sigma^2_{\b_{\n\k}}$, let
\begin{align*}
  \rho^{(1)}_{\n\k} &= \expect{\q(\beta_k|\eta)}{\sum_{m=1}^\ntimepoints \tau_{k}(x_{nm} - \regmatrix_m\mu_\k)} +
  \tau_0 \mu_0 \\
  \rho^{(2)}_{\n\k} &= \ntimepoints \expect{\q(\beta_k|\eta)}{\tau_{k}} + \tau_0,
\end{align*}
where $\mu_0$ and $\tau_0$ are the prior mean and information on $\b_n$, respectively.

The optimal parameters for the Gaussian distribution on $\b_\n$ are given by
\begin{align*}
  \hat\mu_{\b_{\n\k}} &= \rho^{(1)}_{\n\k} / \rho^{(2)}_{\n\k}\\
  \hat\sigma^2_{\b_{\n\k}} &= 1 / \rho^{(2)}_{\n\k}.
\end{align*}

\figref{gene_centroids} shows the inferred smoothers
$\regmatrix \expect{\q}{\mu_\k}$ for selected clusters.
\MiceSmoothers


    \subsection{fastSTRUCTURE}
    \applabel{app_structure}
    The generative process was described in the main text (\secref{results_structure}).
We detail here the variational approximation.
Like in all our examples, the variational distribution is mean-field:
\begin{align*}
\q(\zeta \vert \eta) =
    \left(
    \prod_{\n=1}^{\nindiv}\prod_{\k=1}^{\kmax - 1}
    \q(\nu_{nk} \vert \eta) \right)
    \left(\prod_{\k=1}^{\kmax}\prod_{l=1}^{\nloci}
    \q(\latentpop_{\k l} \vert \eta) \right)
    \left( \prod_{\n=1}^{\N} \prod_{l=1}^{\nloci} \prod_{i=1}^{2} \q(\z_{\n l i} \vert \eta) \right).
\end{align*}
We let all distributions be conditionally conjugate except for the sticks,
which are logit-normal.
Each membership indicator $\z_{\n l i}$ is categorical, and the
allele frequencies $\latentpop_{\k l}$ are Dirichlet distributed.

In this model, we still call $(\beta, \nu)$ the global latent variables, even though they scale
with the number of individuals $\N$;
they do not, however, scale with both the number of individuals and the number of loci
like $\z$ does. Thus, we call $\z$ the local latent variables.
The local variational parameters $\eta_\z$ can be set optimally in
an analagous way as \exref{qz_optimality}, except with the
indices $\n\k$ replaced with $\n\l\i\k$.

The posterior quantities of interest in this application are the admixtures
$\pi_\n$. \figref{stru_init_fit} plots the inferred admixtures
$\expect{\q(\pi_\n \vert \etaopt)}{\pi_\n}$ for all individuals $\n$.

In the approximate posterior with $\alpha_0 = 3$, there appear to be three dominant
latent populations, which we arbitrarily label as populations 1, 2, and 3
(\figref{stru_init_fit}). The inferred admixture proportions generally
correspond with geographic regions: Mbololo individuals are primarily population
1; Ngangao individuals are primarily population 2; and Chawia individuals are a
mixture of populations 1, 2, and 3.

\StructureInitialFit

\section{fastSTRUCTURE Supplemental Results}\applabel{app_structure_results}

\subsection{The expected number of populations}
One posterior quantity of interest is the expected number of
in-sample populations. Define
\begin{align*}
\gclusters(\eta)
&= \expect{\q(\z\vert\eta)}{\sum_{\k=1}^\kmax \ind{
\left(\sum_{\n=1}^{\nindiv}
\sum_{\l=1}^{\nloci}
\sum_{\i=1}^2
\z_{\n \l \i \k}\right) > \tau}},
\end{align*}
which is the expected number of populations in the data set that contains at least $\tau$ loci.
We allow the option of setting $\tau > 0$ in order to count only the populations that comprise a non-negligible fraction of the data set.

The expected number of latent populations is sensitive to $\alpha$
(\figref{stru_alpha_nclusters}). Without any thresholding ($\tau = 0$), the
expected number of populations quickly increases as $\alpha$ increases; in fact,
it nearly saturates at $\kmax = 20$ when $\alpha = 7$. This sensitivity is
due to the fact that the non-thresholded quantity is highly dependent on
the behavior of small, nearly unoccupied populations; even though the
probability of a single locus belonging to these rare populations is small, the
probability that \textit{none} of the $\nindiv \times \nloci \times 2$ observed
genotypes belong to these rare populations is non-negligible.

This motivates the use of thresholding in reporting the number of populations.
We consider two thresholds, $\tau = 20$ and $\tau = 40$, corresponding to
approximately $2\%$ and $4\%$ of the total number of loci in the data set,
respectively. The thresholded estimates for the number of populations is still
moderately sensitive to the value of $\alpha$. When refitting the variational
approximation at $\alpha = 0.5, 1, \ldots, 7$, the thresholded quantities vary
between two and four latent populations.

The linearized variational parameters $\etalinglobal(\t)$ imperfectly captures
the results observed by refitting.
The linearized parameters and
the refitted parameters almost perfectly agree on values of $\gclusters$ with
$\tau = 0$. However, when $\tau = 20$, the linearized parameters underestimated
the true sensitivity of $\gclusters$ found by refitting. In particular, the
linearized parameters failed to produce the reduction to two latent populations
at $\alpha = 0.5$ observed in the refits.

\StructureNClusters

We provide some more intuition concerning the thresholded estimate for the number of populations.
The posterior quantity $\gclusters$ is closely related to the expected number of loci belonging to each population,
defined as
\begin{align*}
\gloci(\eta; k)
&= \expect{\q(\z\vert\eta)}{\sum_{n=1}^{\nindiv}
\sum_{l=1}^{\nloci}
\sum_{i=1}^2
\z_{\n l i \k}}.
\end{align*}

\figref{stru_alpha_cluster_weights} plots $\gloci$ for the first six populations
as $\alpha$ varies.
The expected number of loci at the initial fit, $\gloci(\etaopt(\alpha_0); k)$,
is at least 100 for populations $\k = 1, 2,$ and $3$ and
less than 15
for the remaining populations.
A sample of assignments $\z\sim \q(\z\vert\etaopt(\alpha_0))$ will
almost always have at least $\tau$ loci allocated to populations 1, 2, and 3,
while the allocations to each remaining population will almost always be below $\tau$, for either $\tau = 20$ or $\tau = 40$.
Thus, at $\alpha = \alpha_0$ there then are clearly 3 populations by our definition of $\gclusters$, for either $\tau$.

At $\alpha = 7$, the expected number of loci belonging to population 4
increases to approximately 20,
and a new population emerges above the threshold at $\tau = 20$.
Both the linearized and the refitted variational parameters agree on this shift in allocation to population 4.
On the other hand, under the refitted variational parameters at $\alpha = 0.5$,
the expected number of loci belonging to population 3 decreases to
seven,
below the threshold $\tau = 20$.
Thus, the expected number of latent populations with allocations above
the threshold $\tau = 20$ decreases to two.
The linearized parameters
under-estimated this decrease in allocation to population 3, and
therefore continued to estimate three latent populations even at $\alpha = 0.5$.

\StructureClusterWeights

\subsection{Limitations of local sensitivity}
Recall from \secref{results_structure} and \figref{stru_func_sens_admix}
that the linear approximation
failed to capture the change in the admixture proportion of an individual,
$n = 25$ after a worst-case functional perturbation.

\figref{stru_lin_bad_example} examines individual $n = 25$ more closely.
The bottom row plots this individual's
admixture proportions as $\t$ varies from 0 to 1 in the perturbed prior
$\p(\nu\vert \t) = \p_0(\nuk)\exp(\t\phiworstcase(\nuk))$.
The linearized parameters poorly captured the change in admixture proportions observed after refitting, particularly
for populations 1 and 2, for values of $\t$ close to 1.
Even though we retain non-linearities
in the mapping from variational parameters to the posterior statistic,
for this perturbation, the mapping from prior parameter
$\t$ to the relevant variational parameters
is highly non-linear.
This latter mapping is what we linearize
and what causes our approximation to fail in this case.
Specifically, the variational location parameter on the first stick-breaking proportion is concave as a function of $\t$ --
the location parameter increases for small $\t$,
then decreases as $\t\rightarrow1$.
However,
$\etalin(\t)$ linearizes the relationship between the location parameter and $\t$.
Therefore, the corresponding admixture mixture proportion of
population 1 is over-estimated under the linearized variational parameters.
Furthermore, because our linearized variational parameters
over-estimated the length of the first stick,
and the second admixture proportion is a product of the
remaining stick times the second stick-breaking proportion,
the linearized variational parameters then under-estimates
the admixture proportion of population 2.

\StructureLimitationsA{}

\figref{stru_fully_lin_example} shows a similar situation for individual $n = 74$.
The linearized variational parameters grossly over-estimated the length of the first stick,
resulting in the later admixture proportions being under-estimated.
The third admixture proportion was particularly poorly approximated under the linearized variational parameters.
Given the recursive nature of the relationship between admixtures and stick-breaking proportions, errors at early sticks affect later admixture proportions.
Fully linearizing the mapping $\t\mapsto\g(\etaopt(\t))$ to form the approximation
$\glin(t)$ avoids this problem.
In this example, $\glin(t)$ outperforms $\g(\etalin(t))$, with $\g$ being the admixture proportion of population 3.
In our experience,
computing $\g(\etalin(t))$, and thus retaining non-linearities in the mapping from $\eta\mapsto\g(\eta)$,
is usually beneficial to the quality of the approximation.
It is likely that $\g(\etalin(t))$ outperforms $\glin(\t)$ for most posterior quantities,
though as we see in \figref{stru_fully_lin_example},
this is not guaranteed to always be true.

\StructureLimitationsB{}

\end{appendix}

\end{document}